\documentclass[aps,12pt,tightenlines,amsmath,amssymb,showpacs]{revtex4}
\usepackage{graphicx}
\usepackage{dcolumn}
\usepackage{bm}

\begin{document}

\title{Relevance of axion-like particles for very-high-energy astrophysics}

\author{Alessandro De Angelis}
\affiliation{Dipartimento di Fisica, Universit\`a di Udine, Via delle Scienze 208, I -- 33100 Udine, and INAF, and INFN,  Gruppo Collegato di Udine (Sezione di Trieste), Italy}
\email{deangelis.alessandro@gmail.com}

\author{Giorgio Galanti}
\affiliation{Dipartimento di Fisica, Universit\`a dell'Insubria, Via Valleggio 11, I -- 22100 Como, Italy}
\email{giorgio_galanti@libero.it}

\author{Marco Roncadelli}
\affiliation{INFN, Sezione di Pavia, Via A. Bassi 6, I -- 27100 Pavia, Italy}
\email{marco.roncadelli@pv.infn.it}

\begin{abstract}

{  Several} extensions of the Standard Model {  and in particular superstring theories suggest} the existence of axion-like particles (ALPs), which are very light spin-zero bosons with a two-photon coupling. As a consequence, photon-ALP oscillations occur in the presence of an external magnetic field, and ALPs can lead to observable effects on the measured photon spectrum of astrophysical sources. An intriguing situation arises when blazars are observed in the very-high-energy (VHE) band -- namely above $100 \, {\rm GeV}$ -- as it is the case with the presently operating Imaging Atmospheric Cherenkov Telescopes (IACTs) H.E.S.S., MAGIC, CANGAROO III and VERITAS. The extragalactic background light (EBL) produced by galaxies during cosmic evolution gives rise to a source dimming which becomes important in the VHE band and increases with energy, since hard photons from a blazar scatter off soft EBL photons thereby disappearing into $e^+ e^-$ pairs. This dimming can be considerably reduced by photon-ALP oscillations, and since they are energy-independent the resulting blazar spectra become harder than expected. We consider throughout {  a scenario first proposed by De Angelis, Roncadelli and Mansutti -- to be referred to as  DARMA for short --} in which the above strategy is implemented with photon-ALP oscillations triggered by large-scale magnetic fields, and we systematically investigate its implications for VHE blazars. We find that for ALPs lighter than $5 \cdot 10^{- \, 10} \, {\rm eV}$ the photon survival probability is larger than predicted by conventional physics above a few hundred GeV. Specifically, a boost factor of 10 can easily occur for sources at large distance and large energy, e.g. at $8 \, {\rm TeV}$ for the blazar 1ES 0347-121 at redshift $z = 0.188$. This is a clear-cut prediction which can be tested with the planned Cherenkov Telescope Array and the HAWC water Cherenkov $\gamma$-ray observatory, and possibly with the currently operating IACTs as well as with detectors like ARGO-YBJ and MILAGRO. Moreover, we show that the DARMA scenario offers a new interpretation of the VHE blazars detected so far, according to which the large spread in the values of the observed spectral index is mainly due to the wide spread in the source distances rather than to large variations of their internal physical properties. {  Finally, we stress that ALPs with the right properties to produce the above effects can be discovered by the GammeV experiment at FERMILAB and more likely with the planned photon regeneration experiment ALPS at DESY.}

\end{abstract}


\maketitle


\section{INTRODUCTION}

{  The Standard Model (SM) of strong, weak and electromagnetic interactions with three sequential families of quarks and leptons provides a satisfactory description of the experimental results concerning elementary-particle physics concerning elementary-particle physics at energies up to about the Fermi scale of weak interactions $G_F^{- 1/2} \simeq 250 \, {\rm GeV}$.}

Yet, nobody would seriously regard the SM as the ultimate theory of fundamental processes. Apart from more or less aesthetic reasons and {  actual 
problems like the $g - 2$ value of the muon and the top forward-backward asymmetry} such an expectation is made compelling by the observational evidence for non-baryonic dark matter ultimately responsible for the formation of structure in the Universe as well as for dark energy presumably triggering the present accelerated cosmic expansion.

So, the SM is presently viewed as the low-energy manifestation of some more fundamental and complete theory of all elementary-particle interactions including gravity. Every specific approach to extend the SM in such a way is characterized by a set of new particles along with their mass spectrum and their interactions with the standard world. 

Although it is presently impossible to tell which proposal out of so many ones has any chance to successfully describe Nature, it looks remarkable that attempts along very different directions such as four-dimensional ordinary and supersymmetric models~\cite{susy}, Kaluza-Klein theories~\cite{kaluzaklein, turok} and especially superstring theories~\cite{witten, ringwald, axiverse} all {  suggest the existence} of {\it axion-like particles} (ALPs)~\cite{alprev}. {  A general argument supporting this conclusion will be given in Section II.}

ALPs are very light pseudo-scalar spin-zero bosons $a$ characterized by a two-photon coupling $a\gamma\gamma$. As the name itself suggests, they are a  sort of generalization of the axion, the pseudo-Goldstone boson associated with the Peccei-Quinn symmetry proposed as a natural solution to the strong CP problem~\cite{peccei, WW, invisible, axionrev}. But while the axion enjoys a strict relationship between its mass and the $a\gamma\gamma$ coupling constant, these two parameters are to be regarded as unrelated for ALPs. In fact, depending on the actual values of their mass and $a\gamma\gamma$ coupling constant, ALPs can play an important role in cosmology, either as cold dark matter particles~\cite{cdm} or as quintessential dark energy~\cite{carroll}. 

A remarkable consequence of the $a\gamma\gamma$ coupling is the phenomenon of photon-ALP mixing, which takes place in the presence of an external electromagnetic field and leads to two distinct effects. One is photon-ALP oscillations~\cite{sikivie1983, raffeltstodolsky}, which is quite similar to the 
oscillations of massive neutrinos with different flavours. The other consists in the change of the polarization state of photons traveling in a magnetic field~\cite{mpz, raffeltstodolsky}. 

It turns out that ALPs are extremely elusive in high-energy experiments and the only way to look for them in the laboratory requires very careful polarimetric measurements to be carried out on a laser beam~\cite{pvlas} or alternatively a photon regeneration experiment to be performed~\cite{wall}. Successful detection of ALPs in present-day experiments of this kind is possible in either case for a fairly large $a\gamma\gamma$ coupling. 

Astrophysical manifestations of ALPs appear the best way to discover their existence, since they can give rise to observable effects even for values of the 
$a\gamma\gamma$ coupling constant much smaller than those tested so far in laboratory experiments. Indeed, it is known since a long time that for values of the $a\gamma\gamma$ coupling constant that look hopelessly small to be probed today in the laboratory the stellar evolution would be dramatically altered~\cite{raffeltbook} and this fact sets a strong upper bound on the coupling in question, which is consistent with the negative result of the CAST experiment at CERN~\cite{castcern}.

In the last few years it has been realized that photon-ALP oscillations triggered by intervening cosmic magnetic fields along the line of sight can produce detectable effects in observations of bright $X$-ray and $\gamma$-ray sources~\cite{peloso, fair, dimming, darma, dmpr, bis, Mirizzi2006zy, prada, br, mm, bmr2010}. The effect becomes larger as the distance of the sources increases: blazars, which are Active Galactic Nuclei (AGN) with the beam pointing towards us, constitute the most distant long-lasting gamma-ray sources observed, and are thus the most obvious case study. In order to bring out most simply the relevance of ALPs in the present context, we neglect cosmological effects at this introductory level.

\begin{figure}[h]
\centering
\includegraphics[width=.60\textwidth]{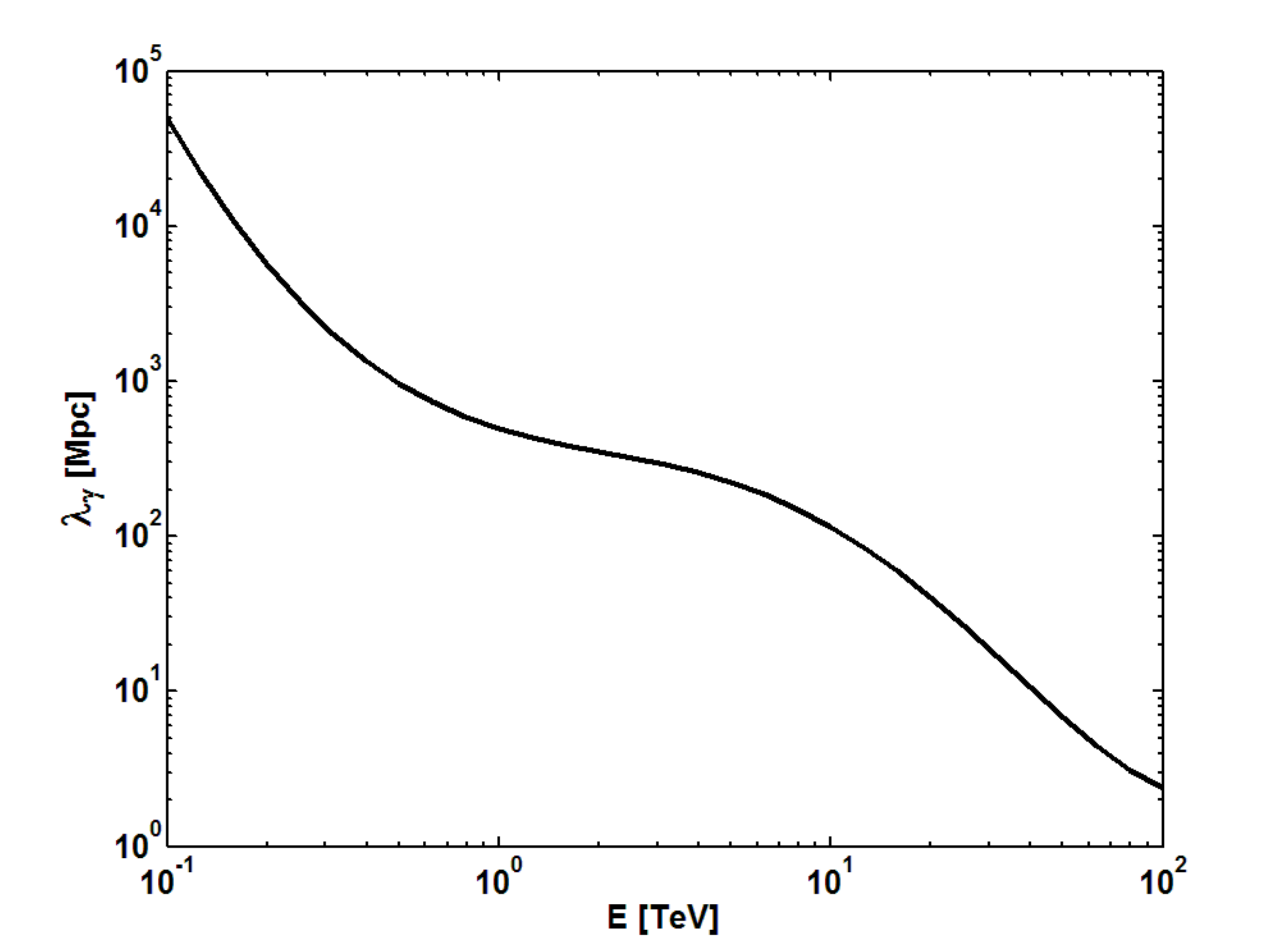}
\caption{\label{fig:france} 
The pair-production mean free path ${\lambda}_{\gamma}$ of a VHE photon is plotted versus its energy $E$ within the EBL model of FRV. Only conventional physics is assumed and in particular the possibility of photon-ALP oscillations is ignored.}
\end{figure}

The mean free path of very-high-energy (VHE) photons -- namely with energy above $100 \, {\rm GeV}$ -- is limited by their interaction with background photons in the Universe through the process $\gamma \gamma \to e^+ e^-$. A high energy photon from a distant blazar has a nonnegligible probability to scatter off background photons in the optical/near infrared band permeating the Universe -- the so-called {\it extragalactic background light} (EBL) -- thereby disappearing into an $e^+ e^-$ pair~\cite{stecker1971,aharonianbook}. The VHE photon mean free path depends on the EBL density. Recently it has become possible to model accurately such density: several parametrizations are available, essentially coincident, and throughout this paper we employ the recent EBL model of Franceschini, Rodighiero and Vaccari (FRV)~\cite{franceschini}). Figure \ref{fig:france} shows the pair-production mean free path ${\lambda}_{\gamma}$ of a VHE photon as a function of its energy $E$ within the FRV model.

The effects of the photon-ALP oscillations on the gamma yield  from blazars can be summarized as follows:
\begin{itemize} 
\item For $E < 100 \, {\rm GeV}$ we infer from Figure \ref{fig:france} that $\lambda_{\gamma} (E)$ is comparable with the Hubble radius, and so EBL absorption is negligible. In such a situation photon-ALP oscillations can only give rise to a source dimming above a certain energy threshold $E_*$. Hence a characteristic {\it distortion} of the source spectrum around $E_*$ is the observable prediction, which can be searched for with the {\it Fermi}/LAT mission if $E_*$ happens to lie in the instrument energy range, namely for $30 \, {\rm MeV} < E_* <  300 \, {\rm GeV}$~\cite{dimming}.
\item For $E > 100 \, {\rm GeV}$ Figure \ref{fig:france} shows that EBL effects become important since now $\lambda_{\gamma} (E)$ quickly decreases as 
$E$ increases. Once emitted, photons can convert into ALPs and next reconvert back into photons before reaching the Earth.  A possibility is that photon-ALP oscillations take place in intergalactic space and the resulting scenario has been called DARMA {  (acronym for De Angelis, Roncadelli and Mansutti)}~\cite{darma, dmpr, mm}. Alternatively, the $\gamma \to a$ conversion can occur inside the blazar while the $a \to \gamma$ reconversion can happen in the Milky Way~\cite{bis}. Of course, also both options can be realized~\cite{prada}. In all cases, photons acquire a split identity, travelling for some time as real photons and for some time as ALPs. However, they suffer EBL absorption {\it only} when they are real photons, which means that the effective photon mean free path $\lambda_{\gamma, {\rm eff}} (E)$ is actually {\it larger} than $\lambda_{\gamma} (E)$ as predicted by conventional physics (see Figure \ref{fig:france}). Since the photon survival probability depends exponentially on minus the optical depth -- which in turn goes like the source distance divided by $\lambda_{\gamma, {\rm eff}} (E)$ -- even a slight increase of $\lambda_{\gamma, {\rm eff}} (E)$ with respect to $\lambda_{\gamma} (E)$ produces a {\it substantial enhancement} of the photon survival probability and so of the observed flux. Note that for a given detector sensitivity a larger photon survival probability just means that a larger distance can be probed, so that the VHE Universe becomes more transparent than generally believed. Actually, since the EBL absorption increases with energy whereas the photon-ALP oscillation probability is energy-independent, the observed flux enhancement gets larger and larger as the energy increases. As a consequence, the observed spectra are harder than currently expected.
\end{itemize} 

Thus -- depending on the values of the free parameters -- a {\it hardening} of the observed blazar spectra is the main prediction of photon-ALP oscillations concerning the VHE band between $100 \, {\rm GeV}$ and $100 \, {\rm TeV}$, which can be probed by Imaging Atmospheric Cherenkov Telescopes (IACTs). More in detail, the presently operating IACTs H.E.S.S., MAGIC, CANGAROO III and VERITAS can reach up $\sim 20 \, {\rm TeV}$ with difficulty, whereas the planned Cherenkov Telescope Array (CTA) and the HAWC water Cherenkov $\gamma$-ray observatory will be able to explore the whole VHE band with a much larger sensitivity. We should add that the VHE range can also be analyzed by other available detectors, like the Extensive Air Shower arrays ARGO-YBJ and MILAGRO.

The aim of the present paper is to investigate the DARMA scenario in great detail by employing the EBL model of FRV~\cite{franceschini}, and systematically working out all its implications for VHE blazar observations. Everything is calculated starting from first principles. 

More specifically, our primary goal is to evaluate the photon survival probability within the DARMA scenario, which allows in turn to quantify the resulting hardening of VHE blazar spectra and to determine how much the Universe becomes more transparent to VHE photons than usually thought. 

We find that these effects can be quite substantial for the free parameters in allowed realistic ranges provided that the ALP mass satisfies the condition $m < 5 \cdot 10^{- 10} \, {\rm eV}$. Actually, our prediction can become spectacular above an energy threshold which is well below the upper detection limit of the CTA and the HAWC observatory. Denoting by $E_{10}$ the energy at which the photon survival probability is 10 times larger than that dictated by conventional physics, $E_{10}$ turns out to decrease with the source redshift $z$. In the most favourable case, we get $E_{10} \simeq 30 \, {\rm TeV}$ for Mrk 421 at $z = 0.030$, $E_{10} \simeq 8  \, {\rm TeV}$ for 1ES 0347-121 at $z = 0.188$, $E_{10} \simeq 2 \, {\rm TeV}$ for 3C 66A at $z = 0.444$ and $E_{10} \simeq 2  \, {\rm TeV}$ for 3C 279 at $z = 0.536$. In addition, it follows from the above considerations that for energies larger than $E_{10}$ the photon survival probability gets even more enhanced.

In spite of the fact that only future observations can provide a clear-cut check of the DARMA scenario -- and so ultimately of the existence of an ALP lighter than $5 \cdot 10^{- 10} \, {\rm eV}$ -- it looks natural to inquire whether available data from IACTs contain some hints in favour of this scenario. As we shall see, the emitted as well as the observed spectra of VHE blazars detected so far have to a good approximation a {  single power-law behaviour~\cite{remste}, so} that they are characterized by their slope, which -- up to a minus sign -- is the emitted $\Gamma_{\rm em}$ and observed $\Gamma_{\rm obs}$ spectral index, respectively. So, we can rephrase the above conclusion by stating that for a fixed value of $\Gamma_{\rm em}$ pertaining to a given source the DARMA scenario predicts that under suitable conditions $\Gamma_{\rm obs}$ should be {\it smaller} than within conventional physics.

In 2006 the H.E.S.S. collaboration reported some evidence that the spectra of the two blazars  H 2356-309 and 1ES 1101-232 have $\Gamma_{\rm obs}$ smaller than expected and this fact was interpreted as strongly suggesting an EBL attenuation lower than currently believed~\cite{aharonian:nature06}. A similar conclusion emerged in 2007 with the discovery of the blazar 3C 279 by the MAGIC collaboration~\cite{3c}. Based on preliminary data~\cite{3cc} and a specific EBL model~\cite{kneiske2004}, in a previous paper~\cite{darma} two of us (A. D. A. and M. R. together with O. Mansutti) have shown for the first time that the mechanism of photon-ALP oscillations can substantially reduce the EBL attenuation for distant blazars and in particular that it can successfully explain the observed spectrum of 3C 279 for allowed realistic values of the free parameters. 

Subsequent developments have demonstrated that realistic EBL models account for VHE blazar observations without the need of any unconventional physics, provided that the large spread in the values of $\Gamma_{\rm obs}$ is fully traced back to an equally large spread in the values of 
$\Gamma_{\rm em}$. Further, far-away sources -- for which EBL absorption is a large effect -- turn out to have energy spectra similar to those of some nearby blazars, for which EBL attenuation is negligible. This means that for distant sources $\Gamma_{\rm em}$ has to be considerably smaller than for nearby ones. Even though a physical explanation for the occurrence of very small values of $\Gamma_{\rm em}$ has recently been achieved~\cite{steckershoks, costamante, bdf}, one is nevertheless led to the {\it cosmic opacity problem}, namely to wonder why these physical effects are important for distant blazars {\it only}. 

We show that within the DARMA scenario the situation is quite different. As a result of the competition between EBL attenuation and photon-ALP oscillations, two important conclusions emerge:
\begin{itemize}
\item The values of $\Gamma_{\rm em}$ for far-away VHE blazars are in the same ballpark of nearby ones, so that the cosmic opacity problem is solved.
\item The observed large spread in the values of $\Gamma_{\rm obs}$ arises mainly from the wide spread in the source distances while the required scatter in the values of $\Gamma_{\rm em}$ is small. 
\end{itemize}

Finally, we stress that ALPs with the right properties to produce the above effects {  can be discovered by the GammeV~\cite{gammaev} experiment at FERMILAB and more likely by the planned photon regeneration experiment ALPS at DESY~\cite{ringwaldfuture}} or with large xenon scintillation detectors developed for dark matter searches~\cite{avignone}. Thus, it looks amazing that the discovery of a new particle -- besides very important in its own right -- would also provide a sort of glasses that allow us to watch much farther out into the $\gamma$-ray Universe.

The paper is organized as follows. In Section II we review the motivation and the main properties of ALPs. Particular attention is paid to the propagation of a photon/ALP beam in the presence of a homogeneous magnetic field, a problem that can be solved exactly. Also the astrophysical and cosmological constraints on the ALP parameters are considered. Section III addresses in great detail the properties of VHE blazar spectra. Everything in this Section is discussed within conventional physics and at this stage ALPs are totally neglected. After a brief account of cosmic opacity from a microscopic point of view, various models of the EBL are cursorily described and compared. Then it is shown that conventional physics provides a correct qualitative understanding of the observed blazar spectra but at the same time it requires some degree of correlation between $\Gamma_{\rm em}$ and $z$ which gives rise to the cosmic opacity problem. It is also argued that a way out of this problem calls for an unconventional photon propagation, and various proposals are briefly reviewed. In Section IV the main conclusions drawn in Sections II and III are combined together to build up the DARMA scenario, which is investigated in detail. The phenomenon of photon-ALP oscillations is considered here in the presence of EBL absorption as well as for a domain-like configuration typical of large-scale magnetic fields. The photon survival probability is ultimately computed by a numerical code and it arises as an average over 5000 random realizations of the beam propagation from the source to us, each corresponding to randomly chosen directions of the magnetic field inside every domain. Sections V, VI and VII are devoted to the discussion of the implications of the DARMA scenario for VHE blazar observations. A detailed analysis of the behaviour of the photon survival probability for some representative values of the free parameters is presented in Section VI, where its relevance for future VHE blazar observations is stressed. Section VII offers a new interpretation of the observed VHE blazars, which solves the cosmic opacity problem and traces the large spread in the values of $\Gamma_{\rm obs}$ mainly to the wide spread in the source distances while the demanded scatter in the values of $\Gamma_{\rm em}$ is small. Finally, we offer our conclusions in Section VIII. A convenient method to solve the eigenvalue problem for a 2 by 2 matrix with complex coefficients is presented in Appendix A, whereas an approximate analytic evaluation of the optical depth within the FRV model of the EBL is reported in Appendix B.

\section{AXION-LIKE PARTICLES (ALPs)}

We review the conceptual motivations in favour of ALPs as well as their properties that are most relevant for our further needs. Natural Lorentz-Heaviside units with $\hbar=c=k_{\rm B}=1$ are employed throughout the paper unless otherwise stated.

\subsection{Motivation}

As already stressed, the Standard Model (SM) of particle physics is presently regarded as the low-energy manifestation of some more fundamental theory 
(FT) characterized by a very large energy scale $\Lambda \gg G_F^{- 1/2}$, with $G_F^{- 1/2} \simeq 250 \, {\rm GeV}$. We collectively denote by $\phi$ the SM particles together with possibly new undetected particles with mass smaller than $G_F^{- 1/2}$, while all particles much heavier than $G_F^{- 1/2}$ that are present in the FT are collectively represented by $\Phi$. Correspondingly, the FT is defined by a Lagrangian of the form ${\cal L}_{\rm FT} (\phi, \Phi)$ and the generating functional for the corresponding Green's functions reads
\begin{equation}
\label{t1}
Z_{\rm FT} [J,K] = N \int {\cal D} \phi \int {\cal D} \Phi \ {\rm exp} \left( i \int d^4x \, \Bigl[ {\cal L}_{\rm FT}  (\phi, \Phi) + 
\phi J+ \Phi K \Bigr] \right)~,
\end{equation}
where $J$ and $K$ are external sources and $N$ is a normalization constant. The resulting low-energy effective theory then emerges by integrating out the heavy particles in $Z_{\rm FT} [J,K]$, and so the low-energy effective Lagrangian ${\cal L}_{\rm eff} (\phi)$ is defined by
\begin{equation}
\label{t2}
{\rm exp} \left( i \int d^4x \, {\cal L}_{\rm eff} (\phi) \right) \equiv \int {\cal D} \Phi \, 
{\rm exp} \left( i \int d^4x \, {\cal L}_{\rm FT} (\phi, \Phi) \right)~.
\end{equation}
Evidently, the SM Lagrangian is contained in ${\cal L}_{\rm eff} (\phi)$, and -- in the absence of any new physics below $G_F^{- 1/2}$ -- it will differ from 
${\cal L}_{\rm eff} (\phi)$ only by non-renormalizable terms involving the $\phi$ particles alone, that are suppressed by inverse powers of $\Lambda$.

In any theory with a sufficiently rich gauge structure -- which is certainly the case of the FT -- some global symmetry ${\cal G}$ invariably shows up as an accidental consequence of gauge invariance. Since the Higgs fields which spontaneously break gauge symmetries carry nontrivial global quantum numbers, it follows that the group ${\cal G}$ undergoes spontaneous symmetry breaking as well. As a consequence, some Goldstone bosons  -- which are collectively denoted by $a$ if ${\cal G}$ is non-abelian -- are expected to appear in the physical spectrum and their interactions are described by the low-energy effective Lagrangian, in spite of the fact that ${\cal G}$ is an invariance of the FT. We stress that Goldstone bosons are necessarily pseudo-scalar particles~\cite{gelmini}. 

{  As far as our main line of development is concerned, the FT is supposed to describe quantum gravitational effects and it is a common lore that they always explicitly break global symmetries~\cite{kallosh}. In fact, this point can be understood in an intuitive fashion. Since black holes do not possess any definite global charges, global symmetries are violated in any scattering process involving black holes. So, we end up with the general conclusion that provided that the Lagrangian of the FT possesses some spontaneously broken global symmetry then pseudo-Goldstone bosons with mass much smaller than 
$G_F^{- 1/2}$ are necessarily present in the low-energy effective Lagrangian.} 

Therefore, by splitting up the set $\phi$ into the set of SM particles ${\phi}_{\rm SM}$ plus the pseudo-Goldstone bosons $a$, the low-energy effective Lagrangian has the structure
\begin{equation}
\label{t2231209}
{\cal L}_{\rm eff} (\phi_{\rm SM}, a) = {\cal L}_{\rm SM} ({\phi}_{\rm SM}) + {\cal L}_{\rm nonren} ({\phi}_{\rm SM}) + {\cal L}_{\rm ren} (a) + 
{\cal L}_{\rm ren} ({\phi}_{\rm SM}, a) + {\cal L}_{\rm nonren} ({\phi}_{\rm SM}, a)~,
\end{equation}
where ${\cal L}_{\rm ren} ({\phi}_{\rm SM}, a)$ stands for renormalizable soft-breaking terms that can be present whenever ${\cal G}$ is not an automatic symmetry of the low-energy effective theory~\cite{automatic}.

Needless to say, it can well happen that between $G_F^{- 1/2}$ and ${\Lambda}$ other relevant mass scales ${\Lambda}_1$, ${\Lambda}_2$, ... exists. 
In such a situation the above scheme remains true, but then ${\cal G}$ may be spontaneously broken at such an intermediate scale.

{  Finally, we would like to stress that a very thoroughly analysis by Arvanitaki {\it et al.}~\cite{axiverse} in the context of superstring theories and by Turok~\cite{turok} in fundamental theories with compact extra dimensions have made the above conclusion more specific, showing that in either case the pseudo-Goldstone bosons are actually ALPs.}

\subsection{Axion as a prototype}

A characteristic feature of the SM is that non-perturbative effects produce the term $\Delta {\cal L}_{\theta} = \theta  g^2 G_a^{\mu \nu} \tilde 
G_{a \mu \nu} /32 {\pi}^2$ in the QCD Lagrangian, where $\theta$ is an angle, $g$ and $G_a^{\mu \nu}$ are the gauge coupling constant and the gauge field strength of $SU_c (3)$, respectively, and $\tilde G_a^{\mu \nu} \equiv \frac{1}{2} {\epsilon}^{\mu \nu \rho \sigma} G_{a \rho \sigma}$. All values of $\theta$ are allowed and theoretically on the same footing, but nonvanishing $\theta$ values produce a P and CP violation in the strong sector of the SM. An additional source of CP violation comes from the chiral transformation needed to bring the quark mass matrix ${\cal M}_q$ into diagonal form, and so the total strong CP violation is parametrized by $\bar{\theta} = \theta + \, {\rm arg \ Det} \, {\cal M}_q$. Observationally, a nonvanishing $\bar{\theta}$ would show up in an electric dipole moment $d_n$ for the neutron. Consistency with the experimental upper bound $|d_n| < 3 \cdot 10^{- 26} \, {\rm e \, cm}$ requires $|\bar{\theta}| < 10^{- 9}$~\cite{axionrev}. Thus, the question arises as to why $|\bar{\theta}|$ is so unexpectedly small. A natural way out of this fine-tuning problem -- which is the {\it strong CP problem} -- was proposed by Peccei and Quinn~\cite{peccei} over 30 years ago. Basically, the idea is to make the SM Lagrangian invariant under an additional global $U(1)_{\rm PQ}$ symmetry in such a way that the $\Delta {\cal L}_{\theta}$ term can be rotated away. While this strategy can be successfully implemented, it turns out that the $U(1)_{\rm PQ}$ is spontaneously broken and then a Goldstone boson is necessarily present in the physical spectrum. Things are slightly more complicated, because $U(1)_{\rm PQ}$ is also explicitly broken by the same non-perturbative effects which give rise to $\Delta {\cal L}_{\theta}$. Therefore, the would-be Goldstone boson becomes a pseudo-Goldstone boson -- the original axion~\cite{WW} -- with nonvanishing mass given by 
\begin{equation}
\label{a6}
m \simeq 0.6 \left( \frac{10^7 \, {\rm GeV}}{f_a} \right) \, {\rm eV}~,
\end{equation}
where $f_a$ denotes the scale at which $U(1)_{\rm PQ}$ is spontaneously broken. Qualitatively, the axion is quite similar to the pion and it posseses Yukawa couplings to quarks which go like the inverse of $f_a$. Moreover -- just like for the pion -- a two-photon coupling $a\gamma\gamma$ of the axion $a$ is generated at one-loop via the triangle graph with internal fermion lines, which is described by the effective Lagrangian
\begin{equation}
\label{aq5}
{\cal L}_{a \gamma \gamma} = - \frac{1}{4 M} \, F^{\mu \nu} \,  \tilde F_{\mu \nu} \, a = \frac{1}{M} \, {\bf E} \cdot {\bf B} \, a~,
\end{equation}
where $F^{\mu \nu} \equiv ({\bf E}, {\bf B}) \equiv \partial^{\mu} A^{\nu} - \partial^{\nu} A^{\mu}$ is the usual electromagnetic field strength and $\tilde F^{\mu \nu} \equiv \frac{1}{2} {\epsilon}^{\mu \nu \rho \sigma} F_{\rho \sigma}$. The constant $M$ entering Eq. (\ref{aq5}) has the dimension of an energy and is given by
\begin{equation}
\label{a7}
M = 1.2 \cdot 10^{10} \, k \, \left( \frac{f_a}{10^7 \, {\rm GeV}} \right) \, \, {\rm GeV}~,
\end{equation}
with $k$ a model-dependent parameter of order one~\cite{cgn}. Note that $M \propto f_a$ and turns out to be independent of the mass of the fermions running in the loop. Hence, the axion is characterized by a strict relation between its mass and two-photon coupling 
\begin{equation}
\label{a8}
m = 0.7 \, k \, \left( \frac{10^{10} \, {\rm GeV}}{M} \right) \, {\rm eV}~.
\end{equation}  

In the original proposal~\cite{peccei}, $U(1)_{\rm PQ}$ is spontaneously broken by two Higgs doublets which also break $SU_w (2) \times U_y (1)$ spontaneously, so that $f_a \leq G_F^{-1/2}$. Correspondingly, from eq. (\ref{a6}) we get $m \geq 24 \, {\rm KeV}$. In addition, the axion is rather strongly coupled to quarks and induces observable nuclear de-excitation effects~\cite{donnely}. In fact, it was soon realized that the original axion was experimentally ruled out~\cite{zehnder}. 

A slight change in perspective led {  shortly thereafter} to the resurrection of the axion strategy. Conflict with experiment arises because the original axion is too strongly coupled and too massive. But, given the fact that both $m$ and all axion couplings go like the inverse of 
$f_a$ the axion becomes weakly coupled and sufficiently light provided that one arranges $f_a \gg G_F^{-1/2}$. This is straightforwardly achieved by performing the spontaneous breakdown of $U(1)_{\rm PQ}$ with a Higgs field which is a singlet under $SU_w (2) \times U_y (1)$~\cite{invisible}. Note that we are thereby led to the conclusion that the $U(1)_{\rm PQ}$ symmetry has nothing to do with the low-energy effective theory to which the axion belongs, but rather it arises within an underlying more fundamental theory. 

Thus, we see that the axion strategy provides a particular realization of the general scenario outlined in Subsection II-A, with ${\cal G} = U(1)_{\rm PQ}$, ${\Lambda}_1 = f_a$ and ${\cal L}_{\rm nonren} ({\phi}_{\rm SM}, a)$ including ${\cal L}_{a \gamma \gamma}$ among other terms involving the SM fermions. This fact also entails that new physics should lurk around the scale at which $U(1)_{\rm PQ}$ is spontaneously broken. The same conclusion is reached from the recognition that the Peccei-Quinn symmetry is dramatically unstable against a tiny perturbation -- even at the Planck scale -- unless it is protected by some discrete gauge symmetry which can only arise in a more fundamental theory~\cite{lusignoli}.

\subsection{Beyond the axion: ALPs}

A generic feature of many extensions of the SM along the lines discussed in Section I is the prediction of ALPs. Generally speaking, ALPs are a straightforward generalization of the axion but important differences exist between the axion and ALPs mainly because the axion arises in a very specific context while in dealing with ALPs the aim is to bring out their properties in a model-independent fashion as much as 
possible~\cite{alprev}. This attitude has two main consequences:
\begin{itemize}
\item Only ALP-photon interaction terms are taken into account. Therefore, any other possible coupling of ALPs to SM particles is presently discarded and this entails that ${\cal L}_{\rm nonren} ({\phi}_{\rm SM}, a)$ in Eq. (\ref{t2231209}) {\it only} includes ${\cal L}_{a \gamma \gamma}$ as defined by Eq. (\ref{aq5}). Observe that such an ALP coupling to two photons $a \gamma \gamma$ is just supposed to exist without further worrying about its origin.
\item The parameters $m$ and $M$ are to be regarded as {\it unrelated} for ALPs, and it is merely assumed that $m \ll G_F^{- 1/2}$ and $M \gg G_F^{- 1/2}$. 
\end{itemize}

As a result, ALPs are described by the Lagrangian
\begin{equation}
\label{t2230812wx}
{\cal L}_{\rm ALP} =  \frac{1}{2} \, \partial^{\mu} a \, \partial_{\mu} a - \frac{1}{2} \, m^2 \, a^2 - \, \frac{1}{4 M} \, F_{\mu\nu} \tilde{F}^{\mu\nu} a = \frac{1}{2} \, \partial^{\mu} a \, \partial_{\mu} a - \frac{1}{2} \, m^2 \, a^2 + \frac{1}{M} \, {\bf E} \cdot {\bf B}~a.
\end{equation}

\subsection{Photon-ALP mixing}

What ultimately characterizes ALPs is the trilinear $a \gamma \gamma$ vertex in ${\cal L}_{\rm ALP}$, which gives rise to photon-ALP mixing in the presence of an {\it external} magnetic field ${\bf B}$. More specifically, what happens can be described as follows.

In such a situation, an off-diagonal element in the mass matrix for the photon-ALP system shows up. Therefore, the interaction eigenstates differ from the propagation eigenstates and the phenomenon of photon-ALP oscillations shows up~\cite{sikivie1983, raffeltstodolsky}. This is analogous to what takes place in the case of massive neutrinos with different flavours, apart from an important difference. All neutrinos have equal spin, and so neutrino oscillations can freely occur. Instead, ALPs are supposed to have spin zero whereas the photon has spin one, hence one of them can transform into the other only if the spin mismatch is compensated for by an external magnetic field. Note that the strength of this effect depends on the ratio $B/M$ and not on $B$ and $M$ separately.

We denote by ${\bf E}$ the electric field and by ${\bf k}$ the wave vector of a propagating photon at a given space-time point. Further,  let ${\bf B}_L$ and ${\bf B}_T$ be the components of the external magnetic field ${\bf B}$ along ${\bf k}$ and perpendicular to ${\bf k}$, respectively. Because ${\bf E}$ is orthogonal to ${\bf k}$ it follows that only the term ${\bf E} \cdot {\bf B}_T$ survives in ${\cal L}_{\rm ALP}$. We next split up ${\bf E}$ into two components, one ${\bf E}_{\parallel}$ in the plane defined by ${\bf k}$ and ${\bf B}$ and the other ${\bf E}_{\perp}$ perpendicular to that plane. By construction, ${\bf E}_{\perp}$ is orthogonal to ${\bf B}_T$, and so the $a \gamma \gamma$ coupling in ${\cal L}_{\rm ALP}$ goes like $E_{\parallel} \, B_T \, a$, which exhibits two characteristic properties of ALPs. First, the photon-ALP mixing depends only on the transverse component ${\bf B}_T$ of the external magnetic field; for notational simplicity we will write ${\bf B}$ rather than ${\bf B}_T$ in the following. Second, only photons linearly polarized along ${\bf E}_{\parallel}$ actually mix with ALPs, whereas photons with polarization ${\bf E}_{\perp}$ do not mix. As a consequence, the $a \gamma \gamma$ coupling acts like a polarimeter, in the sense that it gives rise to a change of the photon polarization state. This effect can be used to look for ALPs both in high-precision polarimetric measurements performed in the laboratory~\cite{pvlas} and in certain astrophysical observations in which the polarization state of the detected photons can be measured~\cite{bmr2010}. 

\subsection{Photon/ALP beam propagation}

We shall be concerned throughout with a monochromatic, unpolarized photon/ALP beam of energy $E$ and wave vector ${\bf k}$ propagating in a cold medium which is both magnetized and ionized (from now on $E$ denotes the energy, and since the electric field will never be considered again no confusion arises). We suppose for the moment that the external magnetic field ${\bf B}$ is homogeneous and we denote by $n_e$ the electron number density. We employ an orthogonal reference frame with the $y$-axis along ${\bf k}$, while the $x$ and $z$ axes are chosen arbitrarily.

It can be shown that in this case the beam propagation equation following from ${\cal L}_{\rm ALP}$ can be written as~\cite{raffeltstodolsky}
\begin{equation}
\label{k3}
\left( \frac{d^2}{d y^2} + E^2 + 2  E {\cal M}_0 \right) \, \psi (y) = 0
\end{equation}
with
\begin{equation}
\label{k3w1}
\psi (y) \equiv \left(
\begin{array}{c}
A_x(y) \\
A_z(y) \\
a(y) \\
\end{array}
\right)~,
\end{equation}
where $A_x(y)$ and $A_z(y)$ denote the photon amplitudes with polarization (electric field) along the $x$- and $z$-axis, respectively, while $a(y)$ is the amplitude associated with the ALP. It is useful to introduce the basis $\{| {\gamma}_x \rangle, | {\gamma}_z \rangle, |a \rangle \}$ defined by
\begin{equation}
\label{aa7j1}
|{\gamma}_x \rangle \equiv \left(
\begin{array}{c}
1 \\
0 \\
0 \\
\end{array}
\right)~,        
\end{equation} 
\begin{equation}
\label{aa7j2}
|{\gamma}_z \rangle \equiv \left(
\begin{array}{c}
0 \\
1 \\
0 \\
\end{array}
\right)~,        
\end{equation} 
\begin{equation}
\label{aa7j3}
|a \rangle \equiv \left(
\begin{array}{c}
0 \\
0 \\
1 \\
\end{array}
\right)~,              
\end{equation} 
where $| {\gamma}_x \rangle$ and $| {\gamma}_z \rangle$ represent the two photon linear polarization states along the $x$- and $z$-axis, respectively, and 
$|a \rangle$ denotes the ALP state. Accordingly, we can rewrite $\psi (y)$ as
\begin{equation}
\label{aa7}
\psi (y) = A_x (y) \, |{\gamma}_x \rangle + A_z (y) \, |{\gamma}_z \rangle + a (y) \, 
|a \rangle~,        
\end{equation} 
and the real, symmetric  photon-ALP mixing matrix ${\cal M}_0$ entering Eq. (\ref{k3}) has the form
\begin{equation}
\label{aa8}
{\cal M}_0 = \left(
\begin{array}{ccc}
\Delta_{xx} & \Delta_{xz} & \Delta^{x}_{ a \gamma} \\
\Delta_{zx} & \Delta_{zz} & \Delta^{z}_{a \gamma} \\
\Delta^{x}_{a \gamma} & \Delta^{z}_{\rm a \gamma} & \Delta_{a a} \\
\end{array}
\right)~,
\end{equation}
where we have set
\begin{equation}
\label{a13o12a}
\Delta^{x}_{a \gamma} \equiv \frac{B_x}{2 M}~,
\end{equation}
\begin{equation}
\label{a13o12b}
\Delta^{z}_{a \gamma} \equiv \frac{B_z}{2 M}~,
\end{equation}
\begin{equation}
\label{a13o12c}
\Delta_{a a} \equiv - \, \frac{m^2}{2 E}~.
\end{equation}

While the terms appearing in the third row and column of ${\cal M}_0$ are dictated by ${\cal L}_{\rm ALP}$ and have an evident physical meaning, the other 
$\Delta$-terms require some explanation. They reflect the properties of the medium -- which are not included in ${\cal L}_{\rm ALP}$ -- and the off-diagonal 
$\Delta$-terms directly mix the photon polarization states giving rise to Faraday rotation. 

In the present paper we are interested in the situation where the photon/ALP energy is much larger than the ALP mass, namely $E \gg m$. As a consequence, the short-wavelength approximation can be successfully employed and can be implemented as~\cite{raffeltstodolsky}
\begin{equation}
\label{a13o}
\left( \frac{d^2}{d y^2} + E^2 \right) \, \psi (y) = \left(i \frac{d}{d y} + E \right) \left(- i \frac{d}{d y} + E \right) \, \psi (y) = 
2 E \left( i \frac{d}{d y} + E \right) \, \psi (y)~,
\end{equation}
which turns the second-order beam propagation equation (\ref{k3}) into the first-order one
\begin{equation}
\label{k3l}
\left( i \frac{d}{d y} + E + {\cal M}_0 \right) \, \psi (y) = 0~.
\end{equation}

We see that a remarkable picture emerges, wherein the beam looks formally like a three-state nonrelativistic quantum system. Explicitly, they are the two photon polarization states and the ALP state. The evolution of the {\it pure} beam states is then described by the three-dimensional wave function $\psi (y)$ -- with the $y$-coordinate replacing time  -- which obeys the Sch\"odinger-like equation (\ref{k3l}) with Hamiltonian
\begin{equation}
\label{k3la}
H_0 \equiv - \left( E + {\cal M}_0  \right)~.
\end{equation}
Denoting by $U_0(y, y_0)$ the transfer matrix -- namely the solution of Eq. (\ref{k3l}) with initial condition $ U_0(y_0, y_0) = 1$ -- the propagation of a generic wave function can be represented as 
\begin{equation}
\label{k3lasq}
\psi (y) = U_0 (y, y_0) \, \psi (y_0)~. 
\end{equation}
Moreover, we have 
\begin{equation}
\label{k3laq1}
U_0 (y, y_0) = e^{i E (y - y_0)} \, {\cal U}_0 (y, y_0)~,
\end{equation}
where ${\cal U}_0(y, y_0)$ is the transfer matrix associated with the reduced Sch\"odinger-like equation
\begin{equation}
\label{k3l1}
\left( i \frac{d}{d y} + {\cal M}_0 \right) \, \psi (y) = 0~.
\end{equation}

Because ${\bf B}$ is supposed to be homogeneous, we have the freedom to choose the $z$-axis along ${\bf B}$, so that $B_x = 0$. The diagonal 
$\Delta$-terms receive in principle two different contributions. One comes from QED vacuum polarization, but since we will be dealing with very weak magnetic fields this effect is irrelevant~\cite{raffeltstodolsky}. The other contribution arises from the fact that the beam is supposed to propagate in a cold plasma, where charge screening produces an effective photon mass resulting in the plasma frequency 
\begin{equation}
\label{a6zz}
{\omega}_{\rm pl} = \left( \frac{4 \pi \alpha n_e}{m_e} \right)^{1/2}~, 
\end{equation}
where $\alpha$ is the fine-structure constant and $m_e$ denotes the electron mass, which entails
\begin{equation}
\label{a13}
{\Delta}_{\rm pl} = - \, \frac{\omega_{\rm pl}^2}{2 E}~. 
\end{equation}
Finally, the $\Delta_{xz}$, $\Delta_{zx}$ terms account for Faraday rotation, but since we are going to take $E$ in the VHE $\gamma$-ray band Faraday rotation is 
 negligible. Altogether, the mixing matrix becomes 
\begin{equation}
\label{a9}
{\cal M}_0^{(0)} = \left(
\begin{array}{ccc}
{\Delta}_{\rm pl} & 0 & 0 \\
0 & {\Delta}_{\rm pl} & \Delta_{a \gamma} \\
0 & \Delta_{a \gamma} & \Delta_{a a} \\
\end{array}
\right)~,
\end{equation}
with the superscript $(0)$ recalling the present choice of the coordinate system and
\begin{equation}
\label{a13o12d}
\Delta_{a \gamma} \equiv \frac{B}{2M}~.
\end{equation}
We see that $A_x$ decouples away while only $A_z$ mixes with $a$, showing that in the present approximation plasma effects do not change the qualitative features previously found in vacuo.

Application of the discussion reported in Appendix A with ${\cal M} \to {\cal M}_0^{(0)}$ yields for the corresponding eigenvalues
\begin{equation}
\label{a91212a1}
{\lambda}_{0,1} = \Delta_{\rm pl}~,
\end{equation}
\begin{equation}
\label{a91212a2}
{\lambda}_{0,2} = \frac{1}{2} \Bigl(\Delta_{\rm pl} + \Delta_{a a} - {\Delta}_{\rm osc} \Bigr)~, 
\end{equation}
\begin{equation}
\label{a91212a3}
{\lambda}_{0,3} = \frac{1}{2} \Bigl(\Delta_{\rm pl} + \Delta_{a a} + {\Delta}_{\rm osc} \Bigr)~,
\end{equation}
where we have set
\begin{equation}
\label{a17}
{\Delta}_{\rm osc} \equiv \Bigl[ \left( \Delta_{\rm pl} - \Delta_{\rm a a} \right)^2 + 4 \left( \Delta_{a \gamma} \right)^2 \Bigr]^{1/2} = 
\left[\left( \frac{m^2 - {\omega}_{\rm pl}^2}{2  E} \right)^2 + \left( \frac{B}{M} \right)^2 \right]^{1/2}~.
\end{equation}
As a consequence, the transfer matrix associated with Eq. (\ref{k3l1}) with mixing matrix ${\cal M}_0^{(0)}$ can be written with the help of Eq. (\ref{mravvq2abcapp}) as
\begin{equation}
\label{mravvq2abcappZX1}
{\cal U}_0 (y, y_0; 0) = e^{i {\lambda}_1 (y - y_0)} \, T_{0,1} (0) + e^{i {\lambda}_2 (y - y_0)} \, T_{0,2} (0) + e^{i {\lambda}_3 (y - y_0)} \, T_{0,3} (0)~, 
\end{equation}
where the matrices $T_{0,1} (0)$, $T_{0,2} (0) $ and $T_{0,3} (0)$ are just those defined by Eqs. (\ref{mravvq1app}), (\ref{mravvq2app}) and (\ref{mravvq3app}) as specialized to the present situation. Actually, a simplification is brought about by introducing the photon-ALP mixing angle
\begin{equation}
\label{a15}
\alpha = \frac{1}{2} \, {\rm arctg} \left(\frac{2 \, \Delta_{a \gamma} }{\Delta_{\rm pl} - \Delta_{a a}} \right) = 
\frac{1}{2} \, {\rm arctg} \left[\left( \frac{B}{M} \right) \left(\frac{2 E}{m^2 - {\omega}_{\rm pl}^2} \right) \right]~,
\end{equation}
since then simple trigonometric manipulations allow us to express the above matrices in the simpler form
\begin{equation}
\label{mravvq2appZ1a}
T_{0,1} (0) \equiv
\left(
\begin{array}{ccc}
1 & 0 & 0 \\
0 & 0 & 0 \\
0 & 0 & 0
\end{array}
 \right)~,
\end{equation}
\begin{equation} 
\label{mravvq2appZ1b}
T_{0,2} (0) \equiv
\left(
\begin{array}{ccc}
0 & 0 & 0 \\ 
0 & \sin^2 \alpha & - \sin \alpha \cos \alpha \\
0 & - \sin \alpha \cos \alpha & \cos^2 \alpha 
\end{array}
\right)~,
\end{equation} 
\begin{equation}
\label{mravvq3appZ}
T_{0,3} (0) \equiv 
\left(
\begin{array}{ccc}
0 & 0 & 0 \\
0 & \cos^2 \alpha  & \sin \alpha \cos \alpha \\
0 & \sin \alpha \cos \alpha  & \sin^2 \alpha 
\end{array}
\right)~.
\end{equation}

Now, the probability that a photon polarized along the $z$-axis oscillates into an ALP after a distance $y$ is evidently
\begin{equation}
\label{a16hh}
P_{0, {\gamma}_{z} \to a}^{(0)}(y) =  \left|\langle a |{\cal U}_0 (y,0;0)|{\gamma}_z \rangle \right|^2
\end{equation}
and in complete analogy with the case of neutrino oscillations~\cite{raffeltbook} it reads
\begin{equation}
\label{a16}
P_{0, {\gamma}_{z} \to a}^{(0)}(y) = {\rm sin}^2 2 \alpha \  {\rm sin}^2
\left( \frac{\Delta_{\rm osc} \, y}{2} \right)~,
\end{equation}
which shows that ${\Delta}_{\rm osc}$ plays the role of oscillation wave number, thereby implying that the oscillation length is 
$L_{\rm osc} = 2 \pi / {\Delta}_{\rm osc}$. Owing to Eq. (\ref{a15}), Eq. (\ref{a16}) can be rewritten as 
\begin{equation}
\label{a16g}
P_{0, {\gamma}_{z} \to a}^{(0)}(y) = \left(\frac{B}{M \, {\Delta}_{\rm osc}} \right)^2 \,  {\rm sin}^2 \left( \frac{\Delta_{\rm osc} \, y}{2} \right)~,
\end{equation}
which shows that the photon-ALP oscillation probability becomes both maximal and energy-independent for
\begin{equation}
\label{a17k}
{\Delta}_{\rm osc} \simeq \frac{B}{M}~,
\end{equation}
and explicitly reads
\begin{equation}
\label{a16gh}
P_{0, {\gamma}_{z} \to a}^{(0)}(y) \simeq {\rm sin}^2 \left(\frac{B y}{2 M} \right)~.
\end{equation}
This is the {\it strong-mixing regime}, which -- from the comparison of Eqs. (\ref{a17}) and (\ref{a17k}) -- turns out to be characterized by the condition
\begin{equation}
\label{a28012011}
\frac{ | m^2 - {\omega}^2_{\rm pl} | }{2 E} \ll \frac{B}{M}~,
\end{equation}
and so it sets in sufficiently {\it above} the energy threshold
\begin{equation}
\label{a17231209}
E_* \equiv \frac{ | m^2 - {\omega}^2_{\rm pl} | M}{2 B}~.
\end{equation}
Note that the strong-mixing regime is unbounded from above since the contribution to photon-ALP mixing arising from QED vacuum polarization is negligible for the magnetic fields considered in this paper~\cite{br}.

Below $E_*$ the photon-ALP oscillation probability becomes energy-dependent and vanishingly small. 

So far, our discussion was confined to the case in which the beam is in a pure polarization state. This assumption possesses the advantage of making the resulting equations particularly transparent but it has the drawback that it is too restrictive for our analysis. For, photon polarization cannot be measured in the VHE $\gamma$-ray band, and so we have to treat the beam as unpolarized. As a consequence, it will be described by a generalized polarization density matrix 
\begin{equation}
\label{k3lw1212}
\rho (y) = \left(\begin{array}{c}A_x (y) \\ A_z (y) \\ a (y)
\end{array}\right)
\otimes \left(\begin{array}{c}A_x (y) \  A_z (y) \ a (y) \end{array}\right)^{*}
\end{equation}
rather than by a wave function $\psi (y)$. Remarkably, the analogy with non-relativistic quantum mechanics entails that 
$\rho (y)$ obeys the Von Neumann-like equation
\begin{equation}
\label{k3lw}
i \frac{d \rho}{d y} = \left[ \rho, {\cal M}_0 \right]
\end{equation}
associated with Eq. (\ref{k3l1}). Thus, the propagation of a generic $\rho (y)$ is given by
\begin{equation}
\label{k3lwf1}
\rho (y) = {\cal U}_0 (y, y_0) \, \rho (y_0) \, {\cal U}_0^{\dagger}(y, y_0)
\end{equation}
and the probability that a photon/ALP beam initially in the state $\rho_1$ will be found in the state $\rho_2$ after a distance $y$ is
\begin{equation}
\label{k3lwf1g}
P_{0, \rho_1 \to \rho_2} (y) = {\rm Tr} \Bigl( \rho_2 \, {\cal U}_0 (y,0) \, \rho_1 \, {\cal U}_0^{\dagger}(y, 0) \Bigr)~,
\end{equation}
since we are assuming as usual that ${\rm Tr} \rho_1 = {\rm Tr} \rho_2 = 1$. Observe that in Eqs. (\ref{k3lw}), (\ref{k3lwf1}) and (\ref{k3lwf1g}) we have dropped  the superscript $(0)$ in ${\cal M}_0$ and replaced $ {\cal U}_0 (y, y_0;0)$ by $ {\cal U}_0 (y, y_0)$ because they retain their form for an arbitrary choice of the coordinate system.

In view of our subsequent discussion it proves essential to deal with the general case in which ${\bf B}$ is not aligned with the $z$-axis but forms a nonvanishing angle $\psi$ with it. Correspondingly, the mixing matrix ${\cal M}_0$ presently arises from ${\cal M}_0^{(0)}$ through the similarity transformation
\begin{equation}
\label{mr101109qq}
{\cal M}_0 = V^{\dagger} (\psi) \, {\cal M}_0^{(0)} \, V (\psi) 
\end{equation}
operated by the rotation matrix in the $x$--$z$ plane, namely 
\begin{equation}
\label{a91212ay}
V (\psi) =
\left(
\begin{array}{ccc}
\cos \psi & - \sin \psi & 0 \\
\sin \psi & \cos \psi & 0 \\
0 & 0 & 1 \\
\end{array}
\right)~.
\end{equation}
This leads to~\cite{Mirizzi2006zy}
\begin{equation}
\label{aa8MRq}
{\cal M}_0 = \left(
\begin{array}{ccc}
\Delta_{\rm pl} & 0 & \Delta_{a \gamma} \, \sin \psi \\
0 & \Delta_{\rm pl} & \Delta_{a\gamma} \, \cos \psi \\
\Delta_{a \gamma} \, \sin \psi& \Delta_{a \gamma} \, \cos\psi& \Delta_{a a} \\
\end{array}
\right)~,
\end{equation} 
indeed in agreement with Eq. (\ref{aa8}) within the considered approximation. Therefore the transfer matrix reads
\begin{equation}
\label{mravvq1}
{\cal U}_0 (y,y_0 ; \psi) = V^{\dagger} (\psi) \, {\cal U}_0 (y,y_0 ; 0) \, V (\psi)
\end{equation}
and its explicit representation turns out to be
\begin{equation}
\label{mravvq2abc}
{\cal U}_0 (y,y_0 ; \psi) = e^{i {\lambda}_1 (y - y_0)} \, T_{0,1} (\psi) + e^{i {\lambda}_2 (y - y_0)} \, T_{0,2} (\psi) + e^{i {\lambda}_3 (y - y_0)} \, T_{0,3} (\psi)~, 
\end{equation}
with
\begin{equation}
\label{mravvq21a}
T_{0,1} (\psi) \equiv
\left(
\begin{array}{ccc}
\cos^2 \psi & -\sin \psi \cos \psi & 0 \\
- \sin \psi \cos \psi & \sin^2 \psi & 0 \\
0 & 0 & 0
\end{array}
\right)~,
\end{equation}
\begin{equation}
\label{mravvq3}
T_{0,2} (\psi) \equiv 
\left(
\begin{array}{ccc}
\sin^2 \theta \sin^2 \psi & \sin^2 \alpha \sin \psi \cos \psi & - \sin \alpha \cos \alpha \sin \psi \\
\sin^2 \alpha \sin \psi \cos \psi & \sin^2 \alpha \cos^2 \psi & - \sin \alpha \cos \alpha \cos \psi \\
- \sin \alpha \cos \alpha \sin \psi & - \sin \alpha \cos \alpha \cos \psi & \cos^2 \alpha 
\end{array}
\right)~,
\end{equation}
\begin{equation} 
\label{mravvq21b}
T_{0,3} (\psi) \equiv
\left(
\begin{array}{ccc}
\sin^2 \psi \cos^2 \alpha & \sin \psi \cos \psi \cos^2 \alpha & 
\sin \alpha \cos \alpha \sin \psi \\ 
\sin \psi \cos \psi \cos^2 \alpha & \cos^2 \psi \cos^2 \alpha & 
\sin \alpha \cos \alpha \cos \psi \\
\sin \psi \cos \alpha \sin \alpha & \cos \psi \sin \alpha \cos \alpha & \sin^2 \alpha 
\end{array}
\right)~.
\end{equation}

\subsection{Astrophysical and cosmological constraints}

Astrophysics has turned out to be quite effective in setting an upper bound on the $a\gamma \gamma$ vertex in ${\cal L}_{\rm ALP}$, which therefore holds for the axion as well as for ALPs.

In the first place, the failure to detect ALPs emitted by the Sun in the CAST experiment at CERN has led to
\begin{equation}
\label{a4}
M > 1.14 \cdot 10^{10} \, {\rm GeV}
\end{equation}
for $m < 0.02 \, {\rm eV}$~\cite{castcern}.

On the theoretical side, the most reliable method concerns ALP photo-production through the Primakoff process, which takes place when an incoming photon scatters on a charged particle and becomes an ALP upon the exchange of a virtual photon. Hot, dense plasmas in stellar cores are ideal environments wherein the Primakoff process involving thermal photons can occur. Once produced, the ALPs escape because their mean free path is much larger than the stellar radius, thereby carrying off energy. Owing to the virial equilibrium, the stellar core has a negative specific heat. Therefore it reacts to such an energy loss by getting hotter. As a result, the rate of nuclear reactions sharply increases, bringing about a substantial change in the observed properties of stars. Since current models of stellar evolution are in fairly good agreement with observations, $M$ has to be large enough to provide a sufficient suppression of unwanted ALP effects. This argument has been applied systematically in a quantitative fashion to the Sun, to main-sequence stars and to red-giants stars in globular clusters, with the result~\cite{raffeltbook} 
\begin{equation}
\label{a3}
M > 10^{10} \, {\rm GeV}~. 
\end{equation} 

A consequence of photon-ALP oscillations is that a lower bound on $M$ stronger than conditions (\ref{a4}) and (\ref{a3}) -- even if much less robust -- can be derived for $m < 10^{- 10} \, {\rm eV}$. In this connections, two methods have been put forward. One is based on the observation of a time-lag between opposite-polarization modes in pulsar radio emission and yields~\cite{mohanti} 
\begin{equation}
\label{a5}
M > 5 \cdot 10^{10} \, {\rm GeV}~.
\end{equation}
The other involves ALPs emitted by the supernova SN1987A, which would convert them into $\gamma$-rays in the magnetic field of the Galaxy. Using the absence of these photons in the Solar Maximum Mission Gamma-Ray Detector, the lower bound
\begin{equation}
\label{a509022011}
M > 10^{11} \, {\rm GeV}
\end{equation}
has been derived~\cite{raffeltmasso}. We stress however that condition (\ref{a509022011}) is affected by large uncertainties, reflecting the lack of precise knowledge of the Galactic magnetic field as well as of the energy dependence of the detector response.

Let us next turn our attention to the cosmological constraints on ALPs. At variance with the previous astrophysical analysis, the case of the axion differs drastically from that of generic ALPs.

We recall that cosmology sets strong constraints on the axion properties because of their coupling to quarks and gluons (indeed necessary in order to solve the strong CP problem). Basically, both thermal and non-thermal mechanisms can produce axions in the early Universe. Since this issue is not directly relevant for our discussion, we cursorily summarize the main results remarking that the situation is in reality much more complex than sketched here. Recalling that 
$f_a$ denotes the scale at which $U(1)_{\rm PQ}$ is spontaneously broken, only the range
\begin{equation}
\label{a528022011c1}
0.6 \cdot 10^{7} \, {\rm GeV} < f_a < 0.6 \cdot 10^{13} \, {\rm GeV}
\end{equation}
is cosmologically allowed. Thanks to Eqs. (\ref{a6}) and (\ref{a7}), this constraint translates into the conditions
\begin{equation}
\label{a528022011c2}
10^{- 6} \, {\rm eV} < m < 1 \, {\rm eV}
\end{equation}
and
\begin{equation}
\label{a528022011c3}
0.7 \cdot 10^{10} \, {\rm GeV} < M < 0.7 \cdot 10^{16} \, {\rm GeV}~,
\end{equation}
respectively (we have taken for simplicity $k = 1$ in Eq. (\ref{a7})). We stress that these bounds should be regarded merely as order-of-magnitude estimates. Moreover, the axion is a very good candidate for dark matter. More specifically, for $m$ close to $10^{- 6} \, {\rm eV}$ non-thermal production dominates and it behaves as a cold dark matter candidate, whereas for $m$ close to $1 \, {\rm eV}$ thermal production dominates and it is a hot dark matter particle~\cite{cdm}. Searches for axionic dark matter are currently underway with the ADMX experiment~\cite{admx}. Finally, it has recently been realized that cold dark matter axions ought to form a Bose-Einstein condensate~\cite{becsi}.

Clearly, all these considerations do not apply to ALPs, since they are supposed to interact with the rest of the world through the two-photon coupling only. As a consequence, they can be produced in the early Universe only thermally through the processes $e^{\pm} \, \gamma \to e^{\pm} \, a$ and $e^{+} \, e^{-} \to \gamma \, a$. It has been shown that in the case $m \ll 1 \, {\rm eV}$ -- which is the one relevant for us as we shall see later -- ALPs are relativistic today and their abundance is anyway smaller than that of CMB photons~\cite{massotoldra}. Hence, we are led to the conclusion that the ALP considered in this paper are totally unconstrained by cosmology and play no role for the dark matter problem.

\section{VERY-HIGH-ENERGY (VHE) BLAZAR SPECTRA}

Among the many achievements of IACTs is the determination of blazar spectra at energies above $100 \, {\rm GeV}$, and to date this task has been accomplished for about 30 sources with redshift up to $z = 0.536$ for 3C 279. Most of these blazars are listed in Table \ref{tab:a1}.


\begin{table}
\begin{tabular}{cccccc}          
\hline
Source \  & $z$  \ \ \ \ \  & Observed energy range  \ \ \ \  & ${\Gamma}_{\rm obs}$ &  {Reference} &  $\Gamma_{\rm{em}}$ \\
\hline
3C 66B    \  &       0.022   \ \ \ \ &     $120 \, {\rm GeV} < E_0 < 1.8 \,  {\rm TeV}$ \ \ \ \ &  $3.10  \pm 0.37$ &    \cite{3C 66B}  & 3.00\\
Mrk 421  \ \ \ \    &    0.030  \ \ \ \    &  $140 \, {\rm GeV} < E_0 < 6 \,  {\rm TeV}$  \ \ \ \  &   $2.33 \pm 0.22 $  &  \cite{Mrk 421}    & 2.16 \\
Mrk 501  \ \ \ \    &  0.034   \ \ \ \    &        $150 \, {\rm GeV} < E_0 < 6 \,  {\rm TeV}$  \ \ \ \  &    $2.09 \pm 0.20 $      &  \cite{Mrk 501}   & 1.90\\
Mrk 501  \ \ \ \    &  0.034  \ \ \ \     &        $150 \, {\rm GeV} < E_0 < 3 \,  {\rm TeV}$  \ \ \ \  &    $2.20 \pm 0.20 $      &  \cite{Mrk 501}   & 2.03\\
1ES 2344+514  \ \ \ \  & 0.044  \ \ \ \   &        $180 \, {\rm GeV} < E_0 < 4 \,  {\rm TeV}$  \ \ \ \  &       $2.95 \pm 0.23 $    &  \cite{1ES 2344+514}  &  2.70\\
Mrk 180  \ \ \ \   &    0.045  \ \ \ \  &        $180 \, {\rm GeV} < E_0 < 1.4 \,  {\rm TeV}$   \ \ \ \   &      $3.30 \pm 0.70$    &  \cite{Mrk 180}    & 3.07\\
1ES 1959+650  \ \ \ \  & 0.047  \ \ \ \  &        $190 \, {\rm GeV} < E_0 < 6 \,  {\rm TeV}$  \ \ \ \  &       $2.72 \pm 0.24$   &  \cite{1ES 1959+650}    & 2.43\\
BL Lacertae   \ \ \ \  & 0.069   \ \ \ \   &              $170 \, {\rm GeV} < E_0 < 700 \,  {\rm GeV}$  \ \ \ \  & $3.60 \pm 0.54$    &  \cite{BL Lacertae}    & 3.27\\
PKS 0548-322  \ \ \ \   &   0.069  \ \ \ \  &        $440 \, {\rm GeV} < E_0 < 2.2 \,  {\rm TeV}$   \ \ \ \   &    $2.80  \pm 0.32$    &  \cite{PKS 0548-322}   & 2.39\\
PKS 2005-489  \ \ \ \  & 0.071  \ \ \ \   &        $230 \, {\rm GeV} < E_0 < 2.3 \,  {\rm TeV}$  \ \ \ \   &    $4.00 \pm 0.41$    &  \cite{PKS 2005-489}     & 3.59\\
RGB J0152+017  \ \ \ \  & 0.080  \ \ \ \   & $320 \, {\rm GeV} < E_0 < 3 \,  {\rm TeV}$  \ \ \ \  &      $2.95 \pm 0.41$   &  \cite{RGB J0152+017}     & 2.47\\
W Comae     \ \ \ \  &   0.102   \ \ \ \  &        $270 \, {\rm GeV} < E_0 < 1.2 \,  {\rm TeV}$  \ \ \ \  &      $3.81 \pm 0.49$    &  \cite{W Comae}     & 3.18\\
PKS 2155-304  \ \ \ \  & 0.117  \ \ \ \    &    $230 \, {\rm GeV} < E_0 < 3 \,  {\rm TeV}$  \ \ \ \   &   $3.37 \pm 0.12$   &  \cite{PKS 2155-304}      & 2.67\\
RGB J0710+591  \ \ \ \   &  0.125  \ \ \ \  &        ? \ \    &      $2.80  \pm 0.30$    &  \cite{RGB J0710+591}   & ? \\
H 1426+428  \ \ \ \  & 0.129  \ \ \ \  &    $800 \, {\rm GeV} < E_0 < 10 \,  {\rm TeV}$   \ \ \ \    &     $2.60 \pm 0.61$  & \cite{H 1426+428}     & 0.85\\
1ES 0806+524   \ \ \ \  &   0.138   \ \ \ \  &    $320 \, {\rm GeV} < E_0 < 630 \,  {\rm GeV}$  \ \ \ \   &     $3.60  \pm 1.04$  & \cite{1ES 0806+524}     & 2.70\\
1ES 0229+200  \ \ \ \  & 0.140  \ \ \ \  &    $580 \, {\rm GeV} < E_0 < 12 \,  {\rm TeV}$  \ \ \ \   &      $2.50 \pm 0.21$  & \cite{1ES 0229+200}     & 0.41\\
H 2356-309   \ \ \ \  &  0.165  \ \ \ \  &    $220 \, {\rm GeV} < E_0 < 900 \,  {\rm GeV}$   \ \ \ \     &     $3.09 \pm 0.26$   & \cite{H 2356-309}     & 2.06\\
1ES 1218+304  \ \ \ \  & 0.182  \ \ \ \  &    $180 \, {\rm GeV} < E_0 < 1.5 \,  {\rm TeV}$  \ \ \ \  &       $3.08 \pm 0.39$  & \cite{1ES 1218+304}     & 2.00\\
1ES 1101-232  \ \ \ \  & 0.186   \ \ \ \  &    $280 \, {\rm GeV} < E_0 < 3.2 \,  {\rm TeV}$  \ \ \ \  &           $2.94 \pm 0.20$   & \cite{1ES 1101-232}     & 1.72\\
1ES 0347-121  \ \ \ \  & 0.188  \ \ \ \    &    $300 \, {\rm GeV} < E_0 < 3.0 \,  {\rm TeV}$  \ \ \ \  &     $3.10 \pm 0.25$    & \cite{1ES 0347-121}    & 1.87\\
1ES 1011+496  \ \ \ \  &  0.212  \ \ \ \   &      $160 \, {\rm GeV} < E_0 < 600 \,  {\rm GeV}$  \ \ \ \  &      $4.00 \pm 0.54$   & \cite{1ES 1011+496}     & 2.90\\
S5 0716+714  \ \ \ \  &  0.31   \ \ \ \  &      $180 \, {\rm GeV} < E_0 < 680 \,  {\rm GeV}$  \ \ \ \  &       $3.45 \pm 0.58$   & \cite{S5 0716+714}    &  1.60\\
PG 1553+113  \ \ \ \  &  0.40  \ \ \ \  &    $95 \, {\rm GeV} < E_0 < 620 \,  {\rm GeV}$  \ \ \ \  &   $4.27 \pm 0.14$  & \cite{PG 1553+113}     & 2.48\\
PKS 1222+21  \ \ \ \  & 0.432  \ \ \ \  &    $80 \, {\rm GeV} < E_0 < 360 \,  {\rm GeV}$  \ \ \ \ & $3.75 \pm 0.34$ & \cite{PKS 1222+21}     & 2.47\\
3C 66A   \ \ \ \   &    0.444  \ \ \ \   &     $230 \, {\rm GeV} < E_0 < 470 \,  {\rm GeV}$  \ \ \ \   &     $4.10 \pm 0.72$  & \cite{3C 66A}     & 1.28\\
PKS 1424+240  \ \ \ \  &          0.5  \ \ \ \   &     $140 \, {\rm GeV} < E_0 <  500 \, {\rm GeV}$  \ \ \ \  &    $3.80  \pm 0.58$  & \cite{PKS 1424+240}    & 1.16\\
3C 279   \ \ \ \   &    0.536  \ \ \ \  &    $80 \, {\rm GeV} < E_0 < 480 \,  {\rm GeV}$   \ \ \ \   &    $4.10  \pm 0.73$   & \cite{3C 279}     & 2.05\\
$\Delta \Gamma_{\rm em}$    \ \ \ \   &   & & &        &       3.18  \\   
$\langle \Gamma_{\rm em} \rangle$   \ \ \ \    &  & & &        &       2.22  \\   
\end{tabular}
\caption{Blazars observed so far with the IACTs with known redshift $z$, measured energy range, measured spectral index ${\Gamma}_{\rm obs}$, and unfolded spectral index at emission ${\Gamma}_{\rm em}$ using the FRV model of the EBL. Statistical and systematic errors are added in quadrature to produce the total error reported on the measured spectral index. When only statistical errors are quoted, systematic errors are taken to be 0.1 for H.E.S.S. and 0.2 for MAGIC. The last two rows show the spread $\Delta \Gamma_{\rm em}$ of the values of $\Gamma_{\rm em}$ discarding errors and the average value 
$\langle \Gamma_{\rm em} \rangle$ including errors, respectively.}
\label{tab:a1}
\end{table}

In view of our later analysis, we carefully address the propagation of a monochromatic photon beam emitted by a blazar at redshift $z$ and detected at energy $E_0$ within the standard $\Lambda$CDM cosmological model, so that the emitted energy is $ E_0 (1+z)$ owing to the cosmic expansion. Clearly -- regardless of the actual physics responsible for photon propagation -- the observed and emitted differential photon number fluxes -- namely $d N/d E$ -- are related by
\begin{equation} 
\label{a02122010}
\Phi_{\rm obs}(E_0,z) = P_{\gamma \to \gamma} (E_0,z) \, \Phi_{\rm em} \left( E_0 (1+z) \right)~, 
\end{equation}
where $P_{\gamma \to \gamma} (E_0,z)$ is the photon survival probability throughout the whole travel from the source to us. We suppose hereafter that $E_0$ lies in the VHE $\gamma$-ray band, and throughout this Section we employ cgs units for clarity.

\subsection{Conventional photon propagation}\label{sec:conve}

Within conventional physics the photon survival probability $P_{\gamma \to \gamma}^{\rm CP}  (E_0,z)$ is usually parametrized as
\begin{equation} 
\label{a012122010}
P_{\gamma \to \gamma}^{\rm CP} (E_0,z)  = e^{- \tau_{\gamma}(E_0,z)}~,
\end{equation}
where $\tau_{\gamma}(E_0,z)$ is the optical depth, which quantifies the dimming of the source. Note that ${\tau}_{\gamma}(E_0,z)$ increases with $z$, since a greater source distance entails a larger probability for a photon to disappear from the beam. Apart from atmospheric effects, one typically has $\tau_{\gamma}(E_0,z) < 1$ for $z$ not too large, in which case the Universe is optically thin up to the source. But depending on $E_0$ it can happen that $\tau_{\gamma}(E_0,z) > 1$, so that at some point the Universe becomes optically thick along the line of sight to the source. The value $z_h$ such that $\tau_{\gamma}(E_0,z_h) = 1$ defines the $\gamma$-ray horizon for a given $E_0$, and it follows from Eq. (\ref{a012122010}) that sources beyond the horizon tend to become progressively invisible as $z$ further increases past $z_h$. Owing to Eq. (\ref{a012122010}), Eq. (\ref{a02122010}) becomes
\begin{equation} 
\label{a02122010A}
\Phi_{\rm obs}(E_0,z) = e^{- \tau_{\gamma}(E_0,z)} \, \Phi_{\rm em} \left( E_0 (1+z) \right)~.
\end{equation}

Whenever dust effects can be neglected, photon depletion arises solely when hard beam photons of energy $E$ scatter off soft background photons of energy $\epsilon$ permeating the Universe and produce $e^+ e^-$ pairs through the standard $\gamma \gamma \to e^+ e^-$ process. Needless to say, in order for this process to take place enough energy has to be available in the centre-of-mass frame to create an $e^+ e^-$ pair. Regarding $E$ as an independent variable, the process is kinematically allowed for
\begin{equation} 
\label{eq.sez.urto01012011}
\epsilon > {\epsilon}_{\rm thr}(E,\varphi) \equiv \frac{2 \, m_e^2 \, c^4}{ E \left(1-\cos \varphi \right)}~,
\end{equation}
where $\varphi$ denotes the scattering angle and $m_e$ is the electron mass. Note that $E$ and $\epsilon$ change along the beam in proportion of $1 + z$. The corresponding Breit-Wheeler cross-section is~\cite{heitler}
\begin{equation} 
\label{eq.sez.urto}
\sigma_{\gamma \gamma}(E,\epsilon,\varphi)  \simeq 1.25 \cdot 10^{-25} \left(1-\beta^2 \right) \left[2 \beta \left( \beta^2 -2 \right) 
+ \left( 3 - \beta^4 \right) \, {\rm ln} \left( \frac{1+\beta}{1-\beta} \right) \right] {\rm cm}^2~,
\end{equation}
which depends on $E$, $\epsilon$ and $\varphi$ only through the dimensionless parameter 
\begin{equation} 
\label{eq.sez.urto01012011q}
\beta(E,\epsilon,\varphi) \equiv \left[ 1 - \frac{2 \, m_e^2 \, c^4}{E \epsilon \left(1-\cos \varphi \right)} \right]^{1/2}~, 
\end{equation}
and the process is kinematically allowed for ${\beta}^2 > 0$. The cross-section $\sigma_{\gamma \gamma}(E,\epsilon,\varphi)$ reaches its maximum ${\sigma}_{\gamma \gamma}^{\rm max} \simeq 1.70 \cdot 10^{- 25} \, {\rm cm}^2$ for $\beta \simeq 0.70$. Assuming head-on collisions for definiteness ($\varphi = \pi$), it follows that $\sigma_{\gamma \gamma}(E,\epsilon,\pi)$ gets maximized for the background photon energy 
\begin{equation} 
\label{eq.sez.urto-1}
\epsilon (E) \simeq \left(\frac{500 \, {\rm GeV}}{E} \right) \, {\rm eV}~,
\end{equation}
where $E$ and $\epsilon$ correspond to the same redshift.

Within the standard $\Lambda$CDM cosmological model ${\tau}_{\gamma}(E_0,z)$ arises by first convolving the spectral number density $n_{\gamma}({\epsilon}(z), z)$ of background photons at a generic redshift with ${\sigma}_{\gamma \gamma} (E(z), {\epsilon}(z), \varphi)$ along the line of sight for fixed values of $z$, $\varphi$ and ${\epsilon}(z)$, and next integrating over all these variables~\cite{stecker1971}. Hence, we have
\begin{eqnarray}
\label{eq:tau}
\tau_{\gamma}(E_0, z) = \int_0^{z} {\rm d} z ~ \frac{{\rm d} l(z)}{{\rm d} z} \, \int_{-1}^1 {\rm d}({\cos \varphi}) ~ \frac{1- \cos \varphi}{2} \ 
\times \\
\nonumber
\times  \, \int_{\epsilon_{\rm thr}(E(z) ,\varphi)}^\infty  {\rm d} \epsilon(z) \, n_{\gamma}(\epsilon(z), z) \,  
\sigma_{\gamma \gamma} \bigl( E(z), \epsilon(z), \varphi \bigr)~, \ \ 
\end{eqnarray}
where the distance travelled by a photon per unit redshift at redshift $z$ is given by
\begin{equation}
\label{lungh}
\frac{d l(z)}{d z} = \frac{c}{H_0} \frac{1}{\left(1 + z \right) \left[ {\Omega}_{\Lambda} + {\Omega}_M \left(1 + z \right)^3 \right]^{1/2}}~,
\end{equation}
with Hubble constant $H_0 \simeq 70 \, {\rm Km} \, {\rm s}^{-1} \, {\rm Mpc}^{-1}$, while ${\Omega}_{\Lambda} \simeq 0.7$ and ${\Omega}_M \simeq 0.3$ represent the average cosmic density of matter and dark energy, respectively, in units of the critical density ${\rho}_{\rm cr} \simeq 0.97 \cdot 10^{- 29} \, 
{\rm g} \, {\rm cm}^{ - 3}$.

Once $n_{\gamma}(\epsilon(z), z) $ is known, $\tau_{\gamma}(E_0, z)$ can be computed exactly, even though in general the integration over $\epsilon (z)$ in Eq. (\ref{eq:tau}) can only be performed numerically.   

Finally, in order to get an intuitive insight into the physical situation under consideration it may be useful to discard cosmological effects (which evidently makes sense for $z$ small enough). Accordingly, $z$ is best expressed in terms of the source distance $D = c z /H_0$ and the optical depth becomes 
\begin{equation}
\label{lungh26122010S}
\tau_{\gamma} (E,D) = \frac{D}{{\lambda}_{\gamma}(E)}~,
\end{equation}
where ${\lambda}_{\gamma}(E)$ is the photon mean free path for $\gamma \gamma \to e^+ e^-$ referring to the present cosmic epoch. As a consequence, Eq. (\ref{a012122010}) becomes
\begin{equation} 
\label{a012122010W}
P_{\gamma \to \gamma}^{\rm CP} (E,D) = e^{- D/{\lambda}_{\gamma}(E)}~,
\end{equation}
and so Eq. (\ref{a02122010A}) reduces to
\begin{equation}
\label{a1Z}
\Phi_{\rm obs}(E,D) = e^{- D/{\lambda}_{\gamma}(E)} \ \Phi_{\rm em}(E)~.
\end{equation}
Note that we have dropped the subscript $0$ for simplicity.

\subsection{Extragalactic Background Light (EBL)}

Blazars detected or detectable in the near future with IACTs lie in the VHE range $100 \, {\rm GeV} < E_0 < 100 \, {\rm TeV}$, and so from Eq. (\ref{eq.sez.urto-1}) it follows that the resulting dimming is expected to be maximal for a background photon energy in the range 
$0.005 \, {\rm eV} < \epsilon_0 < 5 \, {\rm eV}$ (corresponding to the frequency range $1.21 \cdot 10^{3} \, {\rm GHz} < \nu_0 < 1.21 \cdot 10^{5} \, {\rm GHz}$ and to the wavelength range $2.48 \, {\mu}{\rm m} < \lambda_0  < 2.48 \cdot 10^{2} \, {\mu}{\rm m}$), extending from the ultraviolet to the far-infrared. This is just the EBL. We stress that at variance with the case of the CMB, the EBL has nothing to do with the Big Bang. It is instead produced by stars in galaxies during the whole history of the Universe and possibly by a first generation of stars formed before galaxies were assembled. Therefore, a lower limit to the EBL level can be derived from integrated galaxy counts~\cite{madau}.

Determining the spectral number density $n_{\gamma}(\epsilon(z), z) $ of the EBL is a very difficult task. It is affected by large uncertainties, arising mainly from foreground contamination  produced by zodiacal light which is various orders of magnitude larger than the EBL itself~\cite{dwekrev}. While it is beyond the scope of this paper to review the various attempts in this direction, we briefly summarize below the present situation. Basically, six different approaches have been pursued:
\begin{itemize}
\item {\it Forward evolution} -- This is the most ambitious approach, since it begins from first principles, namely from semi-analytic models of galaxy formation in order to predict the time evolution of the galaxy luminosity function~\cite{primack, gilmore}. 
\item {\it Backward evolution} -- This starts from observations of the present galaxy population and extrapolates the galaxy luminosity function backward in time. Among others, this strategy has been followed by Stecker, Malkan and Scully (SMS)~\cite{steckerebl} and by Franceschini, Rodighiero and Vaccari (FRV)~\cite{franceschini}.
\item {\it Inferred evolution} -- This models the EBL by using quantities like the star formation rate, the initial mass function and the dust extinction as inferred from observations~\cite{kneiske2002, kneiske2004}
\item {\it Minimal EBL model} -- {This relies upon the same strategy underlying the previous item but with the parameters tuned in order to reproduce the EBL lower limits from galaxy counts}~\cite{kneiske2010}.
\item {\it Observed evolution} -- This method has the advantage to rely only upon observations by using a very rich sample of galaxies extending over the redshift range $0 \leq z \leq 1$~\cite{dominguez}.
\item {\it Compared observations} -- This technique has been implemented in two ways. One consists in comparing observations of the EBL itself with blazar observations with IACTs and deducing the EBL level from the VHE photon dimming~\cite{mazinraue}. Another starts from some $\gamma$-ray observations of a given blazar below $100 \, {\rm GeV}$ where EBL absorption is negligible and infers the EBL level by comparing the IACT observations of the same blazar with the source spectrum as extrapolated from the former observations~\cite{dwek2011}. In the latter case the main assumption is that the emission mechanism is presumed to be known with great accuracy. In either case, the crucial unstated assumption is that photon propagation in the VHE band is governed by conventional physics.
\end{itemize}

As it is evident, the latter approach does not apply to the DARMA scenario, and so it will not be considered any further. As far as the backward evolution approach is concerned, the models of SMS predict a much higher EBL level as compared to the model of FRV. Recently, the SMS models have been ruled out by Fermi/LAT observations~\cite{abdo}. On the other hand, a remarkable agreement exists among the FRV model and the other models based on forward evolution, inferred evolution and observed evolution. Throughout this paper, we adopt the FRV model mainly because it  supplies a very detailed numerical evaluation of the optical depth based on Eq. (\ref{eq:tau}), which will henceforth be denoted by $\tau_{\gamma}^{\rm FRV}(E_0, z)$~\cite{frvtable}. Regretfully, the errors affecting $\tau_{\gamma}^{\rm FRV}(E_0, z)$ are unknown.

\subsection{Understanding observed VHE blazar spectra}

As a preliminary step to find out the potential relevance of the DARMA scenario for available observations of VHE blazars, we consider in some detail the energy range $0.2 \, {\rm TeV} < E_0 <  2 \, {\rm TeV}$ where most of the blazars in question have been detected. It follows from Eq. (\ref{eq.sez.urto-1}) that the EBL energy band where $\sigma_{\gamma \gamma}(E_0,\epsilon_0,\pi)$ becomes maximal is $0.25 \, {\rm eV} < \epsilon_0 < 2.5 \, {\rm eV}$ (corresponding to $6.07 \cdot 10^4 \, {\rm GHz} < \nu_0 < 6.07 \cdot 10^5 \, {\rm GHz}$ and $0.50 \, \mu {\rm m} < \lambda_0 < 4.94 \, \mu {\rm m}$). 

{  So far, two specific processes have been proposed which can give rise to the emission of VHE photons~\cite{aharonianbook}}. 

\begin{itemize}

\item {  One is the {\it synchro-self-Compton} (SSC) mechanism, according to which relativistic electrons first emit X-ray photons by spiralling in the source magnetic field, which are subsequently boosted into the VHE $\gamma$-ray band by inverse Compton scattering off the parent electrons. In some cases, also external electrons can substantially contribute to the inverse Compton.} 

\item {  The competing mechanism is {\it hadronic pion production} (HPP) in proton-proton scattering, with neutral pions immediately decaying into VHE $\gamma$-ray pairs.}

\end{itemize} 

It turns out that both mechanisms lead to emission spectra which are so far observationally indistinguishable, and in particular within the energy range $0.2 \, {\rm TeV} < E < 2 \, {\rm TeV}$ they both predict a single power-law behaviour for blazar emitted spectra
\begin{equation} 
\label{a006012011A1}
\Phi_{\rm em}(E) = K \, E^{- \Gamma_{\rm em}}~,
\end{equation}
where $K$ is a suitable constant. 

\begin{figure}
\centering
\includegraphics[width=.80\textwidth]{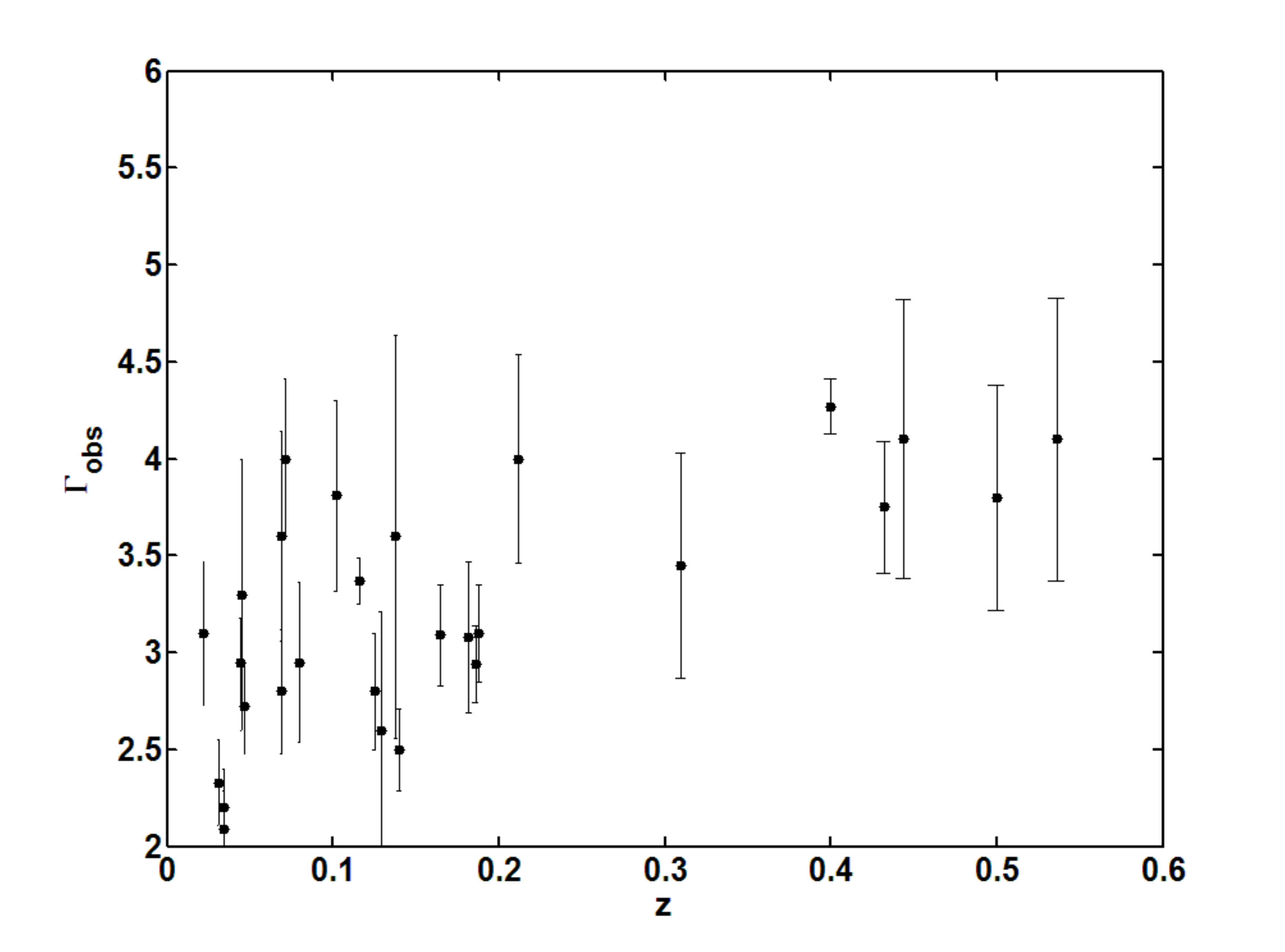}
\caption{\label{Slope-1.pdf} 
The observed values of the observed spectral index $\Gamma_{\rm obs}$ versus the source redshift for all blazars detected so far in the VHE band are represented by dots and corresponding error bars.}
\end{figure}

We next turn our attention to the observed energy spectra. It follows directly from observations that blazar spectra are successfully fitted by a single power law~\cite{remarkA}
\begin{equation} 
\label{a006012011A}
\Phi_{\rm obs}(E_0,z) = K \, E_0^{- \Gamma_{\rm obs} (z)}~.
\end{equation}
As a consequence, the observed spectra of all blazars detected so far are characterized by the observed spectral index $\Gamma_{\rm obs}$, which is 
reported in Table \ref{tab:a1} for every source. It is also very useful to plot $\Gamma_{\rm obs}$ versus the source redshift $z$ for all detected VHE blazars in Figure \ref{Slope-1.pdf} (blobs with error bars).

Let us first try to understand what Figure \ref{Slope-1.pdf} is telling us {  leaving aside any theoretical prejudice}. A striking feature is that the horizontal strip $3.5 < {\Gamma}_{\rm obs} < 4.5$ is almost uniformly populated for all considered redshifts, which would suggest that ${\Gamma}_{\rm obs}$ is independent of $z$. However, things are different for the lower strip $2.5 < {\Gamma}_{\rm obs} < 3.5$. Because it is populated only up to $z \simeq 0.2$ (with the exception of a single source close to $z \simeq 0.3$), the above interpretations is ruled out and we are forced to conclude that ${\Gamma}_{\rm obs}$ {\it correlates} with $z$. Actually, when looking at  Figure \ref{Slope-1.pdf} from this viewpoint a simple trend is easily recognized: ${\Gamma}_{\rm obs}$ increases linearly from 2.5 -- 3 at $0.1 < z < 0.2$ to roughly 3.5 -- 4 at $0.3 < z < 0.6$. Similarly, also for ${\Gamma}_{\rm obs} > 3$ a linear increase is found -- even if with a different slope -- but the number of sources with ${\Gamma}_{\rm obs} > 3$ following this behaviour decreases as $z$ increases until it vanishes for $z > 0.25$. 

A qualitative understanding of this situation emerges naturally by taking the EBL attenuation into account. We stress in the first place that rather nearby blazars -- such as those at $z < 0.05$ -- do not practically suffer EBL absorption at the energies probed so far, thereby implying that the shape of their observed VHE spectra should be the same as that of the emitted spectra, namely $\Gamma_{\rm obs} \simeq \Gamma_{\rm em}$. This is an important fact, since it allows us to see directly the blazar spectra at emission. In addition, we show in Appendix B that an approximate analytic expression for the optical depth within the FRV model is given by
\begin{equation}
\label{pndq1x20122010QP} 
\tau^{\rm app}_{\gamma}(E_0, z) \simeq 2.25 \, \alpha \left(\frac{E_0}{500 \, {\rm GeV}} \right)^{0.85} \, I(z)~,
\end{equation}
with $0.9 \leq \alpha \leq 3.6$ and roughly $ I(z) \sim z$. Hence, by combining Eqs. (\ref{a02122010A}), (\ref{a006012011A1}) and (\ref{pndq1x20122010QP}) the expected observed flux is
\begin{equation} 
\label{a01012011AA}
\Phi_{\rm obs}^{\rm app}(E_0,z) = K \, {\rm exp} \left\{ - \, 2.25 \, \alpha \left(\frac{E_0}{500 \, {\rm GeV}} \right)^{0.85} \, I(z) \right\} \, 
E_0^{- {\Gamma}_{\rm em}} \, \left(1 + z \right)^{- {\Gamma}_{\rm em}}~. 
\end{equation}
Now, Eq. (\ref{a01012011AA}) possesses to two conceptually distinct implications:
\begin{itemize}
\item $\Phi_{\rm obs}^{\rm app}(E_0,z)$ is exponentially damped as the energy increases, thereby entailing that it gets much softer than the emitted flux.
\item $\Phi_{\rm obs}^{\rm app}(E_0,z)$ is exponentially suppressed as the distance increases, so that sufficiently far-away sources tend to become invisible. 
\end{itemize}
Although Eq. (\ref{a01012011AA}) holds up to $2 \, {\rm TeV}$ only, one can check that these conclusions remain true under the replacement $\tau^{\rm app}_{\gamma}(E_0, z) \to \tau_{\gamma}^{\rm FRV}(E_0, z)$ up to $100 \, {\rm TeV}$.

We can relate the expected observed spectral index $\Gamma^{\rm exp}_{\rm obs} (z)$ to $\Gamma_{\rm em}$ and $z$ by best-fitting the l.h.s. of Eq. (\ref{a01012011AA}) to the power-law expression (\ref{a006012011A}) over the energy range where the considered source is observed, which is reported in Table \ref{tab:a1}. 

Because $I(z)$ is independent of $E_0$, it is unaffected by the considered best-fitting procedure, and so we have $\Gamma^{\rm exp}_{\rm obs} (z) \sim I(z) \sim z$ up to logarithmic corrections. This indeed explains in a qualitative fashion both why nearby sources with $2.5 < \Gamma_{\rm obs} < 3$ get replaced by sources with $3.5 < \Gamma_{\rm obs} < 4$ at larger redshift according to a linear trend and why nearby sources with $\Gamma_{\rm obs} > 3$ follow a similar linear behaviour up to a point where $\Gamma_{\rm obs}$ would be so large that the source becomes invisible at sufficiently large distances, thereby disappearing from Figure \ref{Slope-1.pdf} for $z$ large enough.

{  An intrinsic correlation -- due to an observational bias -- between the spectral index and the distance, however, cannot be excluded on the basis of the experimental data. In addition, given the blazar sequence~\cite{bs}, the fact that the Inverse Compton bump moves according to luminosity might give a bias related to the fact that for the same energy range we are actually sampling different regions of the spectral energy distribution. Finally, the fact that the upper limit of the energy sampled decreases with energy might introduce in itself a bias. However, a direct search for spectral index hardening associated with blazar variability gave no evidence~\cite{vp}.  Attempts are presently done (see for example~\cite{tav}) to analyze individually blazars and derive their spectral energy distribution from multi-wavelength data. We hope that in a near future this work will be made more precise. However, the models have  presently a large uncertainty, since, in order to have reasonably constrained fits, one must assume a purely leptonic emission and a 1-zone SSC emission mechanism, while we have indications that the situation can be more complicated for most blazars we know in detail.}

In order to derive the exact value of $\Gamma_{\rm em}$ for the various blazars from observations the use of Eq. (\ref{a01012011AA}) with 
$\Phi_{\rm obs}^{\rm app}(E_0,z) \to \Phi_{\rm obs}(E_0,z)$ combined with Eq. (\ref{a006012011A}) would be unsuited because of its approximate character. A better strategy consists in first de-absorbing $\Gamma_{\rm obs}$ for every source by employing Eq. (\ref{a02122010A}) with 
$\tau_{\gamma} (E_0, z) \to \tau_{\gamma}^{\rm FRV}(E_0, z)$ combined with Eq. (\ref{a006012011A}), and next inferring 
$\Gamma_{\rm em}$ by best-fitting the resulting $\Phi_{\rm em} (E)$ to the power-law expression (\ref{a006012011A1}) over the energy range where the considered source is observed (see Table \ref{tab:a1}). Observe that since $\Gamma_{\rm em}$ depends linearly on $\Gamma_{\rm obs}$, the derived values of $\Gamma_{\rm em}$ have the {\it same} error bars of $\Gamma_{\rm obs}$ as reported in Table \ref{tab:a1} to the extent that errors in $\tau_{\gamma}^{\rm FRV}(E_0, z)$ are neglected (they are actually unknown because they are not quoted by the authors). We are of course well aware that the correct procedure would be to first de-absorb each point of the observed spectrum of a given source and next best-fit these points to a power-law. Unfortunately, the observed energy points with related error bars are not available from published papers, and this explains why we have simply de-absorbed $\Gamma_{\rm obs}$ -- hence $\Gamma_{\rm em}$ has to be understood as the average emitted spectral index for the source in question -- but we want to remark that for our purposes this simplified approach is adequate. The same strategy has been used for a different model of the EBL~\cite{gilmore}. Our results are listed in Table \ref{tab:a1}, along with the spread $\Delta \Gamma_{\rm em}$ of the value of $\Gamma_{\rm em}$ discarding errors and the average value $\langle \Gamma_{\rm em} \rangle$ including errors.

\subsection{The cosmic opacity problem}

It is evident from Table \ref{tab:a1} that the values of ${\Gamma}_{\rm em}$ for some far-away sources are considerably smaller than those for nearby blazars, with the exception of the two sources H 1426+428 and 1ES 0229+200. 

As a consequence, the cosmic opacity problem arises concerning the physical mechanism responsible for such a behaviour involving $\Gamma_{\rm em}$ and $z$.

Certainly cosmology does not help, because no important evolutionary effect is expected to take place for redshifts up to $z \simeq 0.54$ at which the most distant blazar has been detected.

Alternatively, one might guess that it is due to a volume selection effect, since intrinsically brighter sources are the exception rather than the rule. However, the emitted flux depends not only on the slope but also on the normalization factor, which varies by three orders of magnitude over the sample of considered sources. Actually, the existence of the two rather nearby blazars H 1426+428 and 1ES 0229+200 with the hardest emitted spectrum explicitly shows that small $\Gamma_{\rm em}$ does not mean large $z$. So, also this attempt is unsatisfactory.

Yet another possible explanation consists in assuming that far-away blazars are intrinsically different from nearby ones, but to the best of our knowledge no convincing explanation of this circumstance has been put forward so far. 

One might also argue that a solution could come from the fact that some observed blazars are in a quiescent state whereas others are flaring. In fact, because of EBL absorption we might be seeing progressively more distant blazars only during stronger flares (but not all distant blazars are flaring, like e.g. PG 1553+113 which has been observed to have nearly the same luminosity for five years). As a consequence -- working within the SSC mechanism for definiteness -- we could run the risk to compare $\Gamma_{\rm em}$ for different sources at different positions on the Compton peak, since flaring causes this peak which normally lies below $100 \, {\rm GeV}$ to slightly shift towards higher energies~\cite{ghisello}. Clearly, the slope near the bottom of the pick is steeper than close to the tip, and this circumstance would produce a harder emission spectrum for flaring sources. However, such a possibility seems to us quite unlikely. For, the observed energy range of flaring sources is generally considerably wider that the width of the peak~\cite{ghisello} and observations above $100 \, {\rm GeV}$ invariably show that a {\it single} power law behaviour provides an excellent fit to the data. Hence, we see that we are inferring the spectral index well below the pick whether or not a flare takes place.

In conclusion, no satisfactory explanation for the considered behaviour involving $\Gamma_{\rm em}$ and $z$ seems to emerge. 

As a matter of fact, shifting from the astronomical to the physical point of view makes the issue more clear-cut. It is known since a long time that 
${\Gamma}_{\rm em} = 1.5$ arises from the first-order Fermi acceleration mechanism with newtonian shocks for an electron injection spectrum equal to 2~\cite{fermi1}. For this reason, when the cosmic opacity problem was first perceived in 2006 it was thought that the inferred values of $\Gamma_{\rm em}$ were too small to agree with conventional physics assuming current EBL models~\cite{aharonian:nature06}, and indeed values ${\Gamma}_{\rm em} < 1.5$ were considered unphysical e.g. by the H.E.S.S. collaboration. However, it has recently been shown that the required low values of ${\Gamma}_{\rm em}$ can be achieved in the presence of strong relativistic shocks~\cite{steckershoks}, because of photon self-absorption inside the source~\cite{costamante} or by the inverse Compton scattering of CMB photons by shock-accelerated electrons in the jet~\cite{bdf}.  

While these results are gratifying, one still wonders why this kind of physical effects are important for distant blazars {\it only}. Just as before, an answer seems hardly in sight. 

So, either way it is argued no simple solution to the cosmic opacity problem emerges within conventional physics.

Now, what the discussion in Subsection III-C has shown is that the origin of this cosmic opacity problem is not qualitative -- the $z$-dependence of 
$\Gamma_{\rm obs}(z)$ comes out right -- but purely numerical, namely because the EBL level predicted by conventional physics is too high. Were the EBL level somewhat smaller, the cosmic opacity problem would automatically disappear. 

A way out of the cosmic opacity problem appears therefore to call for some sort of {\it unconventional} photon propagation which ultimately reduces the cosmic opacity arising from the EBL, thereby bringing the values of ${\Gamma}_{\rm em}$ for distant blazars in the same ballpark of those for close ones.

Various attempts at reducing the cosmic opacity along these lines have been proposed and they are schematically summarized below:

\begin{itemize}
\item A revolutionary option contemplates a breakdown of Lorentz invariance~\cite{kifune}. 
\item An alternative possibility concerns the emission of cosmic rays from blazars -- rather than photons -- with energy smaller than $50 \, {\rm EeV}$. These cosmic rays can travel unimpeded over cosmological distances and they can interact with the EBL well before reaching our galaxy. In such an interaction secondary photons are produced, that are ultimately detected by the IACTs~\cite{kusenko}.
\item A different proposal relies upon photon-ALP oscillations, which requires the presence of magnetic fields somewhere along the line of sight. As already pointed out, two concrete realizations of this idea have been investigated. One of them -- the DARMA scenario~\cite{darma, dmpr, mm} -- assumes that photon-ALP oscillations take place during propagation in intergalactic space, where large-scale magnetic fields in the nano-Gauss range are supposed to exist. Large-scale magnetic fields of this strength are consistent with current upper bounds and even with the results of the AUGER observatory (more about this, later). The other is in a sense complementary, because it presupposes a $\gamma \to a$ conversion inside the blazar and a $a \to \gamma$ conversion in the Milky way~\cite{bis}. Although the properties of the Galactic magnetic field are rather well known, those of the magnetic field in the blazar are not, and so it is not clear whether the first step of this mechanism actually takes place and if so how large is its efficiency~\cite{br}.
\end{itemize}

The present paper is devoted to a careful investigation of the solution based on the DARMA scenario, even if its scope is by far more general.

\section{DARMA SCENARIO}

Our aim is to offer a detailed description of the structure of the DARMA scenario, and in particular to show how the photon survival probability 
$P^{\rm DARMA}_{\gamma \to \gamma}(E_0,z)$ can be computed in terms of the properties of the intergalactic medium in which the photon/ALP beam propagates.

\subsection{An intuitive insight}

We find it instructive to restate in a slightly different fashion the reason why the mechanism of photon-ALP oscillations allows to substantially reduce the EBL absorption. We neglect here cosmological effects for simplicity.

We suppose that VHE photons are both emitted and detected as usual, but that along their way to us they convert into ALPs and back into photons. Accordingly, the number $N_c$ of either $\gamma \to a$ or $a \to \gamma$ conversions must necessarily be even, and we may schematically regard the beam propagation in large-scale magnetic fields as a succession of such conversions. Assuming ideally that each conversion occurs suddenly at some space point, the source distance $D$ gets divided into a number $N_c + 1$ of steps of equal length $L$, over which a beam particle behaves either as a real photon or as an ALP. Hence, a beam particle exhibits an overall behaviour as a real photon over a total length equal to
\begin{equation}
\label{a1bisa13022011}
D_{\gamma} = \frac{N_c + 2}{2 \left( N_c +1 \right)} \, D = \left(1 - \frac{N_c}{2 N_c + 2} \right) D~.
\end{equation}
We intuitively expect $N_c$ to increase with the photon-ALP oscillation probability -- and so with $B/M$ -- which leads in turn to a slight decrease of $D_{\gamma}$ starting from $D$. Correspondingly, since ALPs do not suffer EBL absorption Eq. (\ref{a012122010W}) gets presently replaced by
\begin{equation} 
\label{a012122010Wp}
P^{\rm DARMA}_{\gamma \to \gamma} (E,D) = e^{- D_{\gamma}/{\lambda}_{\gamma}(E)} = e^{- \left(1 - \frac{N_c}{2 N_c + 2} \right) D/{\lambda}_{\gamma}(E)}~, 
\end{equation}
and thanks to its exponential dependence on $D_{\gamma}/{\lambda}_{\gamma}(E)$ even a small decrease of $D_{\gamma}$ starting from $D$ produces a large enhancement of $P^{\rm DARMA}_{\gamma \to \gamma} (E,D)$ as compared to $P^{\rm CP}_{\gamma \to \gamma} (E,D)$ referring to conventional physics and given by Eq. (\ref{a012122010W})~\cite{remark2}.

\subsection{General strategy}

Our ultimate goal consists in the evaluation of the photon survival probability $P^{\rm DARMA}_{\gamma \to \gamma}(E_0,z)$ from a blazar at redshift $z$ to us when allowance is made for photon-ALP oscillations as well as for EBL photon absorption. It is indeed clear that the considerations developed in Section III can be extended to account for photon-ALP oscillations by the replacement $P^{\rm CP}_{\gamma \to \gamma}(E_0,z) \to P^{\rm DARMA}_{\gamma \to \gamma}(E_0,z)$.

An exact treatment would however be impossible because of the large uncertainty affecting the configuration of the magnetic field ultimately responsible for photon-ALP oscillations. 

As a matter of fact, the line of sight to a distant blazar is expected to traverse magnetic fields extending over a variety of scales. A magnetic field is certainly present inside the source~\cite{ghisel}. Furthermore, the Milky Way magnetic field can give a nontrivial contribution to the effect under consideration~\cite{mwmf}, and the same is true if the line of sight happens to cross a cluster of galaxies because intracluster magnetic fields are known to exist with a strength similar to that of the Galactic field~\cite{intracluster}. Finally, large-scale magnetic fields can play a key role~\cite{lsmagfields1, lsmagfields2}. Here, our attention will be restricted to magnetic fields of the latter sort.

Unfortunately, almost nothing is known about the morphology of large-scale magnetic fields, which reflects both their origin and the evolutionary history of baryonic matter. While it is evident that their coherence length cannot be arbitrarily large, no reliable estimate of its value is presently available. As far as our analysis is concerned, this means that we cannot suppose that large-scale magnetic fields are homogeneous over the whole distance to the source, but their spatial dependence is largely unknown. The usual way out of this difficulty amounts to suppose that large-scale magnetic fields ${\bf B}$ have a {\it domain-like structure}. That is, ${\bf B}$ is assumed to be homogeneous over a domain of size $L_{\rm dom}$ equal to its coherence length, with ${\bf B}$ randomly changing its direction from one domain to another but keeping approximately the same strength~\cite{lsmagfields1, lsmagfields2}. 

Thus, the whole propagation process of the considered photon/ALP beam can be recovered by iterating the propagation over a single domain as many times as the number of domains crossed by the beam, taking each time a random value for the angle $\psi$ between ${\bf B}$ and a fixed fiducial direction equal for all domains. In this way, we are effectively led to the much easier problem of photon-ALP oscillations in a {\it homogeneous} magnetic field (solved exactly in  Subsection II-E). 

What still remains to be done at this point is to take photon absorption into account. This is easy because photon absorption is {\it independent} of the properties of the photon-ALP oscillation mechanism and vice-versa. 

So, our strategy can be implemented through the following steps~\cite{peloso}:
\begin{itemize}
\item We work within the {\it strong-mixing regime} so as to ensure that the photon-ALP oscillation probability is both maximal and energy-independent. We have seen that such a condition requires $E > E_*$, with the energy threshold $E_*$ defined by Eq. (\ref{a17231209}). But demanding the strong-mixing regime to take place for $E > 100 \, {\rm GeV}$ evidently requires $E_* < 100 \, {\rm GeV}$, which sets an upper bound on the ALP mass.
\item We evaluate the transfer matrix across the generic $n$-th domain ${\cal U}_n (E_0, \psi_n)$, where $\psi_n$ accounts for the random orientation of 
${\bf B}$ in the domain in question. Note that ${\cal U}_n (E_0, \psi_n)$ depends on $E_0$ only because of the energy-dependence of EBL absorption.
\item Iteration of the latter result over the total number $N_d$ of domains crossed by the beam from the blazar to us yields the total transfer matrix  
${\cal U} (E_0, z; \psi_1, ... ,\psi_{N_d})$, from which the photon survival probability $P_{\gamma \to \gamma}(E_0,z; \psi_1, ... ,\psi_{N_d})$ can be computed for fixed values of the angles $\psi_1, ... ,\psi_{N_d}$ in every domain.
\item Finally, $P^{\rm DARMA}_{\gamma \to \gamma}(E_0,z)$ emerges by averaging $P_{\gamma \to \gamma}(E_0,z; \psi_1, ... ,\psi_{N_d})$ over all angles $\psi_1, ... ,\psi_{N_d}$.
\end{itemize}

Our discussion is framed within the $\Lambda$CDM cosmological setting, and so the redshift $z$ is the obvious parameter to express distances. Because the proper length per unit redshift at redshift $z$ is still given by Eq. (\ref{lungh}), a generic proper length extending over the redshift interval 
$[z_a,z_b]$ ($z_a < z_b$) is 
\begin{equation}
\label{lungh14012011}
L (z_a,z_b) = \int_{z_a}^{z_b} d z~ \frac{d l(z)}{d z}  \simeq 4.29 \cdot 10^3 \int_{z_a}^{z_b} \frac{d z}{\left(1 + z \right) \left[ 0.7 + 0.3 \left(1 + z \right)^3 \right]^{1/2}} \ {\rm Mpc}~,
\end{equation}
which approximately reads
\begin{equation}
\label{lunghK}
L (z_a,z_b) \simeq 2.96 \cdot 10^3 \, {\rm ln} \left( \frac{1 + 1.45 \, z_b}{1 + 1.45 \, z_a} \right) \, {\rm Mpc}~.
\end{equation}
This result will be applied in particular to evaluate the size of the magnetic domains.

\subsection{Photon absorption}

We proceed to extend the discussion in Subsection II-E so as to take EBL absorption into account. This task is greatly facilitated by the fact that the latter effect is independent of the photon-ALP conversion mechanism.

We have seen that the propagation of a monochromatic photon/ALP beam is formally described as a three-level non-relativistic quantum system with Hamiltonian $H_0$ given by Eq. (\ref{k3la}) and expressed in terms of the mixing matrix ${\cal M}_0$. Taking advantage from this fact, the inclusion of EBL absorption amounts to suppose that the photon/ALP beam is actually analogous to an {\it unstable} quantum system with decay probability 
\begin{equation}
\label{a1dop28012011}
P_{\rm decay} = e^{- y/{\lambda}_{\gamma}(E)}~,
\end{equation}
where ${\lambda}_{\gamma} (E)$ denotes the photon mean free path. As is well known, such a decay probability arises from the inclusion of an absorbitive term 
$- \Delta_{\rm abs}$ into the Hamiltonian, with
\begin{equation}
\label{h1el}
{\Delta}_{\rm abs} \equiv \frac{i}{2 \, {\lambda}_{\gamma}(E)}~.
\end{equation}
More specifically, since photons undergo absorption but ALPs do not, ${\cal M}_0^{(0)}$ in Eq. (\ref{a9}) becomes
\begin{equation}
\label{aa8MRq28012011}
{\cal M}^{(0)} = \left(
\begin{array}{ccc}
\Delta_{\rm pl} + {\Delta}_{\rm abs} & 0 & 0 \\
0 & \Delta_{\rm pl} + {\Delta}_{\rm abs} & \Delta_{a\gamma}  \\
0 & \Delta_{a \gamma} & \Delta_{a a} \\
\end{array}
\right)~,
\end{equation} 
where -- in parallel with the treatment of Subsection II-E -- we are first supposing that ${\bf B}$ lies along the $z$-axis.

As we said, we work throughout within the strong-mixing regime and therefore condition (\ref{a28012011}) has to be met. Recalling the explicit expression for the various $\Delta$-terms entering Eq. (\ref{aa8MRq28012011}) and defined in Subsection II-E, ${\cal M}^{(0)}$ takes the simpler form
\begin{equation}
\label{aa8MRq28012011a}
{\cal M}^{(0)} = \left(
\begin{array}{ccc}
{\Delta}_{\rm abs} & 0 & 0 \\
0 &  {\Delta}_{\rm abs} & \Delta_{a\gamma}  \\
0 & \Delta_{a \gamma} & 0 \\
\end{array}
\right)~,
\end{equation} 
which is denoted by the same symbol for notational simplicity (only Eq. (\ref{aa8MRq28012011a}) will be used hereafter). Note that $m$ and 
${\omega}_{\rm pl}$ presently drop out of ${\cal M}^{(0)}$. Just as before, use of the results contained in Appendix A with ${\cal M} \to {\cal M}^{(0)}$ directly gives the corresponding eigenvalues
\begin{equation}
\label{a91212a1P}
{\lambda}_{1} = \frac{i}{2 \, {\lambda}_{\gamma}(E)}~,
\end{equation}
\begin{equation}
\label{a91212a2P}
{\lambda}_{2} = \frac{i}{4 \, {\lambda}_{\gamma}(E)} \left(1 - \sqrt{1 - 4 \, \delta^2} \right)~, 
\end{equation}
\begin{equation}
\label{a91212a3P}
{\lambda}_{3} = \frac{i}{4 \, {\lambda}_{\gamma}(E)}  \left(1 + \sqrt{1 - 4 \, \delta^2} \right)~,
\end{equation}
where we have set
\begin{equation}
\label{a17P}
{\delta} \equiv \frac{B \,  {\lambda}_{\gamma}(E)}{M}~,
\end{equation}
roughly measuring the ratio of the photon mean free path to the photon-ALP oscillation length. Hence, the transfer matrix associated with the reduced Sch\"odinger-like equation (\ref{k3l1}) with ${\cal M}_0 \to {\cal M}^{(0)}$ reads
\begin{equation}
\label{mravvq2abcappZ}
{\cal U} (y, y_0; 0) = e^{i {\lambda}_1 (y - y_0)} \, T_1 (0) + e^{i {\lambda}_2 (y - y_0)} \, T_2 (0) + e^{i {\lambda}_3 (y - y_0)} \, T_3 (0)~, 
\end{equation}
with the matrices $T_1 (0)$, $T_2 (0) $ and $T_3 (0)$ dictated by Eqs. (\ref{mravvq1app}), (\ref{mravvq2app}) and (\ref{mravvq3app}) as specialized to the present case. Explicitly
\begin{equation}
\label{mravvq2appZP1c}
T_1 (0) \equiv
\left(
\begin{array}{ccc}
1 & 0 & 0 \\
0 & 0 & 0 \\
0 & 0 & 0
\end{array}
 \right)~,
\end{equation}
\begin{equation} 
\label{mravvq2appZP1d}
T_2 (0) \equiv
\left(
\begin{array}{ccc}
0 & 0 & 0 \\ 
0 & \frac{- 1 + \sqrt{1 - 4 {\delta}^2}}{2 \sqrt{1 - 4 {\delta}^2}} & \frac{i \delta}{\sqrt{1 - 4 {\delta}^2}} \\
0 & \frac{i \delta}{\sqrt{1 - 4 {\delta}^2}} & \frac{ 1 + \sqrt{1 - 4 {\delta}^2}}{2 \sqrt{1 - 4 {\delta}^2}}
\end{array}
\right)~,
\end{equation} 
\begin{equation}
\label{mravvq3appZP}
T_3 (0) \equiv 
\left(
\begin{array}{ccc}
0 & 0 & 0 \\
0 & \frac{ 1 + \sqrt{1 - 4 {\delta}^2}}{2 \sqrt{1 - 4 {\delta}^2}} & - \frac{i \delta}{\sqrt{1 - 4 {\delta}^2}}   \\
0 & - \frac{i \delta}{\sqrt{1 - 4 {\delta}^2}}   & \frac{- 1 + \sqrt{1 - 4 {\delta}^2}}{2 \sqrt{1 - 4 {\delta}^2}} 
\end{array}
\right)~.
\end{equation}

Clearly, we will need the generalization of this result to the case in which ${\bf B}$ forms an arbitrary angle $\psi$ with the $z$-axis. Proceeding exactly 
like in Subsection II-E, we find
\begin{equation}
\label{mr101109qqP}
{\cal M} = V^{\dagger} (\psi) \, {\cal M}^{(0)} \, V (\psi)~, 
\end{equation}
where the matrix $V (\psi)$ is given by Eq. (\ref{a91212ay}). This yields
\begin{equation}
\label{aa8MRqP}
{\cal M} = \left(
\begin{array}{ccc}
\Delta_{\rm abs} & 0 & \Delta_{a \gamma} \, \sin \psi \\
0 & \Delta_{\rm abs} & \Delta_{a\gamma} \, \cos \psi \\
\Delta_{a \gamma} \, \sin \psi& \Delta_{a \gamma} \, \cos\psi& 0 \\
\end{array}
\right)~,
\end{equation} 
and now the resulting transfer matrix evidently reads
\begin{equation}
\label{mravvq1P}
{\cal U} (y,y_0 ; \psi) = V^{\dagger} (\psi) \, {\cal U} (y,y_0 ; 0) \, V (\psi)~,
\end{equation}
whose explicit form arises by inserting Eq. (\ref{mravvq2abcappZ}) into Eq. (\ref{mravvq1P}). We obtain 
\begin{equation}
\label{mravvq2abcQ}
{\cal U} (y,y_0 ; \psi) = e^{i {\lambda}_1 (y - y_0)} \, T_1 (\psi) + e^{i {\lambda}_2 (y - y_0)} \, T_2 (\psi) + e^{i {\lambda}_3 (y - y_0)} \, T_3 (\psi)~, 
\end{equation}
with
\begin{equation}
\label{mravvq2Q1aX2}
T_1 (\psi) \equiv
\left(
\begin{array}{ccc}
\cos^2 \psi & -\sin \psi \cos \psi & 0 \\
- \sin \psi \cos \psi & \sin^2 \psi & 0 \\
0 & 0 & 0
\end{array}
\right)~,
\end{equation}
\begin{equation}
\label{mravvq3Q1a}
T_2 (\psi) \equiv 
\left(
\begin{array}{ccc}
\frac{- 1 + \sqrt{1 - 4 {\delta}^2}}{2 \sqrt{1 - 4 {\delta}^2}} \sin^2 \psi & \frac{- 1 + \sqrt{1 - 4 {\delta}^2}}{2 \sqrt{1 - 4 {\delta}^2}} \sin \psi \cos \psi & \frac{i \delta}{\sqrt{1 - 4 {\delta}^2}} \sin \psi \\
\frac{- 1 + \sqrt{1 - 4 {\delta}^2}}{2 \sqrt{1 - 4 {\delta}^2}} \sin \psi \cos \psi & \frac{- 1 + \sqrt{1 - 4 {\delta}^2}}{2 \sqrt{1 - 4 {\delta}^2}} \cos^2 \psi & \frac{i \delta}{\sqrt{1 - 4 {\delta}^2}} \cos \psi \\
\frac{i \delta}{\sqrt{1 - 4 {\delta}^2}} \sin \psi & \frac{i \delta}{\sqrt{1 - 4 {\delta}^2}} \cos \psi & \frac{ 1 + \sqrt{1 - 4 {\delta}^2}}{2 \sqrt{1 - 4 {\delta}^2}}
\end{array}
\right)~,
\end{equation}
\begin{equation} 
\label{mravvq2Q1bX3}
T_3 (\psi) \equiv
\left(
\begin{array}{ccc}
\frac{ 1 + \sqrt{1 - 4 {\delta}^2}}{2 \sqrt{1 - 4 {\delta}^2}} \sin^2 \psi  & \frac{ 1 + \sqrt{1 - 4 {\delta}^2}}{2 \sqrt{1 - 4 {\delta}^2}} \sin \psi \cos \psi  & \frac{- i \delta}{\sqrt{1 - 4 {\delta}^2}}    \sin \psi \\ 
\frac{ 1 + \sqrt{1 - 4 {\delta}^2}}{2 \sqrt{1 - 4 {\delta}^2}} \sin \psi \cos \psi  & \frac{ 1 + \sqrt{1 - 4 {\delta}^2}}{2 \sqrt{1 - 4 {\delta}^2}} \cos^2 \psi  & \frac{- i \delta}{\sqrt{1 - 4 {\delta}^2}} \cos \psi \\
\frac{- i \delta}{\sqrt{1 - 4 {\delta}^2}} \sin \psi  & \frac{- i \delta}{\sqrt{1 - 4 {\delta}^2}} \cos \psi  &  \frac{- 1 + \sqrt{1 - 4 {\delta}^2}}{2 \sqrt{1 - 4 {\delta}^2}}   
\end{array}
\right)~.
\end{equation} 

We stress that due to the imaginary nature of ${\Delta}_{\rm abs}$ the mixing matrix ${\cal M}$ is not self-adjoint, and so the transfer matrix 
${\cal U} (y,y_0 ; \psi)$ fails to be unitary. In addition, the Von Neumann-like equation (\ref{k3lw}) becomes
\begin{equation}
\label{k3lwY}
i \frac{d \rho}{d y} = \rho \, {\cal M}^{\dagger} - {\cal M} \rho~.
\end{equation}
Still, it is straigthforward to check that Eq. (\ref{k3lwf1}) remains valid with ${\cal U}_0 (y,y_0) \to {\cal U} (y,y_0 ; \psi)$ in spite of the fact that ${\cal M}^{\dagger} \neq {\cal M}$. Since we now have $\rho (y)^{\dagger} \neq \rho (y)$, it follows that in general ${\rm Tr} \rho (y) \neq 1$, and so the probability that a photon/ALP beam initially in the state $\rho_1$ will be found in the state $\rho_2$ after a distance $y$ is presently given by
\begin{equation}
\label{k3lwf1w1}
P_{\rho_1 \to \rho_2} (y) = {{\rm Tr} \left( {\rho}_2 \, {\cal U} (y,0 ; \psi) \, {\rho_1} \, {\cal U}^{\dagger} (y,0 ; \psi) \right)}~,
\end{equation}
where we assume ${\rm Tr} \rho_1 = {\rm Tr} \rho_2 = 1$ as before.

\subsection{Intergalactic medium (IGM)}

As is well known, the absence of the Gunn-Peterson effect~\cite{gp} is generally regarded as evidence that the IGM is ionized, and from the resulting high electrical conductivity it follows that the electron number density $n_e (z)$ traces the cosmic mass distribution. Because of this fact, we have
\begin{equation}
\label{k3lwf1w110022011}
n_e (z) = {\bar n}_{e,0} (1 + \delta (z)) (1 + z)^3~,
\end{equation}
where
$\delta(z) \equiv (\rho(z) - {\bar \rho} (z)) / {\bar \rho} (z)$ is the mass density contrast and ${\bar n}_{e,0}$ is the average electron number density. As a consequence, Eq. (\ref{a6zz}) entails for the plasma frequency
\begin{equation}
\label{mag210022011}
{\omega}_{\rm pl} (z) = {\bar \omega}_{{\rm pl},0} (1 + \delta(z))^{1/2} \left(1 + z \right)^{3/2}~,
\end{equation}
with ${\bar \omega}_{{\rm pl},0}$ obviously corresponding to ${\bar n}_{e,0}$.

Observations of the primordial abundance of the light elements yields ${\bar n}_{e,0} \simeq 1.8 \cdot 10^{- 7} \, {\rm cm}^{- 3}$, but it has been argued that in the $z < 1$ Universe which is relevant for us $n_e (z)$ ought to be smaller than $ {\bar n}_{e,0}$ by a factor 15~\cite{csaki2}. Correspondingly, from Eqs. (\ref{a6zz}) and (\ref{mag210022011}) we get
\begin{equation}
\label{mag210022011q}
{\omega}_{\rm pl} (z) \simeq 4.04 \cdot 10^{- 15} \left( 1 + z \right)^{3/2} \, {\rm eV}~,
\end{equation}
where the $(1 + \delta(z))^{1/2}$ factor has been dropped because irrelevant.

A crucial issue concerns the large-scale magnetic fields traversed by the beam, whose origin and structure is still unknown to a large extent. A possibility is that very small magnetic fields present in the early Universe were subsequently amplified by the process of structure formation~\cite{dolag}. An alternative option is that the considered magnetic fields have been generated in the low-redshift Universe by energetic quasar outflows~\cite{furlanetto}. Finally, it has been suggested that large-scale magnetic fields originated from the so-called Biermann battery effect~\cite{biermann}, namely from electric currents driven by merger shocks during the structure formation processes. Presumably, all these mechanisms can take place, even if it is presently impossible to assess their relative importance~\cite{lsmagfields1, lsmagfields2}. At any rate, we suppose that magnetic fields already exist out to the redshift $z = 1$.

Owing to the high conductivity of the IGM, the magnetic flux lines can be thought as frozen inside the IGM. Therefore, flux conservation during the cosmic expansion entails that $B$ scales like the volume to the power $2/3$, thereby implying the magnetic field strength in a domain at redshift $z$ 
is~{  \cite{lsmagfields2}
\begin{equation}
\label{mag1}
B = B_0 \, \left(1 + z \right)^2~.       
\end{equation}

{  In 2007 the AUGER collaboration reported positive evidence for a correlation between charged cosmic rays and candidate sites for emission~\cite{auger1}. More in detail, the AUGER collaboration found that 20 out of the 27 recorded events with energy larger than $57 \, {\rm EeV}$ are located within $3.1^{\circ}$ of an AGN closer than $75 \, {\rm Mpc}$ from Earth. The conclusion drown from the AUGER collaboration is that such a result is inconsistent with the hypothesis of an isotropic distribution of these cosmic rays with at least a $99 \, \%$ confidence level from a prescribed a-priori test. As explained elsewhere, this fact  supports the existence of large-scale magnetic fields with coherence length $L_{\rm dom}$ in the range $1 - 10 \, {\rm Mpc}$ and strength $B_0$ in the range $0.1- 1 \, {\rm nG}$ at $z = 0$~\cite{dpr}. However, such a correlation has become considerably weaker when a larger data set (69 events, including the events on which the previous publication was based) has been recorded and analyzed by the AUGER colaboration~\cite{auger2}. The fraction of events correlated to a nearby AGN is in the most recent publication of $38 \, \%$ -- to be compared with an expected value of $21 \, \%$ in the case of no correlation -- and no a-priori probability estimate is provided in the new paper. The present situation appears to us unclear, even though the conclusions obtained from the first AUGER results are still statistically consistent with the more recent ones.}} 

For this reason, we prefer { to avoid committing} ourselves with any conclusion relying upon the AUGER data and we consider only well-established upper bounds. They depend on the size of their domain-like structure $L_{\rm dom}$ and within the current cosmological setting they take the form~\cite{blasi}
\begin{equation}
\label{mag210022011w}
B_0 < 3.8 \, {\rm nG}   \ \ {\rm for}   \ \ L_{\rm dom} = 50 \, {\rm Mpc}~,
\end{equation}

\begin{equation}
\label{mag210022011e}
B_0 < 6.3 \, {\rm nG}   \ \ {\rm for}   \ \ L_{\rm dom} = 1 \, {\rm Mpc}~.
\end{equation}
It is usually supposed that $1 \, {\rm Mpc} \leq L_{\rm dom} \leq 10 \, {\rm Mpc}$, and so we will assume throughout
\begin{equation}
\label{mag210022011f}
B_0 < 6 \, {\rm nG}~.
\end{equation}

Within the cosmological context, the overall structure of the cellular configuration of large-scale magnetic fields is naturally described by a {\it uniform} mesh in redshift space with elementary step $\Delta z$, which can be constructed as follows. The magnetic domain closest to us and labelled by 
$n = 1$ extends from 0 to $\Delta z$. Hence, its size $L_{\rm dom}^{(1)}$ can also be written as
\begin{equation}
\label{mag2Z}
L_{\rm dom}^{(1)} = L (0, \Delta z) = \left( \frac{L_{\rm dom}^{(1)}}{5 \, {\rm Mpc}} \right) 5 \, {\rm Mpc}~,
\end{equation}
where $L (0, \Delta z)$ is the domain's proper size and the second equality stresses our preferred choice for $L_{\rm dom}^{(1)}$. By combining Eqs. (\ref{lunghK}) and (\ref{mag2Z}) we infer
\begin{equation}
\label{mag2ZH}
\Delta z \simeq 1.17 \cdot 10^{- 3} \left( \frac{L_{\rm dom}^{(1)} }{5 \, {\rm Mpc}} \right)~,
\end{equation}
indeed in agreement with the linear Hubble law. Because our mesh in redshift space is uniform, $\Delta z$ sets the redshift size of {\it all} magnetic domains. Therefore for a source at redshift $z$ the total number $N_d$ of magnetic domains crossed by the beam can be estimated as
\begin{equation}
\label{mag2ZI}
N_d \simeq \frac{z}{\Delta z} \simeq 0.85 \cdot 10^3 \left( \frac{5 \, {\rm Mpc}}{L_{\rm dom}^{(1)} } \right) \, z
\end{equation}
and since we are assuming $z < 0.54$ we have $N_d \leq 0.46 \cdot 10^3 (5 \, {\rm Mpc} / L_{\rm dom}^{(1)})$. Furthermore, the $n$-th domain  extends from $z = (n - 1) \Delta z$ to $z= n \Delta z$ and its proper size can be written as $L_{\rm dom}^{(n)} = L ((n - 1) \Delta z, n \Delta z)$. Thanks again to Eq. (\ref{lunghK}), it reads
\begin{eqnarray}
\label{mag2ZHx}
L_{\rm dom}^{(n)} \simeq 2.96 \cdot 10^3 \, {\rm ln} \left( 1 + \frac{1.45 \, \Delta z}{1 + 1.45 \, (n - 1) \Delta z} \right) \, {\rm Mpc} \simeq \\\nonumber 
\simeq \frac{4.29 \cdot 10^3 \Delta z}{1 + 1.45 \, (n - 1) \Delta z} \, {\rm Mpc}~,   \ \ \ \ \ \ \ \ \ \ \ \ \ \ \ \ \ \ 
\end{eqnarray}
where the last equality is justified by the fact that our analysis is confined to $z < 0.54$ in conjunction with Eq. (\ref{mag2ZI}).

\subsection{Propagation over a single domain}

We are now in position to describe the propagation of the considered photon/ALP beam across the $n$-th magnetic domain. 

The transfer matrix is directly supplied by Eq. (\ref{mravvq2abcQ}), which for notational convenience we rewrite as
\begin{equation}
\label{mravvq2abcQQ}
{\cal U}_n (E_n, \psi_n) \equiv e^{i \left({\lambda}^{(n)}_1 \, L_{\rm dom}^{(n)} \right)} \, T_1 (\psi_n) + e^{i \left({\lambda}^{(n)}_2  \, L_{\rm dom}^{(n)} \right)} \, T_2 (\psi_n) + e^{i \left({\lambda}^{(n)}_3  \, L_{\rm dom}^{(n)} \right)} \, T_3 (\psi_n)~, 
\end{equation}
with
\begin{equation}
\label{mravvq2Q1a}
T_1 (\psi_n) \equiv
\left(
\begin{array}{ccc}
\cos^2 \psi_n & -\sin \psi_n \cos \psi_n & 0 \\
- \sin \psi_n \cos \psi_n & \sin^2 \psi_n & 0 \\
0 & 0 & 0
\end{array}
\right)~,
\end{equation}
\begin{equation}
\label{mravvq3Q1b}
T_2 (\psi_n) \equiv 
\left(
\begin{array}{ccc}
\frac{- 1 + \sqrt{1 - 4 {\delta}_n^2}}{2 \sqrt{1 - 4 {\delta}_n^2}} \sin^2 \psi_n & \frac{- 1 + \sqrt{1 - 4 {\delta}_n^2}}{2 \sqrt{1 - 4 {\delta}_n^2}} \sin \psi_n \cos \psi_n & \frac{i \delta_n}{\sqrt{1 - 4 {\delta}_n^2}} \sin \psi_n \\
\frac{- 1 + \sqrt{1 - 4 {\delta}_n^2}}{2 \sqrt{1 - 4 {\delta}_n^2}} \sin \psi_n \cos \psi_n & \frac{- 1 + \sqrt{1 - 4 {\delta}_n^2}}{2 \sqrt{1 - 4 {\delta}_n^2}} \cos^2 \psi_n & \frac{i \delta_n}{\sqrt{1 - 4 {\delta}_n^2}} \cos \psi_n \\
\frac{i \delta_n}{\sqrt{1 - 4 {\delta}_n^2}} \sin \psi_n & \frac{i \delta_n}{\sqrt{1 - 4 {\delta_n}^2}} \cos \psi_n & \frac{ 1 + \sqrt{1 - 4 {\delta}_n^2}}{2 \sqrt{1 - 4 {\delta}_n^2}}
\end{array}
\right)~,
\end{equation}
\begin{equation} 
\label{mravvq2Q1b}
T_3 (\psi_n) \equiv
\left(
\begin{array}{ccc}
\frac{ 1 + \sqrt{1 - 4 {\delta}_n^2}}{2 \sqrt{1 - 4 {\delta}_n^2}} \sin^2 \psi_n  & \frac{ 1 + \sqrt{1 - 4 {\delta}_n^2}}{2 \sqrt{1 - 4 {\delta}_n^2}} \sin \psi_n \cos \psi_n  & \frac{- i \delta_n}{\sqrt{1 - 4 {\delta}_n^2}}    \sin \psi_n \\ 
\frac{ 1 + \sqrt{1 - 4 {\delta}_n^2}}{2 \sqrt{1 - 4 {\delta}_n^2}} \sin \psi_n \cos \psi_n  & \frac{ 1 + \sqrt{1 - 4 {\delta}_n^2}}{2 \sqrt{1 - 4 {\delta}_n^2}} \cos^2 \psi_n  & \frac{- i \delta_n}{\sqrt{1 - 4 {\delta}_n^2}} \cos \psi_n \\
\frac{- i \delta_n}{\sqrt{1 - 4 {\delta}_n^2}} \sin \psi_n  & \frac{- i \delta_n}{\sqrt{1 - 4 {\delta}_n^2}} \cos \psi_n  &  \frac{- 1 + \sqrt{1 - 4 {\delta}_n^2}}{2 \sqrt{1 - 4 {\delta}_n^2}}   
\end{array}
\right)~,
\end{equation} 
where $\psi_n$ denotes the angle between ${\bf B}_n$ and the $z$-axis, which is fixed for all domains. Moreover, we have set
\begin{equation}
\label{a91212a1PW}
{\lambda}^{(n)}_{1} \equiv \frac{i}{2 \, {\lambda}^{(n)}_{\gamma} (E_0)}~,
\end{equation}
\begin{equation}
\label{a91212a2PW}
{\lambda}^{(n)}_{2} \equiv \frac{i}{4 \, {\lambda}^{(n)}_{\gamma}} \left(1 - \sqrt{1 - 4 \, \delta^2_n} \right)~, 
\end{equation}
\begin{equation}
\label{a91212a3PW}
{\lambda}^{(n)}_{3} \equiv \frac{i}{4 \, {\lambda}^{(n)}_{\gamma}}  \left(1 + \sqrt{1 - 4 \, \delta^2_n} \right)~,
\end{equation}
with
\begin{equation}
\label{a17PW14022011}
E_n \equiv E_0  \, \Bigl[1 + (n - 1) \, \Delta z) \Bigr]~,
\end{equation}
\begin{equation}
\label{a17PW}
{\delta}_n \equiv \frac{B_{n} \, {\lambda}^{(n)}_{\gamma} (E_0)}{M}
\end{equation}
and we have introduced the shorthand
\begin{equation}
\label{a17PWW}
{\lambda}^{(n)}_{\gamma} (E_0) \equiv {\lambda}_{\gamma} \left(E_n \right)~.
\end{equation}
In addition, Eq. (\ref{mag1}) implies
\begin{equation}
\label{mag1S}
B_n = B_0 \, \Bigl[1 + (n - 1) \, \Delta z) \Bigr]^2~.
\end{equation}

What remains to be done in to evaluate the photon mean free path ${\lambda}^{(n)}_{\gamma} (E_0)$. A convenient procedure is as follows. Let us suppose to observe two hypothetical sources located at both edges of the $n$-th domain. Then we apply Eq. (\ref{a02122010A}) to either source. With the notational simplifications $\Phi_{\rm obs}(E_0,z) \to \Phi (E_0)$ and $\Phi_{\rm em} \left( E_0 (1+z) \right) \to \Phi \left( E_0 (1+z) \right)$, we have
\begin{equation} 
\label{a02122010Am}
\Phi (E_0) = e^{- \tau_{\gamma} \left(E_0, (n - 1) \Delta z \right)} \, \Phi \left(E_n \right)~,
\end{equation}
\begin{equation} 
\label{a02122010An}
\Phi (E_0) = e^{- \tau_{\gamma} \left(E_0, n \Delta z \right)} \, \Phi \left(E_{n + 1} \right)~,
\end{equation}
and so the flux change across the considered domain is
\begin{equation} 
\label{a02122010Ao}
\Phi \left(E_n \right) = e^{- \left[ \tau_{\gamma} \left(E_0, n \, \Delta z \right) - \tau_{\gamma} \left(E_0, (n - 1) \Delta z \right) \right]} \, 
\Phi \left(E_{n + 1} \right)~.
\end{equation}
Now, since $\Delta z \sim 10^{- 3}$ evolutionary effects can be neglected inside a single domain and only accounted for when jumping from one domain to the next. As a consequence -- owing to Eq. (\ref{a1Z}) -- Eq. (\ref{a02122010Ao}) reduces to
\begin{equation} 
\label{a02122010Ap}
\Phi \left( E_n \right) = e^{- L_{\rm dom}^{(n)} / {\lambda}^{(n)}_{\gamma} (E_0)} \, \Phi \left( E_{n + 1} \right)~,
\end{equation}
and the comparison of Eqs. (\ref{a02122010Ao}) and (\ref{a02122010Ap}) yields
\begin{equation} 
\label{a02122010Aq}
{\lambda}^{(n)}_{\gamma} (E_0) = \frac{L_{\rm dom}^{(n)}}{\tau_{\gamma} \left(E_0, n \, \Delta z \right) - \tau_{\gamma} \left(E_0, (n - 1) \Delta z \right)}~.
\end{equation}
Further, by inserting Eq. (\ref{mag2ZHx}) into Eq. (\ref{a02122010Aq}), we get the desired photon mean free path
\begin{equation} 
\label{a02122010Ar}
{\lambda}^{(n)}_{\gamma} (E_0) = \left( \frac{4.29 \cdot 10^3}{1 + 1.45 \, (n - 1) \Delta z} \right) \left( \frac{\Delta z }{\tau_{\gamma} \left(E_0, n \, \Delta z \right) - \tau_{\gamma} \left(E_0, (n - 1) \Delta z \right)} \right) \, {\rm Mpc}~.
\end{equation}

\subsection{Propagation over many domains}

We are finally ready to carry the strategy outlined in Subsection IV-B to completion, namely to evaluate the photon survival probability $P^{\rm DARMA}_{\gamma \to \gamma} (E_0, z)$ for a monochromatic beam emitted by a blazar at redshift $z$ and detected at energy $E_0$.

This task can be accomplished by first noticing that for a considered blazar at redshift $z$ the overall behaviour of the photon/ALP beam is described by the following transfer matrix
\begin{equation}
\label{as1g}
{\cal U} \left(E_0, z; \psi_1, ... , \psi_{N_d} \right) = \prod^{N_d}_{n = 1} \, {\cal U}_n \left(E_n, \psi_n \right)~.
\end{equation}
According to Eq. (\ref{k3lwf1w1}), the probability that a photon/ALP beam emitted by a blazar at $z$ in the state $\rho_1$ will be detected in the state 
$\rho_2$ for fixed orientations $ \psi_1, ... , \psi_{N_d}$ of ${\bf B}$ in every domain is 
\begin{equation}
\label{k3lwf1w1Q}
P_{\rho_1 \to \rho_2} \left(E_0, z; \psi_1, ... , \psi_{N_d} \right)  =       
{{\rm Tr} \left( {\rho}_2 \, {\cal U} \left(E_0, z; \psi_1, ... , \psi_{N_d} \right) \, \rho_1 \, {\cal U}^{\dagger} \left(E_0, z; \psi_1, ... , \psi_{N_d} \right) \right)}~,    
\end{equation}
where it is assumed that ${\rm Tr} \rho_1 = {\rm Tr} \rho_2 = 1$. As a consequence, the actual detection probability for the beam in question emerges by averaging the above expression over all angles, namely
\begin{equation}
\label{k3lwf1w1W}
P_{\rho_1 \to \rho_2} \left(E_0, z \right)  =  \Big\langle P_{\rho_1 \to \rho_2} \left(E_0, z; \psi_1, ... , \psi_{N_d} \right) \Big\rangle_{\psi_1, ... , \psi_{N_d}}~. 
\end{equation}
Because of the fact that the photon polarization cannot be measured at the energies considered here we have to sum this result over the two final polarization 
states
\begin{equation}
\label{a9s11A}
{\rho}_x = \left(
\begin{array}{ccc}
1 & 0 & 0 \\
0 & 0 & 0 \\
0 & 0 & 0 \\
\end{array}
\right)~,
\end{equation}
\begin{equation}
\label{a9s11B}
{\rho}_z = \left(
\begin{array}{ccc}
0 & 0 & 0 \\
0 & 1 & 0 \\
0 & 0 & 0 \\
\end{array}
\right)~.
\end{equation}
Moreover, we suppose for simplicity that the emitted beam consists $100 \, \%$ of unpolarized photons, so that the initial beam state is described by
\begin{equation}
\label{a9s11C}
{\rho}_{\rm unpol} = \frac{1}{2}\left(
\begin{array}{ccc}
1 & 0 & 0 \\
0 & 1 & 0 \\
0 & 0 & 0 \\
\end{array}
\right)~.
\end{equation}
Hence, we ultimately have
\begin{eqnarray}
\label{k3lwf1w1WW}
P^{\rm DARMA}_{\gamma \to \gamma} \left(E_0, z \right)  =  \Big\langle P_{\rho_{\rm unpol} \to \rho_x} \left(E_0, z; \psi_1, ... , \psi_{N_d} \right) \Big\rangle_{\psi_1, ... , \psi_{N_d}} +   \\  \nonumber
+ \, \Big\langle P_{\rho_{\rm unpol} \to \rho_z} \left(E_0, z; \psi_1, ... , \psi_{N_d} \right) 
\Big\rangle_{\psi_1, ... , \psi_{N_d}}~.   \ \ \ \ \ \ \ \ \ \ \ 
\end{eqnarray}

We implement this procedure as follows. In the first place, we arbitrarily choose the angle $\psi_n$ in the n-th domain and we evaluate the corresponding transfer matrix ${\cal U}_n \left(E_n, \psi_n \right)$ for a given value of $E_0$, keeping Eq. (\ref{a17PW14022011}) in mind. Next, the application of Eqs. (\ref{as1g}) and (\ref{k3lwf1w1Q}) yields the corresponding photon survival probabilities entering Eq. (\ref{k3lwf1w1WW}) for a single realization of the propagation process. We repeat these steps $5000$ times, by randomly varying all angles $\psi_n$ each time, thereby generating $5000$ random realizations of the propagation process. Finally, we average the resulting photon survival probabilities over all these realizations of the propagation process, thereby accomplishing the average process in Eq. (\ref{k3lwf1w1WW}). We find in this way the physical photon survival probability $P^{\rm DARMA}_{\gamma \to \gamma} \left(E_0, z \right)$.

\section{DISCUSSION}

Let us proceed to investigate the implications of the DARMA scenario for VHE blazars observations.

We begin by stressing that all its physical predictions depend solely on $B/M$ and not on $B$ and $M$ separately (this was true for photon-ALP oscillations and it remains true in general, because absorption does not depend on these quantities). For this reason, it is quite useful to introduce the dimensionless parameter
\begin{equation}
\label{as1g10022011}
\xi \equiv \left(\frac{B_0}{{\rm nG}} \right) \left(\frac{10^{11} \, {\rm GeV}}{M} \right)~.
\end{equation}
Owing to conditions (\ref{a509022011}) and (\ref{mag210022011f}), it will be assumed
\begin{equation}
\label{as1g10022011q}
\xi < 6
\end{equation}
throughout our discussion. Specifically, we will focus our attention on the representative cases $\xi =5.0$, $\xi =1.0$, $\xi =0.5$, $\xi =0.1$, taking both  $L_{\rm dom} = 4 \, {\rm Mpc}$ and $L_{\rm dom} = 10 \, {\rm Mpc}$ at $z = 0$. Nevertheless, it is important to keep under control which values of $B_0$ and 
$M$ are allowed in each case. From the constraints (\ref{a509022011}) and (\ref{mag210022011f}) we find the allowed ranges reported in Table \ref{tab:ranges}.

\begin{table}
\begin{tabular}{rrr}
\hline
 {$\xi$   \ \ \ \ \ \ \ \ \ \ \ \ \ \ \ \ \ \ \ \         }{}{}   
&  {}{$M$/$(10 ^{11} \, {\rm GeV})$  \ \ \ \ \ \ \ \ \ \ \ \ \ \ \ \        }{} 
&  {}{}{$B_0$/{\rm nG}} \\

\hline
$0.1$  \ \ \ \ \ \ \ \ \ \ \ \ \ \ \ \ \ \ \ \       & 1 -- 60    \ \ \ \ \ \ \ \ \ \ \ \ \ \ \ \ \ \ \ \ \ \ \  & 0.1 -- 6 \\
$0.5$   \ \ \ \ \ \ \ \ \ \ \ \ \ \ \ \ \ \ \ \        & 1 -- 12    \ \ \ \ \ \ \ \ \ \ \ \ \ \ \ \ \ \ \ \ \ \ \   & 0.5 -- 6 \\
$1.0$   \ \ \ \ \ \ \ \ \ \ \ \ \ \ \ \ \ \ \ \       & 1 -- 6.0    \ \ \ \ \ \ \ \ \ \ \ \ \ \ \ \ \ \ \ \ \ \ \   & 1 -- 6 \\
$5.0$   \ \ \ \ \ \ \ \ \ \ \ \ \ \ \ \ \ \ \ \       & 1 -- 1.2     \ \ \ \ \ \ \ \ \ \ \ \ \ \ \ \ \ \ \ \ \ \ \    & 5 -- 6 \\
\hline
\end{tabular}
\caption{Allowed values of $M$ and $B_0$ in the considered cases.}
\label{tab:ranges}
\end{table}

Next, we have to make sure that we stay within the strong-coupling regime all the way up to the source for $E_0 > 100 \, {\rm GeV}$. Therefore, by combining Eq. (\ref{a17231209}) with the requirement $E_* < 100 \, {\rm GeV}$ the resulting upper bound on $m$ can be expressed as
\begin{equation}
\label{upperbound24012011}
| m^2 - \omega^2_{\rm pl} |^{1/2} < 1.97 \cdot 10^{- 10} \left(\frac{B}{\rm nG} \right)^{1/2} \left(\frac{10^{11} \, {\rm GeV}}{M} \right)^{1/2}  \, {\rm eV}~,
\end{equation}
which -- on account of Eqs. (\ref{mag210022011q}) and (\ref{as1g10022011}) -- can be more suitably rewritten in the form
\begin{equation}
\label{upperbound24012011a}
\left| \left( \frac{m}{10^{ - 10} \, {\rm eV}} \right)^2 - \left(1.14 \cdot 10^{- 4} \right)^2 \right|^{1/2} < 1.97 \, {\xi}^{1/2}         
\end{equation}
valid for all sources considered here~\cite{upperb}. Thanks to condition (\ref{as1g10022011q}), we see that within the DARMA scenario ALPs have to be very light, with mass never exceeding $5 \cdot 10^{- 10} \, {\rm eV}$. In particular, the axion needed to solve the strong CP problem is therefore ruled out by several orders of magnitude. Observe that for $m < 1.14 \cdot 10^{- 14} \, {\rm eV}$ the plasma frequency dominates, so that even massless ALPs behave as if their mass where equal to the plasma frequency. The upper bounds on $m$ corresponding to the cases under consideration are reported in 
Table \ref{tab:axi}.

\begin{table}
\begin{tabular}{rr}
\hline
 {Upper bound on $m$   \ \ \ \ \ \ \ \ \ \ \ \ \ \ \ \    }{}   
&  {}{Value of the $\xi$ parameter} \\
\hline
$   4.40 \cdot 10^{- 10} \, {\rm eV}$  \ \ \ \ \ \ \ \ \ \ \ \ \ \ \ \ \ \ \ \        & $ \xi = 5.0$  \ \ \ \ \ \ \ \ \ \ \ \     \\
$   1.97 \cdot 10^{- 10} \, {\rm eV}$ \ \ \ \ \ \ \ \ \ \ \ \ \ \ \ \ \ \ \ \           & $ \xi = 1.0$  \ \ \ \ \ \ \ \ \ \ \ \     \\
$   1.39 \cdot 10^{- 10} \, {\rm eV}$  \ \ \ \ \ \ \ \ \ \ \ \ \ \ \ \ \ \ \ \         & $ \xi = 0.5 $ \ \ \ \ \ \ \ \ \ \ \ \      \\
$   0.62 \cdot 10^{- 10} \, {\rm eV}$  \ \ \ \ \ \ \ \ \ \ \ \ \ \ \ \ \ \ \ \         & $ \xi = 0.1$ \ \ \ \ \ \ \ \ \ \ \ \      \\
\hline
\end{tabular}
\caption{Upper bounds on the ALP mass in the considered cases.}
\label{tab:axi}
\end{table}

As far as EBL absorption is concerned, we will take for the optical depth entering Eq. (\ref{a02122010Ar}) the exact expression $\tau_{\gamma}^{\rm FRV}(E_0, z)$ provided by the FRV model~\cite{frvtable}.

A general expectation is that -- because in the absence of EBL absorption photon-ALP oscillations only produce a dimming~\cite{dimming} -- an enhancement of the photon survival probability with respect to the case of conventional physics shows up only at sufficiently high energy, where EBL absorption becomes substantial. Therefore, close enough to $100 \, {\rm GeV}$ a dimming rather than an enhancement should occur.

\section{PREDICTIONS FOR FUTURE OBSERVATIONS}

The best way to figure out the relevance of the DARMA scenario for future observations to be performed with the CTA and with the HAWC water Cherenkov 
$\gamma$-ray observatory is to compare the photon survival probability $P^{\rm DARMA}_{\gamma \to \gamma} (E_0,z)$ with the one predicted by conventional physics $P_{\gamma \to \gamma}^{\rm CP} (E_0,z)$, with the EBL described in either case by the FRV model.

We do that for a sample of different redshifts, like $z = 0.031$, $z = 0.188$, $z = 0.444$ and $z = 0.536$. We remark that the case of $z = 0.031$ may look somewhat academic, since its location inside the Local Group is likely to make the morphology of the magnetic field crossed by its line of sight more complicated than assumed in this paper. Nevertheless, we include $z = 0.031$ in the present analysis in order to see what happens for a very nearby blazar even if a drastic simplifying assumption is made.

The results are displayed in Figure 3. For each of the selected sources, we consider the above choices for $\xi$, which are represented by a solid black line ($\xi =5.0$), a dotted-dashed line ($\xi =1.0$), a dashed line ($\xi =0.5$) and a dotted line ($\xi =0.1$), while the solid grey line corresponds to conventional physics. We take both $L_{\rm dom} = 4 \, {\rm Mpc}$ and $L_{\rm dom} = 10 \, {\rm Mpc}$ for the domain size at $z = 0$.

\begin{figure}
\begin{center}
\includegraphics[width=.4\textwidth]{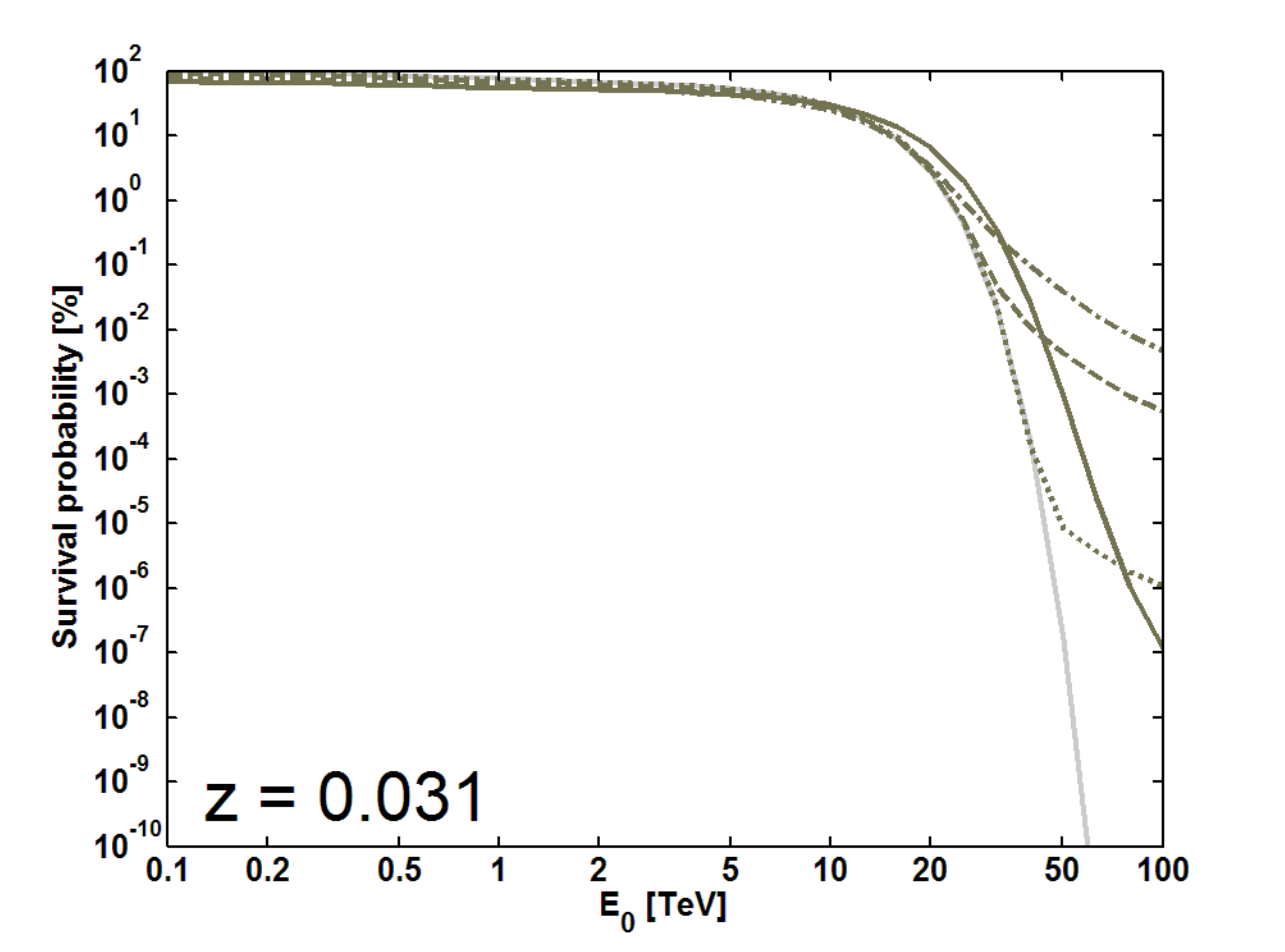}\includegraphics[width=.4\textwidth]{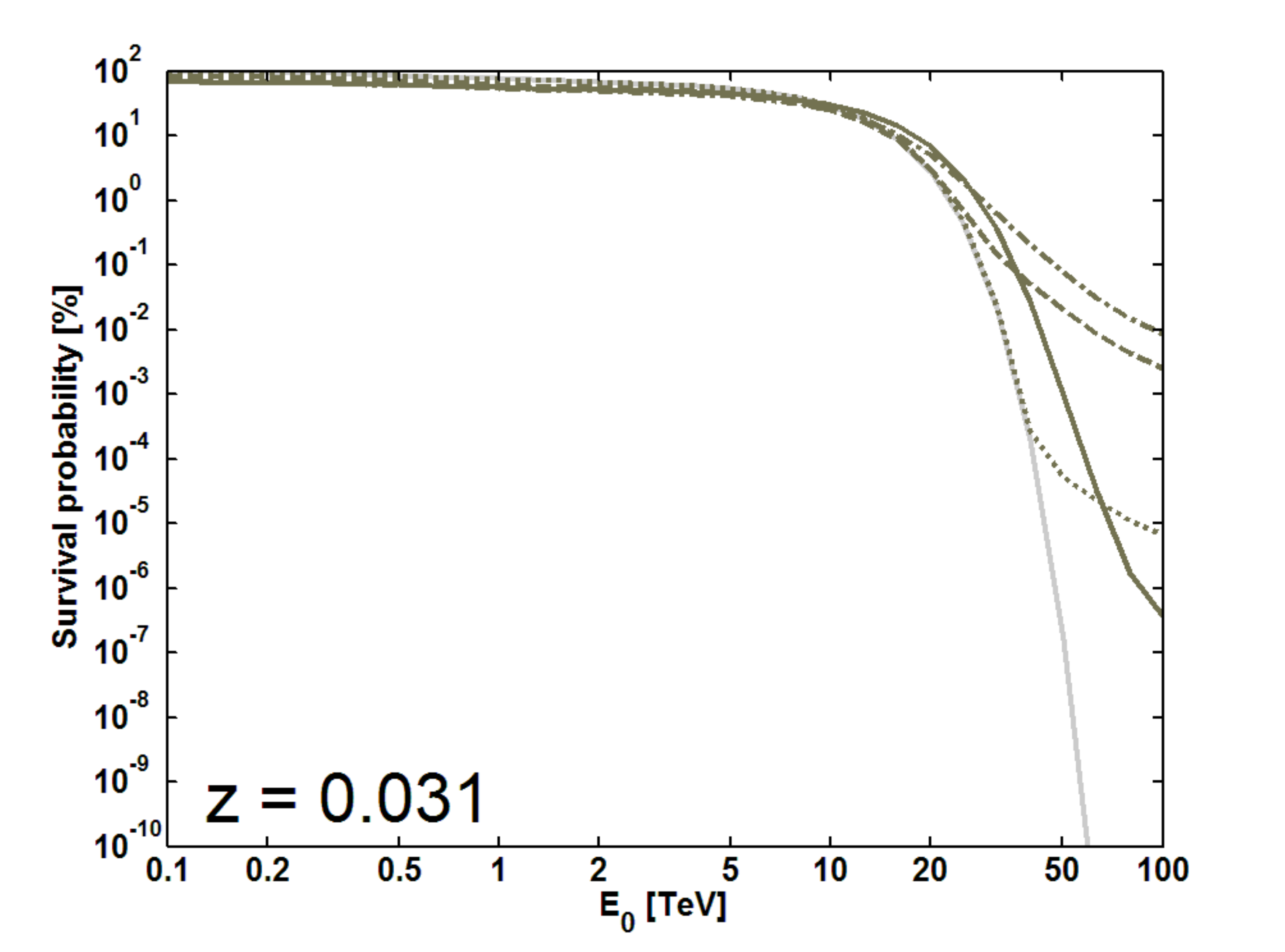}
\includegraphics[width=.4\textwidth]{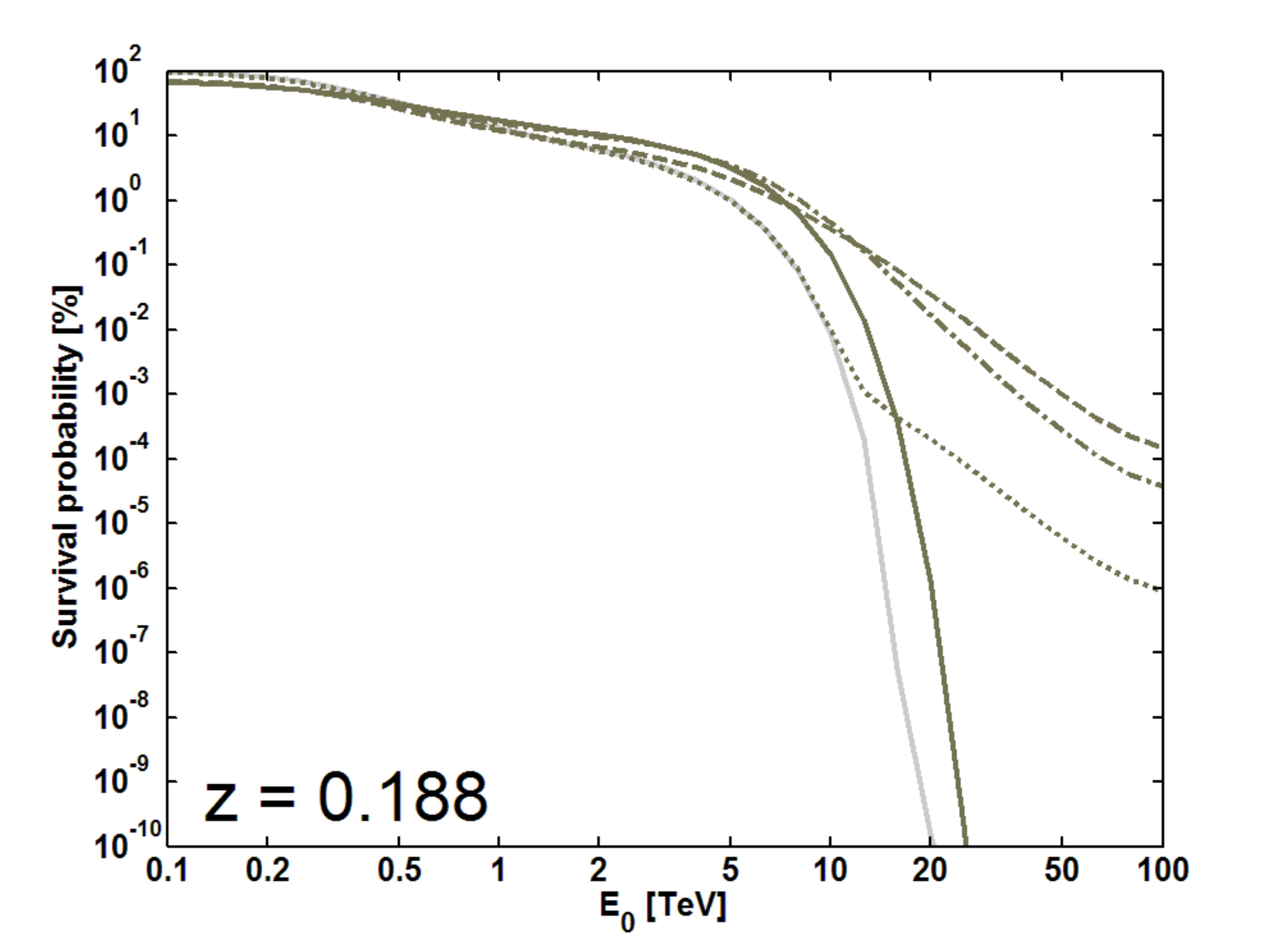}\includegraphics[width=.4\textwidth]{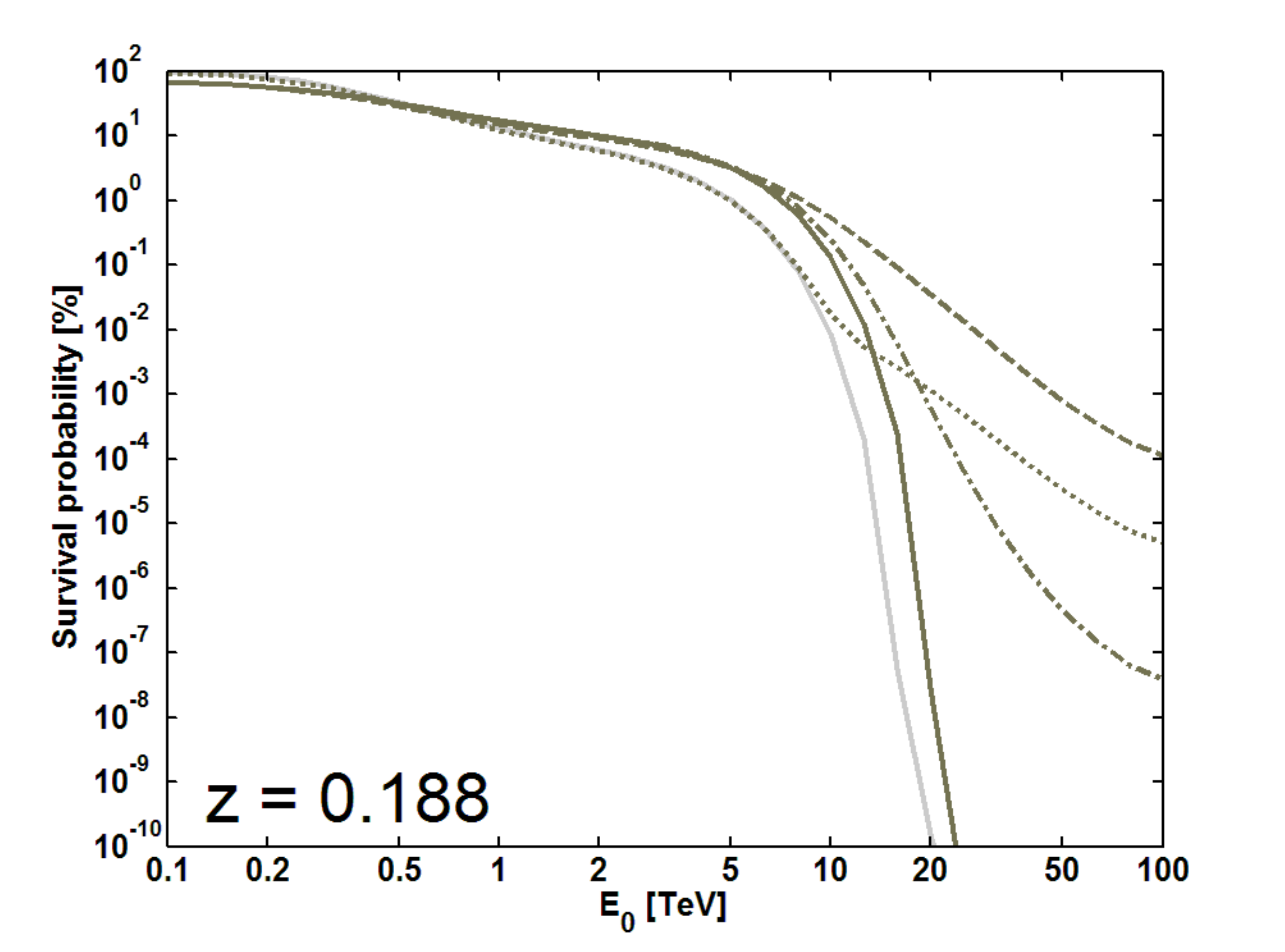}
\includegraphics[width=.4\textwidth]{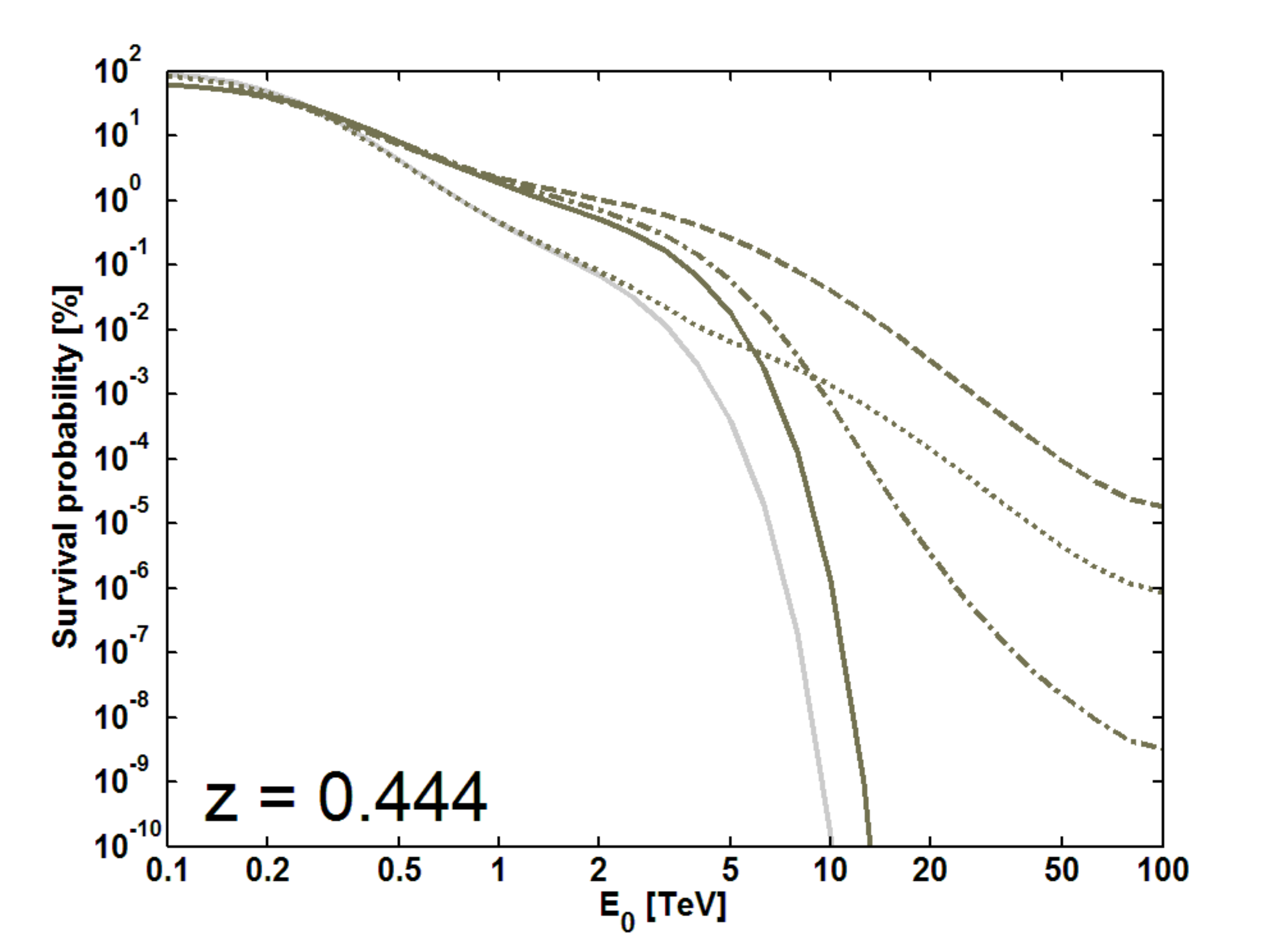}\includegraphics[width=.4\textwidth]{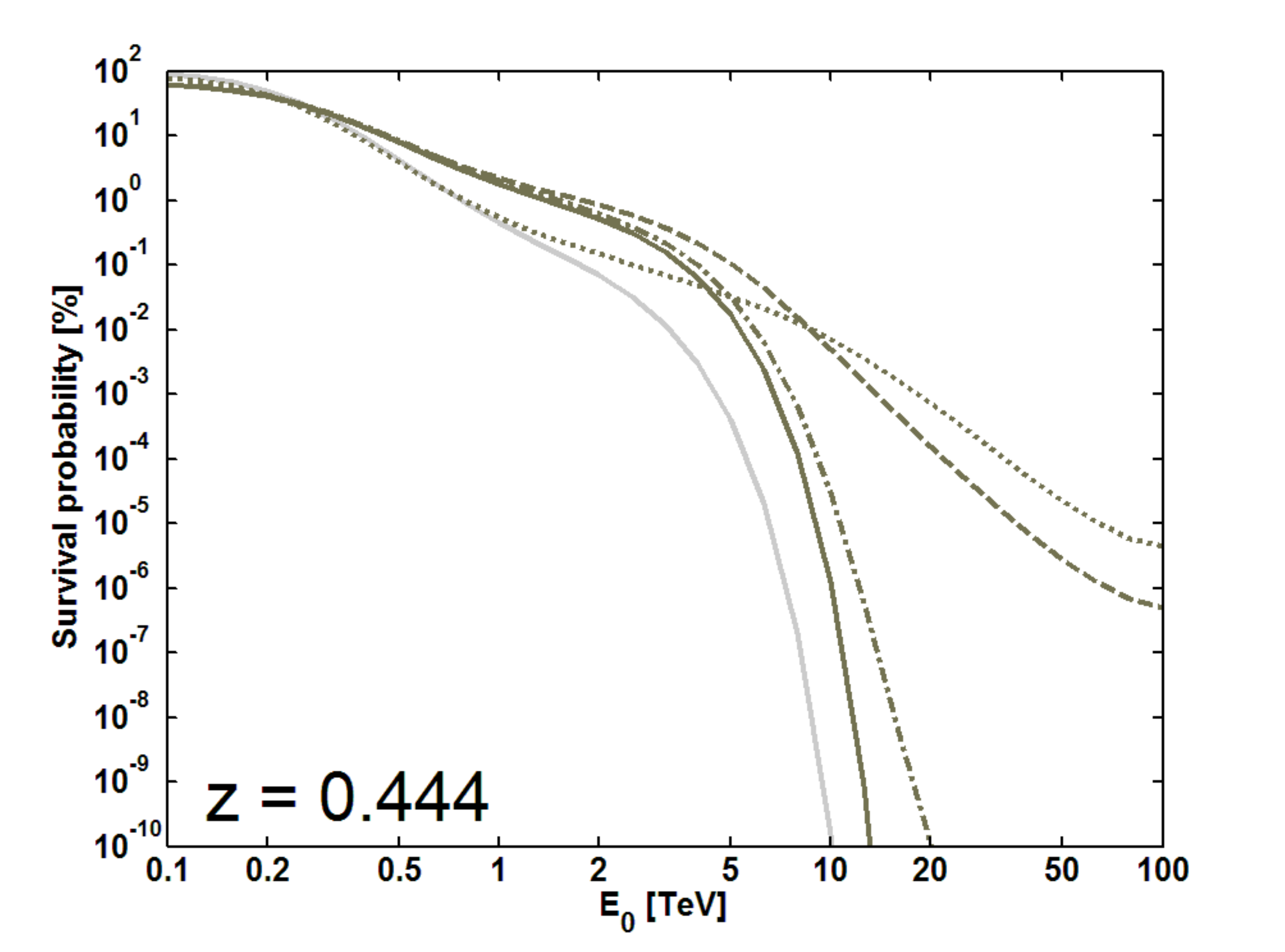}
\includegraphics[width=.4\textwidth]{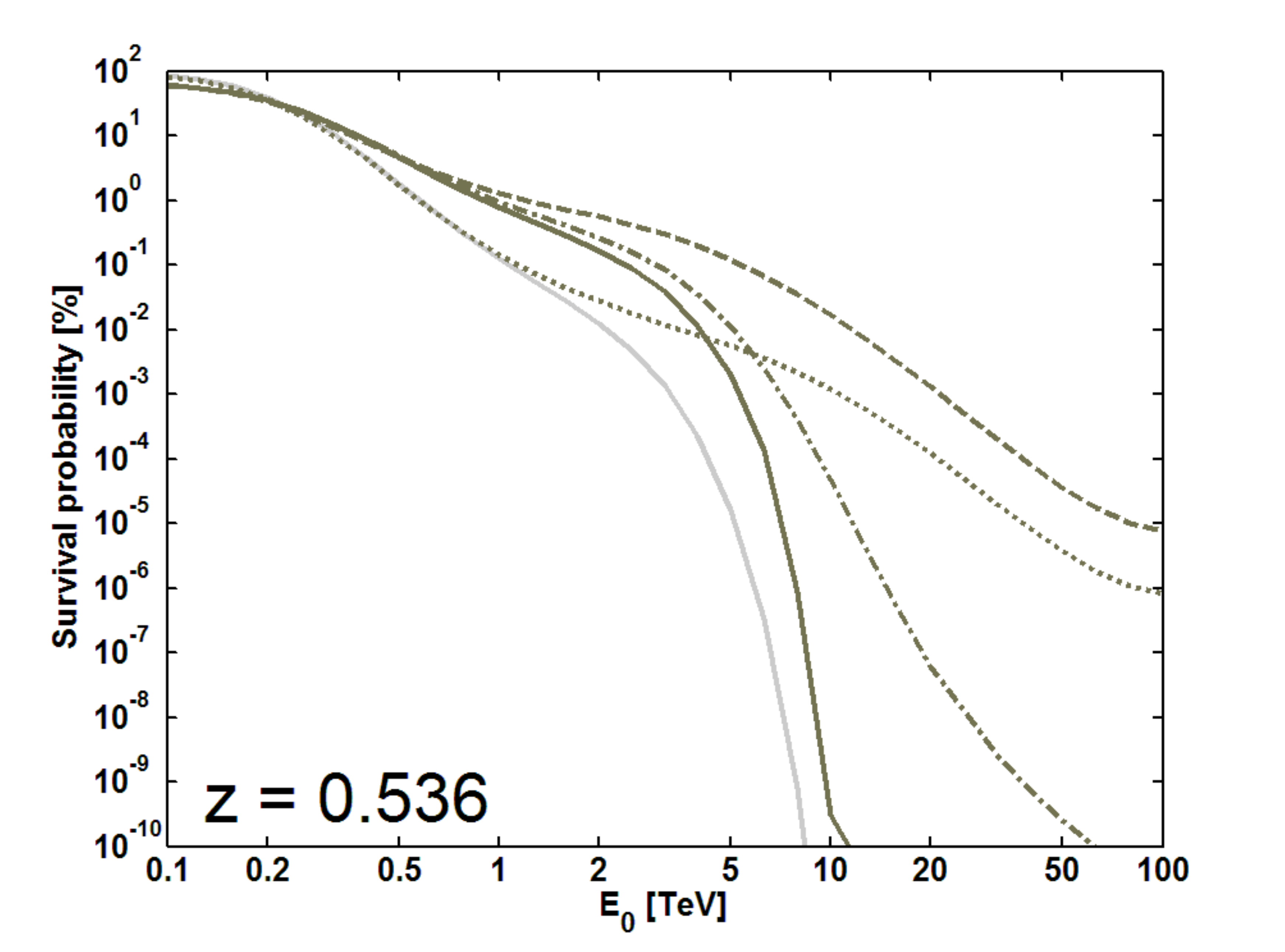}\includegraphics[width=.4\textwidth]{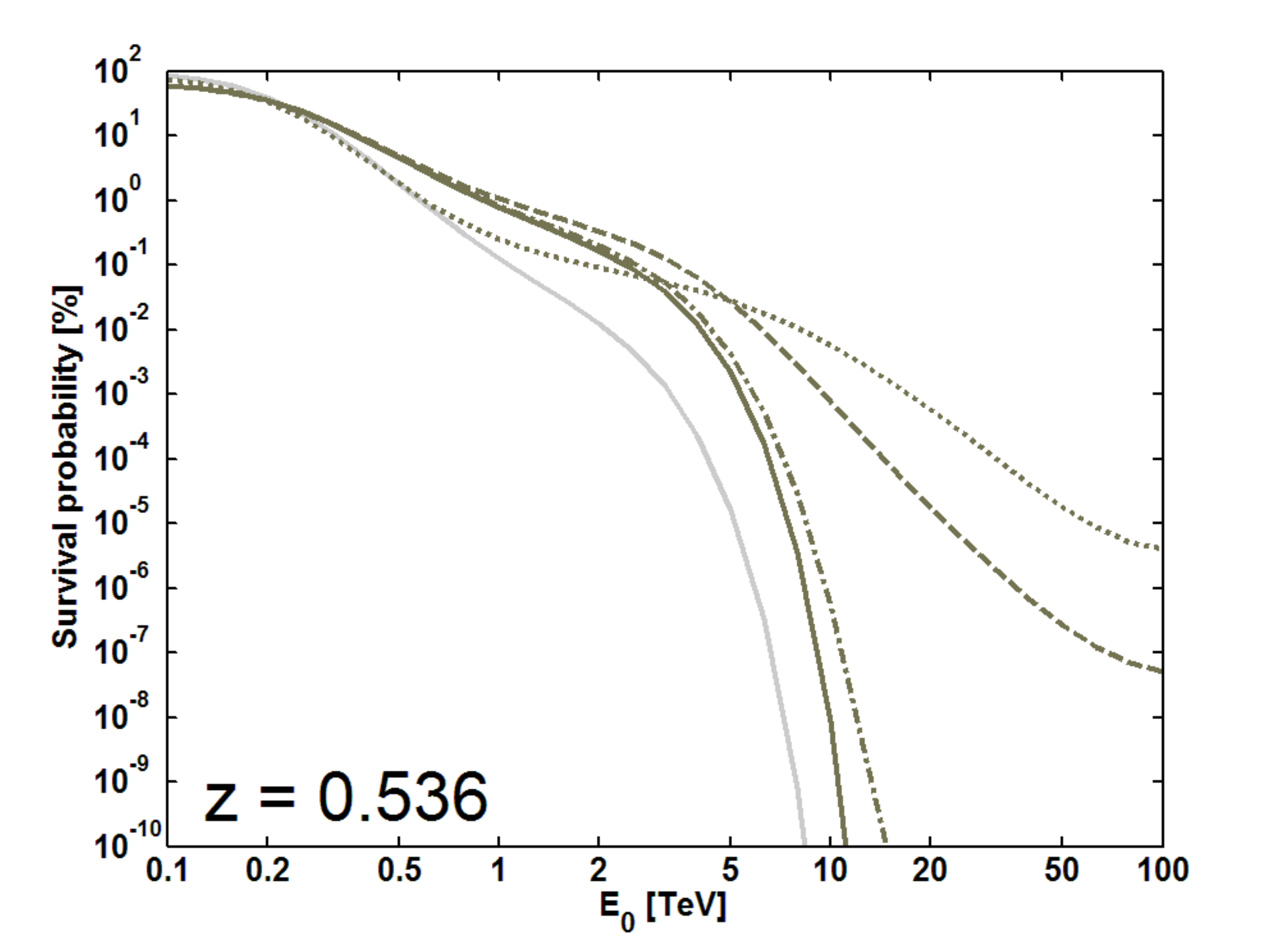}
\end{center}
\caption{\label{SP-Mrk421-BM-L4.pdf} 
Behaviour of $P^{\rm DARMA}_{\gamma \to \gamma}$ versus the observed energy $E_0$ for: $z=0.031$ (top row), $z=0.188$ (second row), $z=0.444$ (third row), $z=0.536$ (bottom row). The solid black line corresponds to $\xi =5.0$, the dotted-dashed line to $\xi =1.0$, the dashed line to $\xi =0.5$, the dotted line to $\xi =0.1$ and the solid grey line to conventional physics. We have taken $L_{\rm dom} = 4 \, {\rm Mpc}$ (left column) and $L_{\rm dom} = 10 \, {\rm Mpc}$ (right column).}
\end{figure}

All plots show one common trend. At energies only slightly in excess of $100 \, {\rm GeV}$, $P_{\gamma \to \gamma}^{\rm CP} (E_0,z)$ is {\it larger} than $P^{\rm DARMA}_{\gamma \to \gamma} (E_0,z)$, indeed in agreement with expectations. As the energy further increases, the situation reverses and $P^{\rm DARMA}_{\gamma \to \gamma} (E_0,z)$ gets progressively larger and larger than $P_{\gamma \to \gamma}^{\rm CP} (E_0,z)$ until the value $100 \, {\rm TeV}$ is attained which is the highest energy value considered in the present analysis. 

A somewhat surprising result emerges at large enough energies. Indeed, since $\xi$ sets the strength of the photon-ALP oscillation mechanism, it would be natural to expect $P^{\rm DARMA}_{\gamma \to \gamma} (E_0,z)$ to monotonically increase with $\xi$. However, this is not the case. More specifically, for $\xi = 5.0$ the behaviour of $P^{\rm DARMA}_{\gamma \to \gamma} (E_0,z)$ as a function of $E_0$ resembles closely that of $P_{\gamma \to \gamma}^{\rm CP} (E_0,z)$ -- apart from an overall shift towards higher energies -- and it is practically independent of $L_{\rm dom}$ apart from the case of {  $z = 0.031$}  which exhibits a mild $L_{\rm dom}$-dependence. Moreover, at sufficiently high energies the values of $P^{\rm DARMA}_{\gamma \to \gamma} (E_0,z)$ corresponding to $\xi = 5.0$ are the {\it lowest} predicted by the DARMA scenario for all sources. The case $\xi = 1.0$ is different, since the resulting values of $P^{\rm DARMA}_{\gamma \to \gamma} (E_0,z)$ are among the highest predicted at low redshift but become among the lowest at high redshift even though they always exceed those corresponding to $\xi = 5.0$. In addition -- with the exception of {  $z = 0.031$} -- $P^{\rm DARMA}_{\gamma \to \gamma} (E_0,z)$ strongly increases as $L_{\rm dom}$ decreases. As $\xi$ decreases the trend exhibits a radical modification. Actually, the case $\xi = 0.5$ shows a mild $L_{\rm dom}$-dependence for all considered sources, and with the exception of $z = 0.031$ it leads to the largest values of $P^{\rm DARMA}_{\gamma \to \gamma} (E_0,z)$ for $L_{\rm dom} = 4 \, {\rm Mpc}$. Finally, the case $\xi = 0.1$ depends more strongly on $L_{\rm dom}$ and -- again with the exception of {  $z = 0.031$} -- for some energy values it can make $P^{\rm DARMA}_{\gamma \to \gamma} (E_0,z)$ larger than in the case $\xi = 0.5$ for $L_{\rm dom} = 10 \, {\rm Mpc}$, but this never occurs for $L_{\rm dom} = 4 \, {\rm Mpc}$. We observe that the different situation found for $z = 0.031$ as compared to the other blazars should not come as a surprise, owing to the above remarks. 

What is the reason for such a behaviour?

Owing to the random structure of the considered magnetic field, coherence is maintained only within one domain and so $P^{\rm DARMA}_{\gamma \to \gamma} (E_0,z)$ is ultimately controlled by two quantities: the photon-ALP conversion probability over a single domain 
$P_{\gamma \to a} (L_{\rm dom})$ and the photon absorption probability in Eq. (\ref{a1dop28012011}). In order to clarify this issue in an intuitive fashion, we argue as follows, discarding cosmological effects for simplicity. 

As far as $P_{\gamma \to a} (L_{\rm dom})$ is concerned, we have seen that it is given by Eqs. (\ref{a16gh}) in the case of photons linearly polarized in the direction parallel to ${\bf B}$. This is not true in the present situation where the beam photons are assumed to be unpolarized, but for the sake of an order-of-magnitude estimate we can still suppose that $P_{\gamma \to a} (L_{\rm dom})$ has the form (\ref{a16gh}) and therefore we write it as
\begin{equation}
\label{a16gh2705a}
P_{\gamma \to a} (L_{\rm dom})  \simeq {\rm sin}^2 \left[1.6 \cdot 10^{- 2} \, \xi \left(\frac{L_{\rm dom}}{{\rm Mpc}} \right) \right]~.
\end{equation}

\begin{figure}[h]
\begin{center}
\includegraphics[width=.4\textwidth]{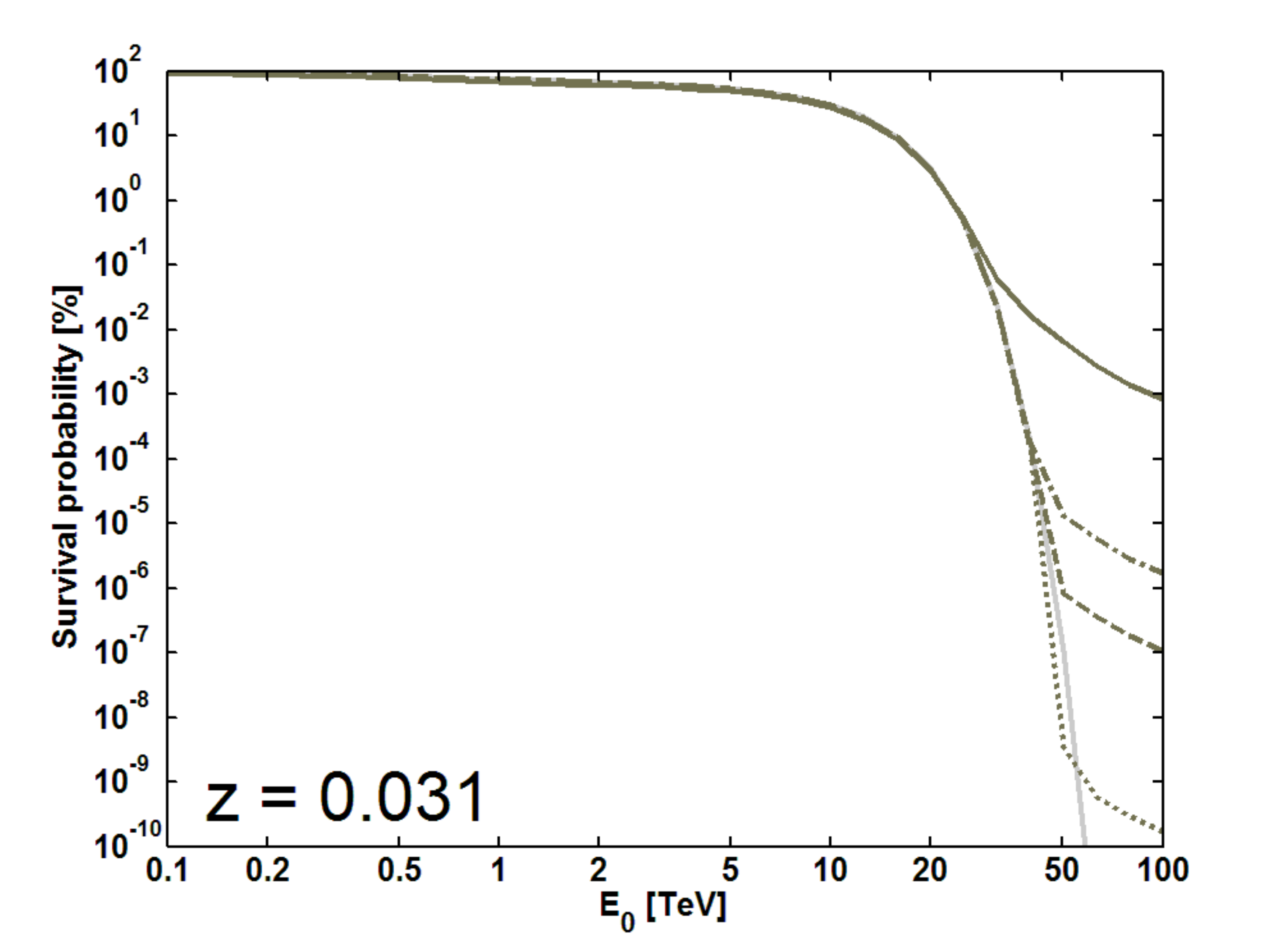}\includegraphics[width=.4\textwidth]{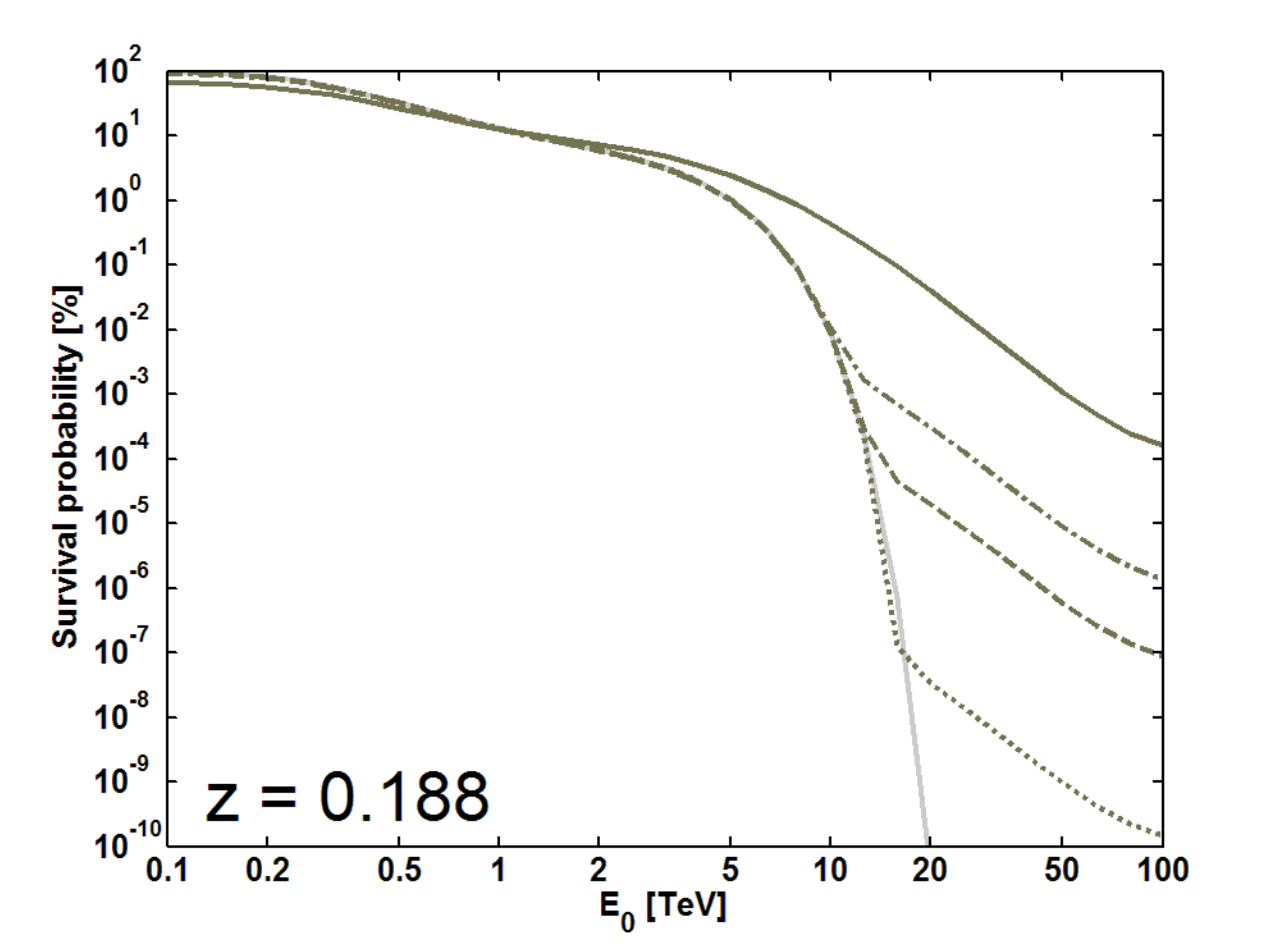}
\includegraphics[width=.4\textwidth]{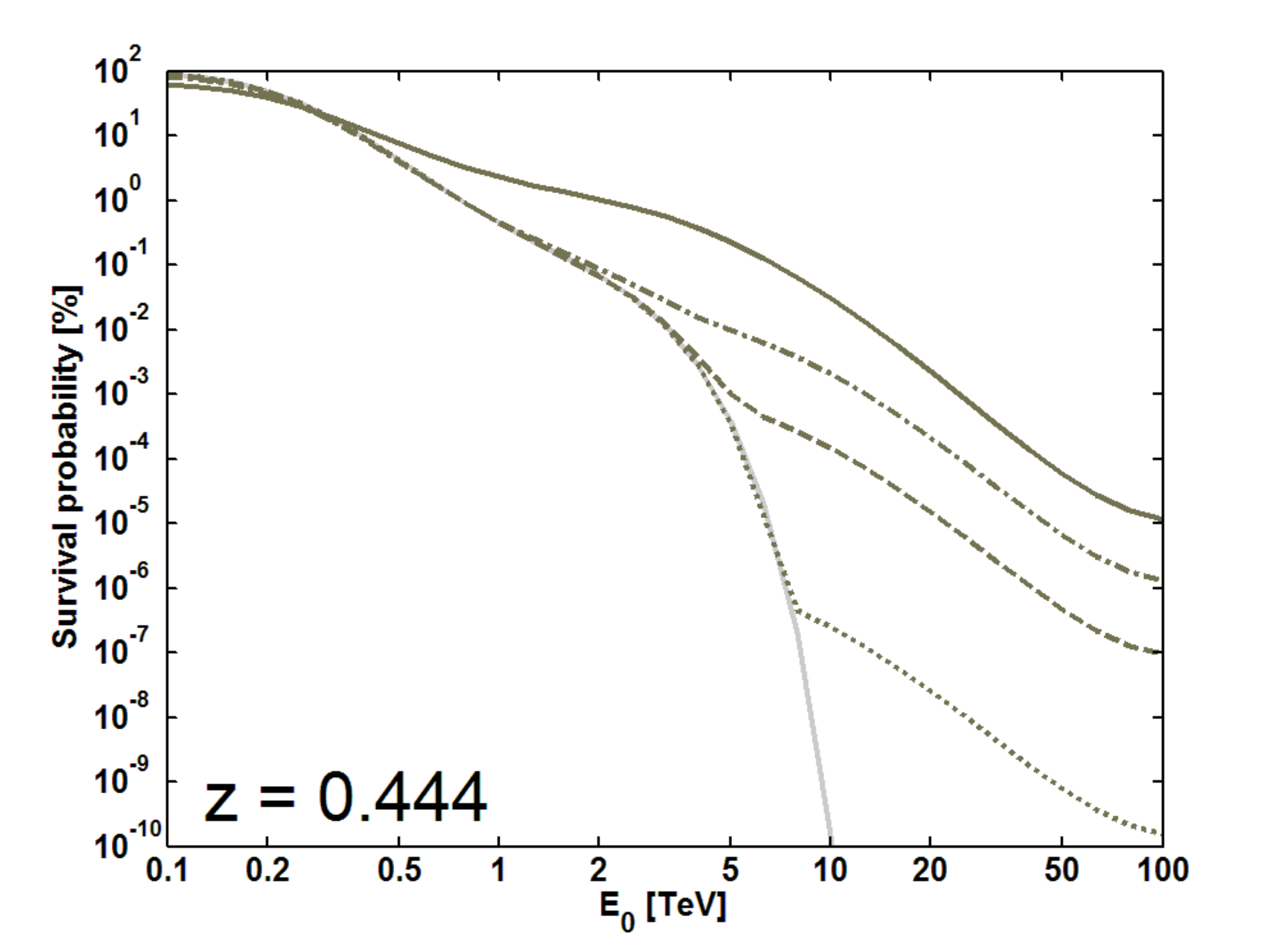}\includegraphics[width=.4\textwidth]{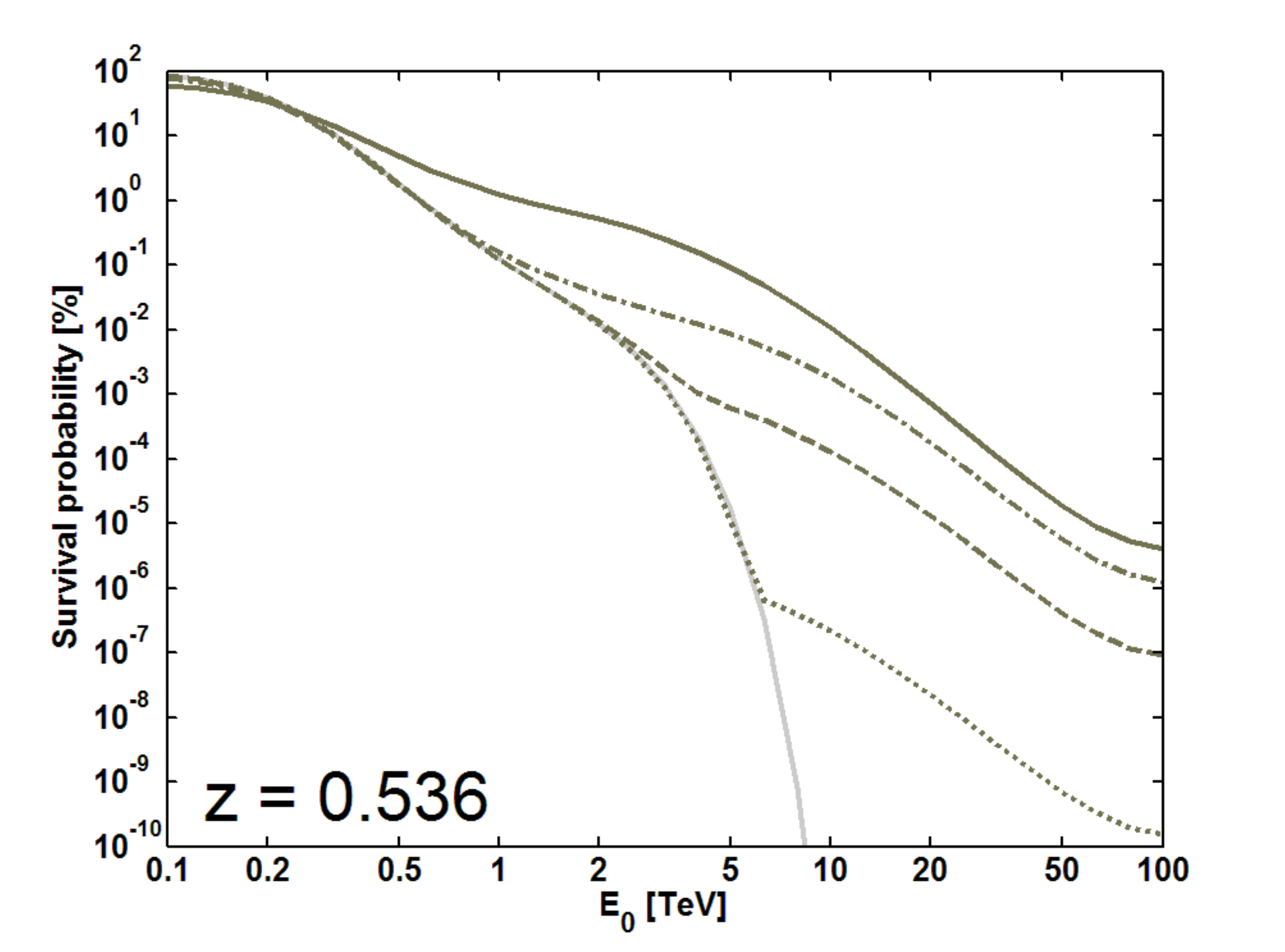}
\end{center}
\caption{\label{Mrk421L0.05.pdf} 
Behaviour of $P^{\rm DARMA}_{\gamma \to \gamma}$ versus the observed energy $E_0$ for: $z=0.031$ (top row left), $z=0.188$ (top row right), $z=0.444$ (bottom row left), $z=0.536$ (bottom row right) for $L_{\rm dom} = 0.05 \, {\rm Mpc}$. The solid black line corresponds to $\xi =5.0$, the dotted-dashed line to $\xi =1.0$, the dashed line to $\xi =0.5$, the dotted line to $\xi =0.1$ and the solid grey line to conventional physics.}
\end{figure}

We distinguish two cases and we discuss them in turn:
\begin{itemize}
\item As long as $\xi \ll 60 \left({{\rm Mpc}}/{L_{\rm dom}} \right)$, Eq. (\ref{a16gh2705a}) yields $P_{\gamma \to a} (L_{\rm dom}) \ll 1$ which entails that the fraction of ALPs produced over a single domain is very small. Since we are supposing the beam to be initially fully made of photons, it takes a length much larger that $L_{\rm dom}$ before a sizeable fraction of the beam consists of ALPs. In the same fashion, once such a situation is realized, a similar long length is needed in order for the beam to contain a sizeable amount of photons. Moreover, it follows from Eq. (\ref{a16gh2705a}) that presently 
$P_{\gamma \to a} (L_{\rm dom})$ becomes a quadratic function of $\xi$ which therefore increases monotonically with $\xi$. Accordingly, the picture outlined in Subsection IV-A is expected to emerge straightforwardly and this is confirmed by a numerical simulation in which $\xi$ takes the above values but we assume $L_{\rm dom} = 0.05 \, {\rm Mpc}$, which yields the behaviour shown in the plots reported in Figure 4.
\item When condition $\xi \ll 60 \left({{\rm Mpc}}/{L_{\rm dom}} \right)$ is not fulfilled the situation becomes considerably more complicated. In the first place, $P_{\gamma \to a} (L_{\rm dom})$ fails to be a monotonically increasing function of $\xi$ and it becomes oscillatory. So, depending on the actual value of $L_{\rm dom}$ it follows that $P_{\gamma \to a} (L_{\rm dom})$ can decrease as $\xi$ increases. As stressed above, Eq. (\ref{a16gh2705a}) can be taken at most to provide an order-of-magnitude estimate but it is clear that condition $\xi \ll 60 \left({{\rm Mpc}}/{L_{\rm dom}} \right)$ fails to be met for $L_{\rm dom} = 4 \, {\rm Mpc}$ and $L_{\rm dom} = 10 \, {\rm Mpc}$ along with the considered values of $\xi$. Hence, for a fixed source distance the behaviour exhibited in the plots in Figure 3 can arise. Still, this is not the end of the story, since in the present situation even after the domain closest to the source a relevant fraction of the beam consists of ALPs. In other words, a large enough number of $\gamma \to a$ and $a \to \gamma$ transitions take place inside a single domain. So, the overall effect is to have a larger number of photons per unit length between the source and us as compared to the previous case. As a consequence, EBL absorption is now more effective thereby giving rise to a {\it smaller} observed photon flux. Moreover, this dimming evidently increases with the source distance, which explains why $P^{\rm DARMA}_{\gamma \to \gamma} (E_0,z)$ tends to decrease as $z$ increases at fixed energy and eventually it behaves like $P^{\rm CP}_{\gamma \to \gamma} (E_0,z)$, which indeed occurs for $\xi = 5.0$. 
\end{itemize}

In conclusion, it is evident from Figures 3 that in the most favourable case a boost factor of 10 in $P^{\rm DARMA}_{\gamma \to \gamma} (E_0,z)$ occurs at progressively lower energies as the source distance increases. Explicitly, for $z = 0.031$, $z = 0.188$, $z = 0.444$, $z = 0.536$ it takes place at $E_{10} \simeq 30 \, {\rm TeV}, 8 \, {\rm TeV}, 2 \, {\rm TeV}, 1.2 \, {\rm TeV}$, respectively. Above $E_{10}$ the boost factor can be much larger.

\section{A NEW INTERPRETATION OF OBSERVED VHE BLAZARS}

Observed VHE blazars provide a great deal of information which can challenge the DARMA scenario. Within conventional physics the values of $\Gamma_{\rm em}$ have to be tuned for every source in such a way to reproduce the corresponding values of $\Gamma_{\rm obs}$. No rational lies behind this procedure and a large spread in the values of $\Gamma_{\rm em}$ is demanded in order to account for the equally large spread in the values of $\Gamma_{\rm obs}$. While this procedure does not pose any technical problem, such a systematic fine-tuning lacks any conceptual appeal and moreover leads to the cosmic opacity problem.

Thus, it looks natural to inquire whether the DARMA scenario sheds some light on this issue.

The most straightforward way to investigate this question is to proceed somehow in parallel with the treatment followed in Subsection III-C, namely to de-absorb the values of $\Gamma_{\rm obs}$ within the present context.

Our starting point is the general relation between the observed and emitted fluxes expressed by Eq. (\ref{a02122010}), which presently reads
\begin{equation} 
\label{a26032011}
\Phi_{\rm obs}(E_0,z) = P^{\rm DARMA}_{\gamma \to \gamma} (E_0,z) \, \Phi_{\rm em} \left( E_0 (1+z) \right)~. 
\end{equation}
Thanks to Eq. (\ref{a006012011A}), we first rewrite Eq. (\ref{a26032011}) as
\begin{equation} 
\label{a26032011q}
K \, E_0^{- \Gamma_{\rm obs} (z)} = P^{\rm DARMA}_{\gamma \to \gamma} (E_0,z) \, \Phi_{\rm em} \left( E_0 (1+z) \right)~, 
\end{equation}
from which we get $\Phi_{\rm em} \left( E_0 (1+z) \right)$ for every detected VHE blazar. We next best-fit this function to the power-law expression (\ref{a006012011A1}), namely
\begin{equation} 
\label{a006012011A1w}
\Phi_{\rm em} \left( E_0 (1+z) \right) = K \, \Bigl[ E_0 (1+z) \Bigr]^{- \Gamma^{\rm DARMA}_{\rm em}}
\end{equation}
over the energy range where the considered source is observed (see Table \ref{tab:a1}). We obtain in this way the values of $\Gamma^{\rm DARMA}_{\rm em}$. 

We stress that for a given choice of $\xi$ and $L_{\rm dom}$ the photon survival probability $P^{\rm DARMA}_{\gamma \to \gamma} (E_0,z)$ is 
uniquely fixed, apart from errors affecting $\tau_{\gamma}^{\rm FRV}(E_0, z)$ which are unknown and therefore again ignored. Since $\Gamma^{\rm DARMA}_{\rm em}$ is linearly related to $\Gamma_{\rm obs}$, the associated error bars are the {\it same} even in the present context (see Table \ref{tab:a1}) and so they will not be explicitly exhibited.

We implement the considered procedure by taking for the free parameters $\xi$ and $L_{\rm dom}$ the same representative values chosen above. The resulting numerical values of $\Gamma^{\rm DARMA}_{\rm em}$ for the various cases are reported in Tables \ref{tab:aaa} and \ref{tab:aaa2}, together with the corresponding spread $\Delta \Gamma^{\rm DARMA}_{\rm em}$ of the value of $\Gamma_{\rm em}$ neglecting errors and the average value $\langle \Gamma^{\rm DARMA}_{\rm em} \rangle$ including errors. 

Let us consider first what happens as $\xi$ increases from $\xi = 0.1$ to $\xi = 5.0$ assuming $L_{\rm dom} = 4 \, {\rm Mpc}$. As long as $\xi = 0.1$ the difference between the DARMA scenario and conventional physics strictly vanishes up to $z = 0.138$ and becomes negligible at larger redshifts. However, as soon as the regime $\xi = 0.5$ is attained the DARMA scenario starts to differ drastically from conventional physics. As $\xi$ increases from 0.5 to 5.0 it is found that $\langle \Gamma^{\rm DARMA}_{\rm em} \rangle$ monotonically increases, even if at a rate that slows down for increasing $\xi$, which entails that no improvement is to be expected for $\xi > 5.0$ regardless of any other consideration. On the other hand, 
$\Delta \Gamma^{\rm DARMA}_{\rm em}$ decreases for $\xi = 0.1 \to \xi = 1.0$ but next increases for $\xi = 1.0 \to \xi = 5.0$.

A somewhat similar pattern is found for $L_{\rm dom} = 10 \, {\rm Mpc}$. Again for $\xi = 0.1$ the difference between the DARMA scenario and conventional physics strictly vanishes up to $z = 0.116$ and remains negligible at larger redshifts. A big difference shows up around $\xi = 0.5$. As before, $\langle \Gamma^{\rm DARMA}_{\rm em} \rangle$ increases monotonically, but its rate slows down for increasing $\xi$ and just vanishes in the step $\xi = 1.0 \to \xi = 5.0$. Here $\Delta \Gamma^{\rm DARMA}_{\rm em}$ still decreases for $\xi = 0.1 \to \xi = 0.5$ but then increases for $\xi = 0.5 \to \xi = 5.0$.

Let us next find out what happens in the change $L_{\rm dom} = 4 \, {\rm Mpc} \to L_{\rm dom} = 10 \, {\rm Mpc}$ at fixed $\xi$. As far as 
$\Delta \Gamma^{\rm DARMA}_{\rm em}$ is concerned, it decreses for $\xi = 0.1$ and $\xi = 0.5$ but it increases for $\xi = 1.0$ while it remains practically unchanged for $\xi = 5.0$. The behaviour of $\langle \Gamma^{\rm DARMA}_{\rm em} \rangle$ is slightly different, since only in the case $\xi = 0.5$ it shows a slight variation.

Physically, all this means that for $\xi = 0.1$ DARMA effects are negligible, but they suddenly become important shortly before $\xi = 0.5$ is reached and they remain more or less unchanged up to $\xi = 5.0$. This conclusion is in remarkable agreement with our previous results concerning the behaviour of $P^{\rm DARMA}_{\gamma \to \gamma} (E_0,z)$. In this connection, two points should be stressed. It follows from Eqs. (\ref{a26032011q}) and (\ref{a006012011A1w}) that at fixed $z$ $\Gamma^{\rm DARMA}_{\rm em}$ depends logarithmically on $P^{\rm DARMA}_{\gamma \to \gamma} (E_0,z)$, which makes its dependence on $\xi$, $E_0$ and $L_{\rm dom}$ much shallower than that of $P^{\rm DARMA}_{\gamma \to \gamma} (E_0,z)$ itself. Consequently, the sharp differences found in Section VI among the cases with different $\xi$ get smoothed out here, apart from one thing: since in the case $\xi = 0.1$ the photon survival probability is nearly the same in conventional physics and within the DARMA scenario over the energy range $0.2 \, {\rm TeV} < E_0 < 2 \, {\rm TeV}$ where most blazars are observed, the same is evidently true for the values of $\Gamma_{\rm em}$. In addition, the marked difference among the cases $\xi = 0.1, 0.5, 1.0, 5.0$ discovered in Section VI takes place at energies considerably larger than those in the presently considered range, which explains why the cases $\xi = 0.5, 1.0, 5.0$ exhibit a fairly similar behaviour for the observed blazars.

\begin{table}
\begin{tabular}{cccccc}          
\hline
 {${\rm Source}$}{}{}{}   
&  {}{${\Gamma}^{\rm CP}_{\rm em}$}{}{} 
&  {}{}{${\Gamma}^{\rm DARMA}_{\rm em}$} 
&  {}{}{${\Gamma}^{\rm DARMA}_{\rm em}$} 
&  {}{}{${\Gamma}^{\rm DARMA}_{\rm em}$}
&  {}{}{${\Gamma}^{\rm DARMA}_{\rm em}$} 
\\
&   
& ${\xi = 0.1}$ 
& ${\xi = 0.5}$
& ${\xi = 1.0}$ 
& ${\xi = 5.0}$ \\
\hline
3C 66B     &  3.00    &  3.00      &  3.00    &  3.00      &  3.03  \\
Mrk 421        &  2.16   &  2.16      &  2.16     &  2.17      &  2.21 \\   
Mrk 501      \    & 1.90        &  1.90      &  1.90     &  1.91      &  1.96  \\    
Mrk 501       \   &  2.03       &  2.03      &  2.03    &  2.04      &  2.08   \\    
1ES 2344+514      \  &  2.70       &  2.70      &  2.71    &  2.73      &  2.78  \\     
Mrk 180      \   &  3.07       &  3.07      &  3.07    &  3.09      &  3.14   \\    
1ES 1959+650      \     & 2.43       &  2.43      &  2.44      &  2.46      &  2.53   \\    
BL Lacertae       \     &  3.27       &  3.27      &  3.28      &  3.32      &  3.38   \\    
PKS 0548-322       \   & 2.39       &  2.39      &  2.40    &  2.45      &  2.52   \\    
PKS 2005-489       \     & 3.59       &  3.59      &  3.60       &  3.66      &  3.73   \\    
RGB J0152+017      \     & 2.47      &  2.47      &  2.48      &  2.56      &  2.63   \\    
W Comae      \    &  3.18        &    3.18       &  3.21    &    3.32       &  3.39   \\    
PKS 2155-304      \  &  2.67             &   2.67      & 2.72      &   2.85      & 2.90   \\ 
RGB J0710+591      \   &   ?           &      ?       & ?        &      ?       & ?    \\ 
H 1426+428      \  &   0.85          &       0.85      &  1.28    &       1.57      &  1.44     \\ 
1ES 0806+524      \  &   2.70              &     2.70      & 2.77   &     2.93      & 3.00    \\ 
1ES 0229+200      \  &  0.41           &     0.42      & 1.15      &     1.37      & 1.13    \\ 
H 2356-309       \  &   2.06            &    2.06       & 2.17    &    2.35       & 2.40      \\ 
1ES 1218+304      \  & 2.00            &      2.00      & 2.15    &      2.32      & 2.36     \\ 
1ES 1101-232      \  &  1.72              &    1.73       & 1.96    &    2.13       & 2.13    \\ 
1ES 0347-121      \  &  1.87            &    1.87        & 2.11    &    2.28        & 2.28     \\ 
1ES 1011+496      \  & 2.90             &      2.90       &  3.06   &      3.22       &  3.26    \\ 
S5 0716+714       \  &   1.60           &   1.61        & 2.07    &   2.22        & 2.22    \\ 
PG 1553+113      \  & 2.48            &  2.49      & 3.00     &  3.08      & 3.08   \\ 
PKS 1222+21      \  & 2.47          &  2.47     & 2.80     &  2.88     & 2.90       \\ 
3C 66A       \    &  1.28               &     1.30      & 2.19    &     2.25      & 2.22   \\ 
PKS 1424+240      \   &  1.16           &     1.18      & 2.03   &     2.06      & 2.04     \\ 
3C 279       \    &    2.05              &  2.06       & 2.71     &  2.74       & 2.73      \\ 
$\Delta \Gamma_{\rm em}$    \ \ \  & 3.18        &  3.17         &  2.45               &  2.29         &  2.60          \\
$\langle \Gamma_{\rm em} \rangle $    \ \ \  & 2.22          &   2.23              & 2.41     &   2.51              & 2.52         \\
\hline
\end{tabular}
\caption{We have inferred the emitted spectral index ${\Gamma}_{\rm em}$ by de-absorbing within the DARMA scenario the observed value of
 ${\Gamma}_{\rm obs}$ for every source neglecting errors. This procedure has been carried out for the choice of parameters $\xi = 0.1$, 
$\xi = 0.5$, $\xi = 1.0$ and $\xi = 5.0$. In all cases, we have taken $L_{\rm dom} = 4 \, {\rm Mpc}$. The similar values obtained in Subsection III-C within conventional physics have been quoted for comparison and are denoted by ${\Gamma}^{\rm CP}_{\rm em}$. The last two lines report the spread $\Delta \Gamma_{\rm em}$ of the value of $\Gamma_{\rm em}$ discarding errors and the average value $\langle \Gamma_{\rm em} \rangle $ including errors, respectively, for the various cases.}
\label{tab:aaa}
\end{table}

\begin{table}
\begin{tabular}{cccccc}          
\hline
 {${\rm Source}$}{}{}{}   
&  {}{${\Gamma}^{\rm CP}_{\rm em}$}{}{} 
&  {}{}{${\Gamma}^{\rm DARMA}_{\rm em}$} 
&  {}{}{${\Gamma}^{\rm DARMA}_{\rm em}$} 
&  {}{}{${\Gamma}^{\rm DARMA}_{\rm em}$}
&  {}{}{${\Gamma}^{\rm DARMA}_{\rm em}$} 
\\
&   
& ${\xi = 0.1}$ 
& ${\xi = 0.5}$
& ${\xi = 1.0}$ 
& ${\xi = 5.0}$ \\
\hline
3C 66B       \   &  3.00       &  3.00      &  3.03    &  3.01      &  3.03     \\
Mrk 421      \   &  2.16       &  2.16      &  2.16   &  2.18      &  2.21      \\   
Mrk 501      \    & 1.90        &  1.90      &  1.90   &  1.93      &  1.96      \\    
Mrk 501       \   &  2.03       &  2.03      &  2.03     &  2.06      &  2.08   \\    
1ES 2344+514      \  &  2.70       &  2.70      &  2.72    &  2.76      &  2.79     \\     
Mrk 180      \   &  3.07       &  3.07      &  3.08        &  3.11      &  3.14    \\    
1ES 1959+650      \     & 2.43       &  2.43      &  2.45    &  2.50      &  2.53   \\    
BL Lacertae       \     &  3.27       &  3.27      &  3.30      &  3.36      &  3.38    \\    
PKS 0548-322       \   & 2.39       &  2.39      &  2.43       &  2.50      &  2.52        \\    
PKS 2005-489       \     & 3.59       &  3.59      &  3.63      &  3.70      &  3.73      \\    
RGB J0152+017      \     & 2.47      &  2.47      &  2.53    &  2.61      &  2.63     \\    
W Comae      \    &  3.18        &    3.18       &  3.28     &    3.37       &  3.39      \\     
PKS 2155-304      \  &  2.67             &   2.67      & 2.81         &   2.89      & 2.90    \\ 
RGB J0710+591      \   &   ?           &      ?       & ?         &      ?       & ?    \\ 
H 1426+428      \  &   0.85          &       0.86      &  1.53     &       1.53      &  1.44   \\ 
1ES 0806+524      \  &   2.70              &     2.71      & 2.88       &     2.98      & 3.00   \\ 
1ES 0229+200      \  &  0.41           &     0.48      & 1.36     &     1.27      & 1.12  \\ 
H 2356-309       \  &   2.06            &    2.06       & 2.31    &    2.39       & 2.40    \\  
1ES 1218+304      \  & 2.00            &      2.00      & 2.29     &      2.35      & 2.36     \\  
1ES 1101-232      \  &  1.72              &    1.73       & 2.10    &    2.14       & 2.13    \\ 
1ES 0347-121      \  &  1.87            &    1.88        & 2.26       &    2.29        & 2.28    \\ 
1ES 1011+496      \  & 2.90             &      2.90       &  3.19      &      3.25       &  3.26   \\ 
S5 0716+714       \  &   1.60           &   1.62        & 2.20     &   2.22        & 2.22    \\ 
PG 1553+113      \  & 2.48            &  2.52      & 3.07               &  3.08      & 3.08       \\  
PKS 1222+21      \  & 2.47          &  2.49     & 2.87      &  2.89     & 2.90     \\ 
3C 66A       \    &  1.28               &     1.36      & 2.25   &     2.23      & 2.22    \\  
PKS 1424+240      \   &  1.16           &     1.26      & 2.07     &     2.05      & 2.04    \\  
3C 279       \    &    2.05              &  2.13       & 2.74       &  2.74       & 2.73   \\  
$\Delta \Gamma_{\rm em}$    \ \ \  & 3.18        &  3.11         &  2.27         &  2.43         &  2.61        \\
$\langle \Gamma_{\rm em} \rangle $    \ \ \  & 2.22          &   2.24              & 2.49   &   2.52              & 2.52            \\
\hline
\end{tabular}
\caption{Same as Table \ref{tab:aaa} but with $L_{\rm dom} = 10 \, {\rm Mpc}$.}
\label{tab:aaa2}
\end{table}

\subsection{Solution of the cosmic opacity problem}

A glance at Tables \ref{tab:aaa} and \ref{tab:aaa2} shows that the values of $\Gamma^{\rm DARMA}_{\rm em}$ for {\it all} VHE blazars happen to be in the {same} ballpark, thereby implying that within the DARMA scenario { {the observations can be explained with the same physical mechanism operating in all blazars -- as a consequence, there is no cosmic opacity problem.}}

\subsection{Fitting individual sources}

The foregoing analysis has shown that the gist of the DARMA scenario for the observed VHE blazars is to drastically reduce the spread in the values of 
$\Gamma_{\rm em}$ as compared with what happens in conventional physics, thereby tracing the large spread in the values of $\Gamma_{\rm obs}$ to the wide spread in the blazar distances. 

It seems therefore worthwhile to investigate this point in a quantitative fashion according the following strategy:
\begin{itemize}
\item As a zero-order approximation, we suppose that all blazars have the {\it same} value of $\Gamma^{\rm DARMA}_{\rm em}$, which for definiteness is taken to be the average value over all observed sources $\langle \Gamma^{\rm DARMA}_{\rm em} \rangle$ for a given choice of $\xi$ and $L_{\rm dom}$.
\item As a first-order correction -- which is meant to improve on the above idealized situation -- we allow for a small spread around $\langle \Gamma^{\rm DARMA}_{\rm em} \rangle$, which we tentatively take to be $\pm \, 0.2$.
\end{itemize}

In order to keep the situation under control, we focus our attention on the single case $\xi = 1.0$ and $L_{\rm dom} = 4 \, {\rm Mpc}$ which we regard as the most favourable one not only because $\Delta \Gamma^{\rm DARMA}_{\rm em}$ is very small -- the case $\xi = 0.5$ and $L_{\rm dom} = 10 \, {\rm Mpc}$ would be even better in this respect -- but also because we feel that $L_{\rm dom} = 4 \, {\rm Mpc}$ is more realistic than $L_{\rm dom} = 10 \, {\rm Mpc}$. This amounts to take $\langle \Gamma^{\rm DARMA}_{\rm em} \rangle = 2.51$, which entails in turn $ 2.31 < \Gamma^{\rm DARMA}_{\rm em} < 2.71$ for all observed VHE blazars. The value 2.51 is close to the value 2.40 that we used in a previous discussion of the DARMA scenario~\cite{dmpr}, as well as to 2.47 which is the average value for the observed VHE blazars with $z < 0.05$ that undergo a negligible EBL attenuation. 

{  Next, we evaluate for every source the {\it expected} observed spectral index $\Gamma_{\rm obs}^{\rm exp} (z)$. Basically, this amounts to run backwards the same procedure whereby we have got the values of $\Gamma_{\rm em}^{\rm DARMA}$ reported in Table \ref{tab:aaa} for $\xi = 1.0$. Explicitly, by combining Eqs. (\ref{a26032011}) and (\ref{a006012011A1w}) we can write 
\begin{equation} 
\label{a26032011M}
\Phi_{\rm obs}^{\rm exp}(E_0,z) = P^{\rm DARMA}_{\gamma \to \gamma} (E_0,z) \, K \, \Bigl[ E_0 (1+z) \Bigr]^{- 2.51}~. 
\end{equation}
Since $P^{\rm DARMA}_{\gamma \to \gamma} (E_0,z)$ is known, $\Phi_{\rm obs}^{\rm exp}(E_0,z)$ can be computed exactly. Then we best-fit this function to the power-law expression (\ref{a006012011A}), namely
\begin{equation} 
\label{a006012011AM}
\Phi_{\rm obs}^{\rm exp}(E_0,z) = K \, E_0^{- \Gamma_{\rm obs}^{\rm exp} (z)}
\end{equation}
over the energy range where each source is observed. We find in this way the values of $\Gamma_{\rm obs}^{\rm exp} (z)$ for every source. As repeatedly stressed, the observed and emitted spectral indices are linearly related, and so they have the same error bars.}

\begin{figure}
\begin{center}
\includegraphics[width=.49\textwidth]{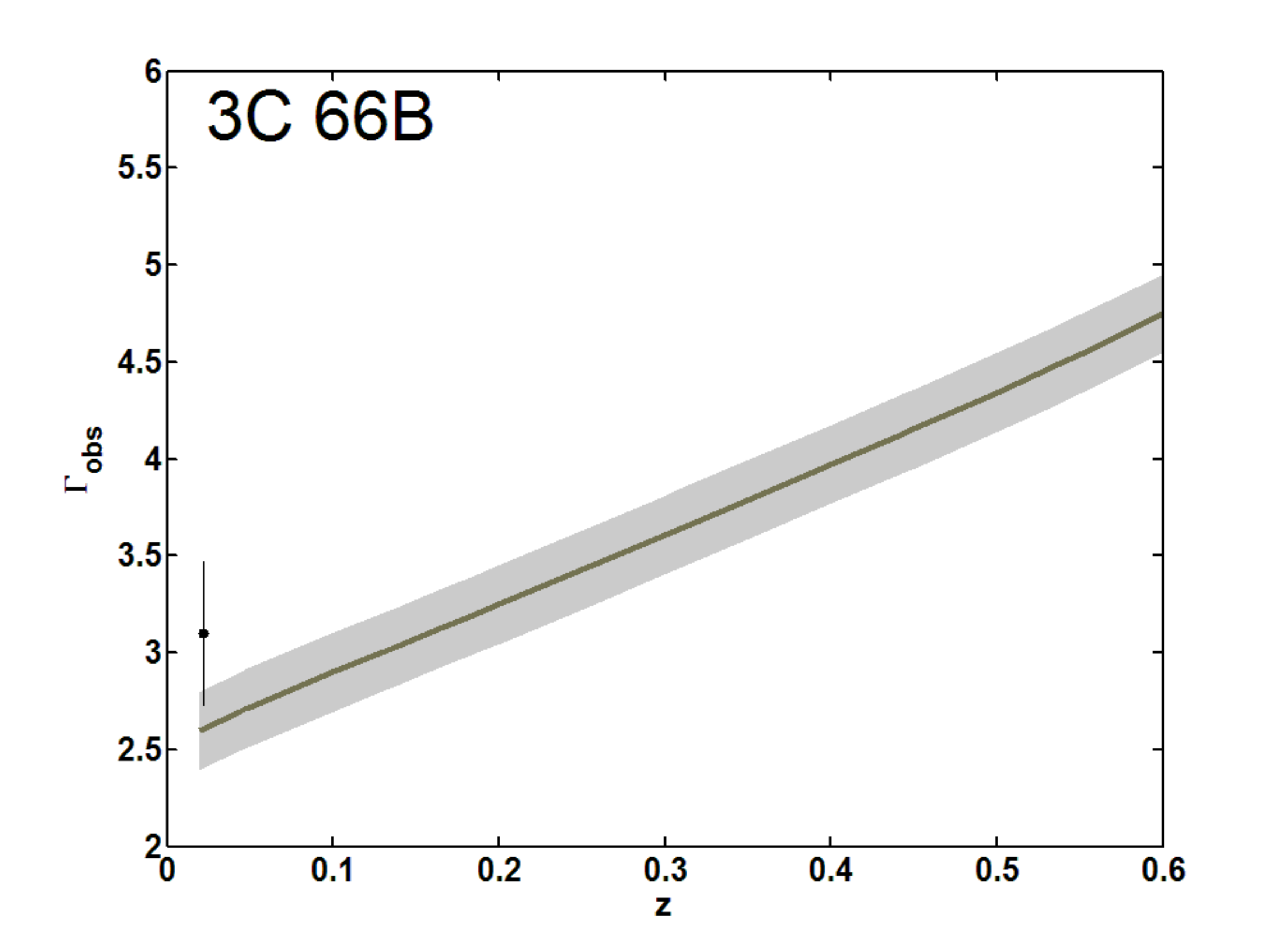}\includegraphics[width=.49\textwidth]{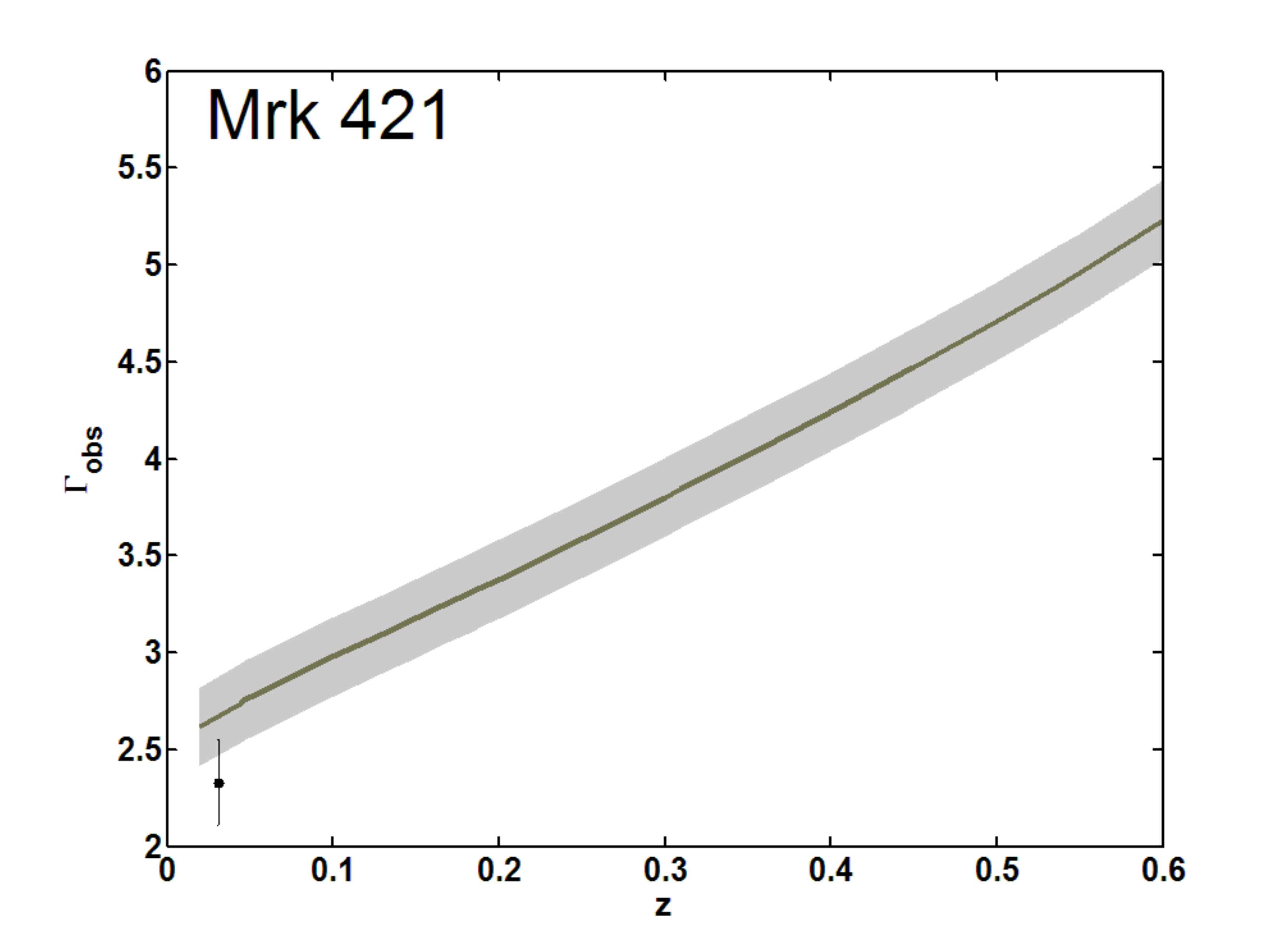}
\includegraphics[width=.49\textwidth]{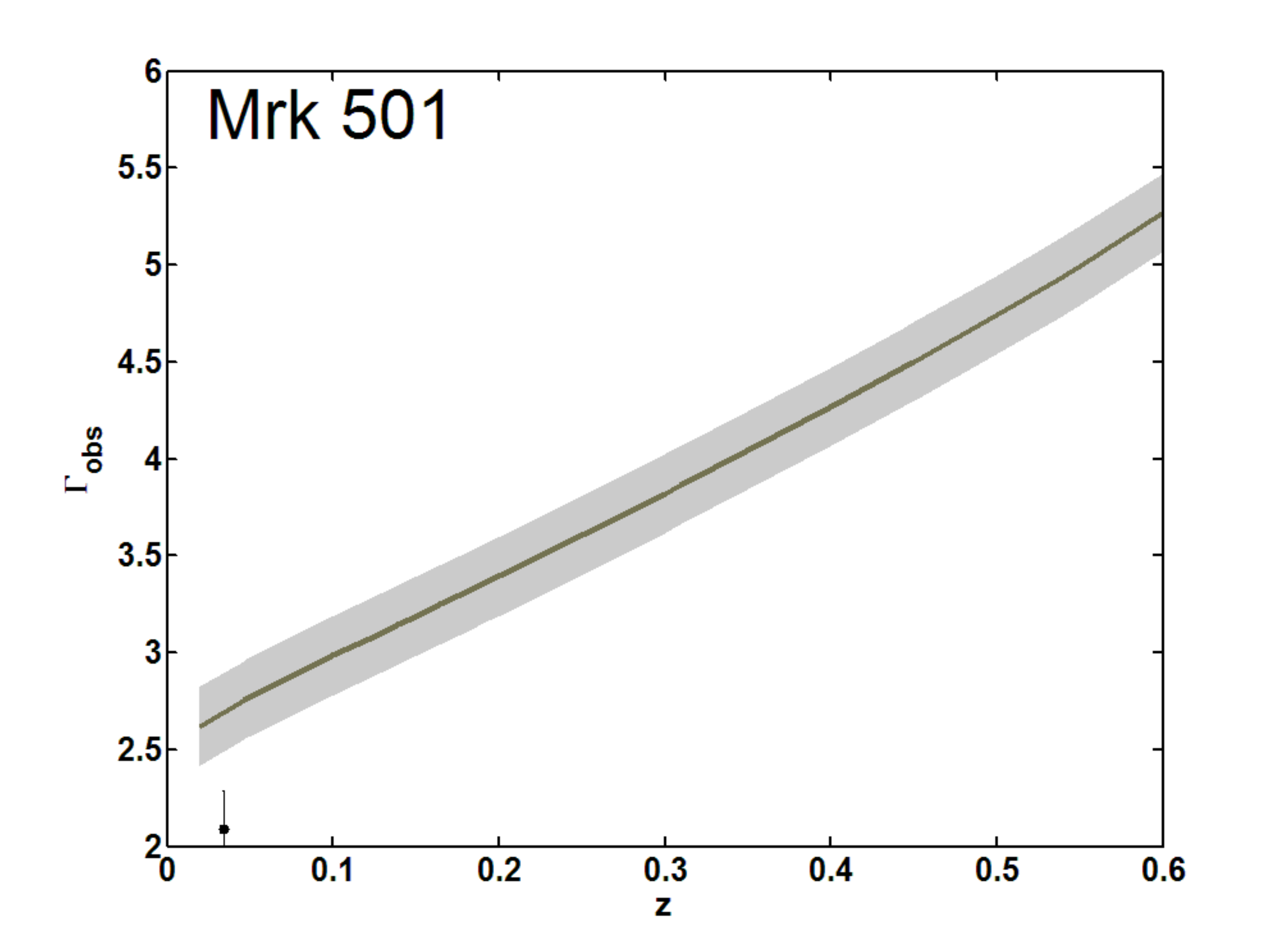}\includegraphics[width=.49\textwidth]{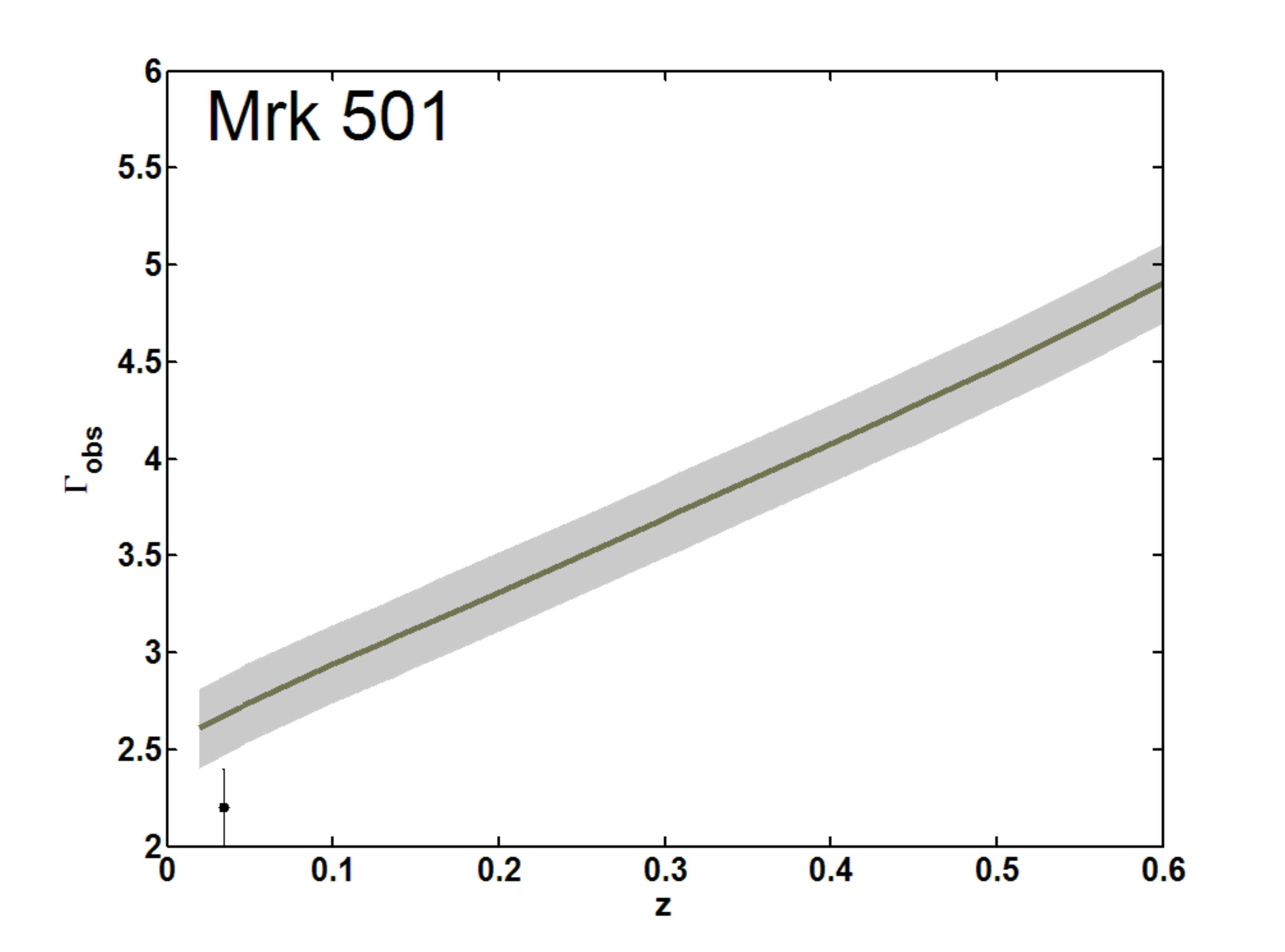}
\includegraphics[width=.49\textwidth]{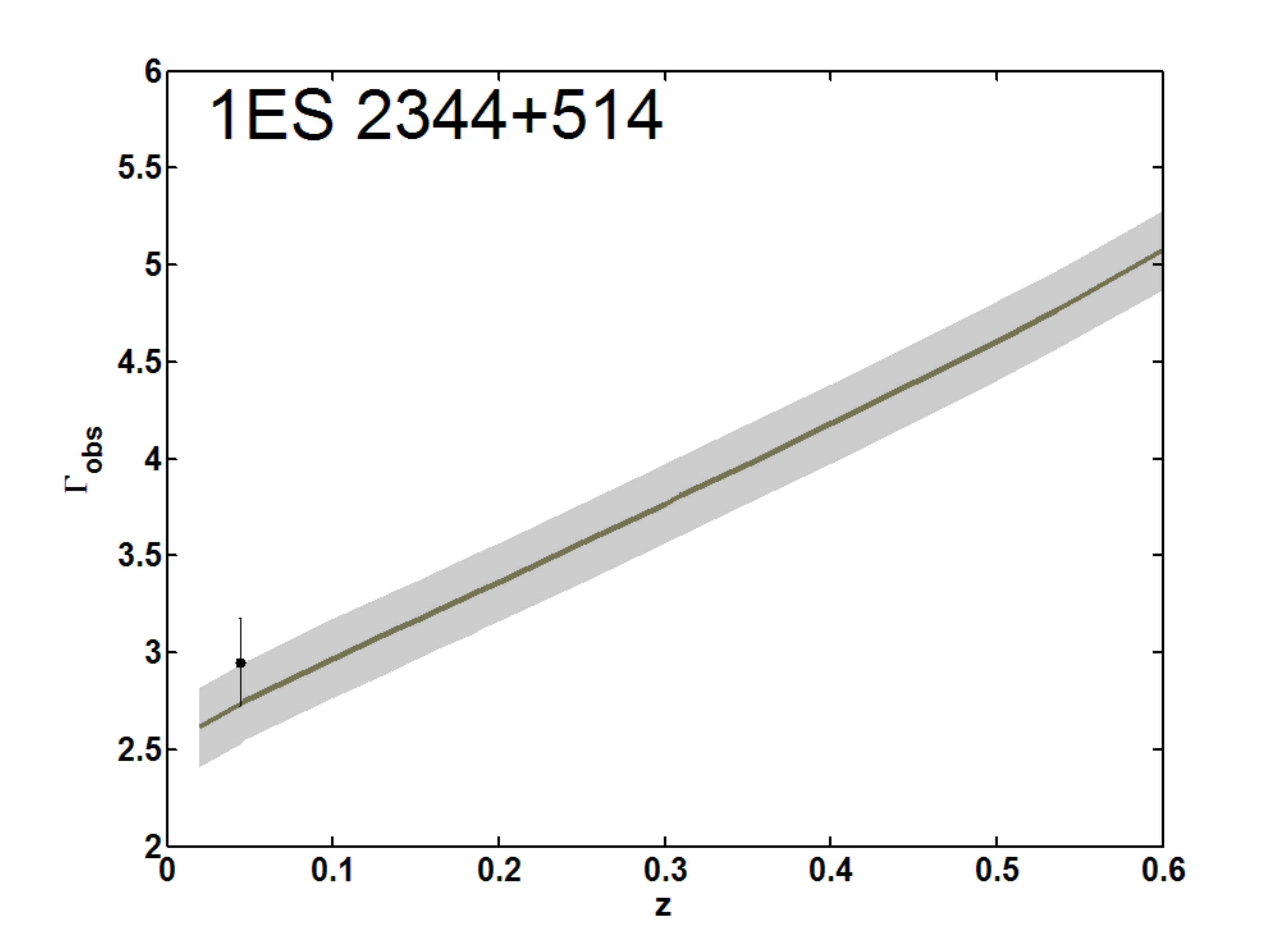}\includegraphics[width=.49\textwidth]{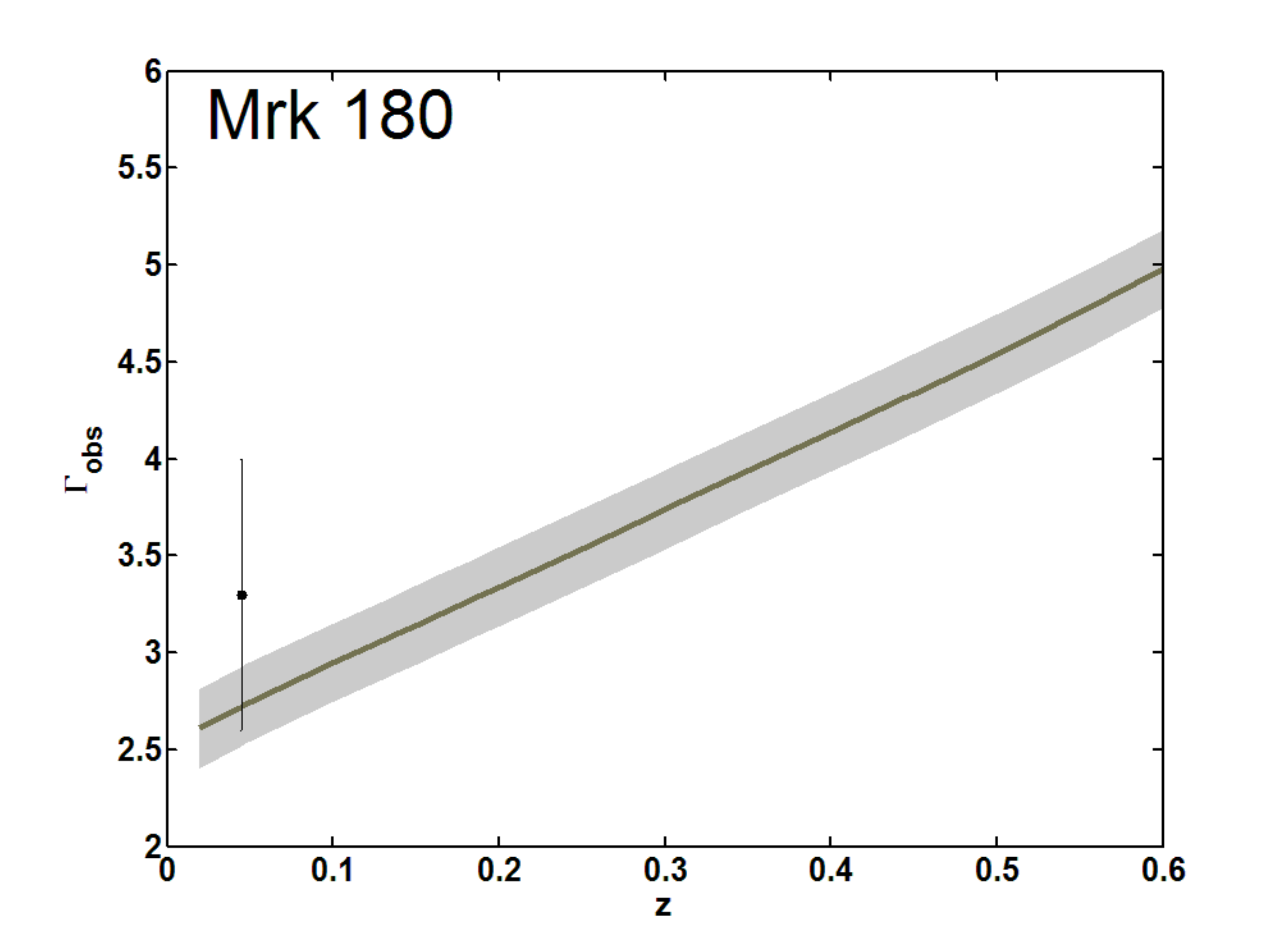}
\end{center}
\caption{\label{pendenze1.pdf} 
Behaviour of $\Gamma^{\rm DARMA}_{\rm obs}$ for the blazars 3C 66B, Mrk 421, Mrk 501 (with the two measurements of $\Gamma_{\rm obs}$ in the literature), 1ES 2344+514 and Mrk 180. The solid black line corresponds to $\Gamma^{\rm DARMA}_{\rm em} = 2.51$ and the grey strip represents the range $2.31 < \Gamma^{\rm DARMA}_{\rm em} < 2.71$.}
\end{figure}

\begin{figure}       
\begin{center}
\includegraphics[width=.49\textwidth]{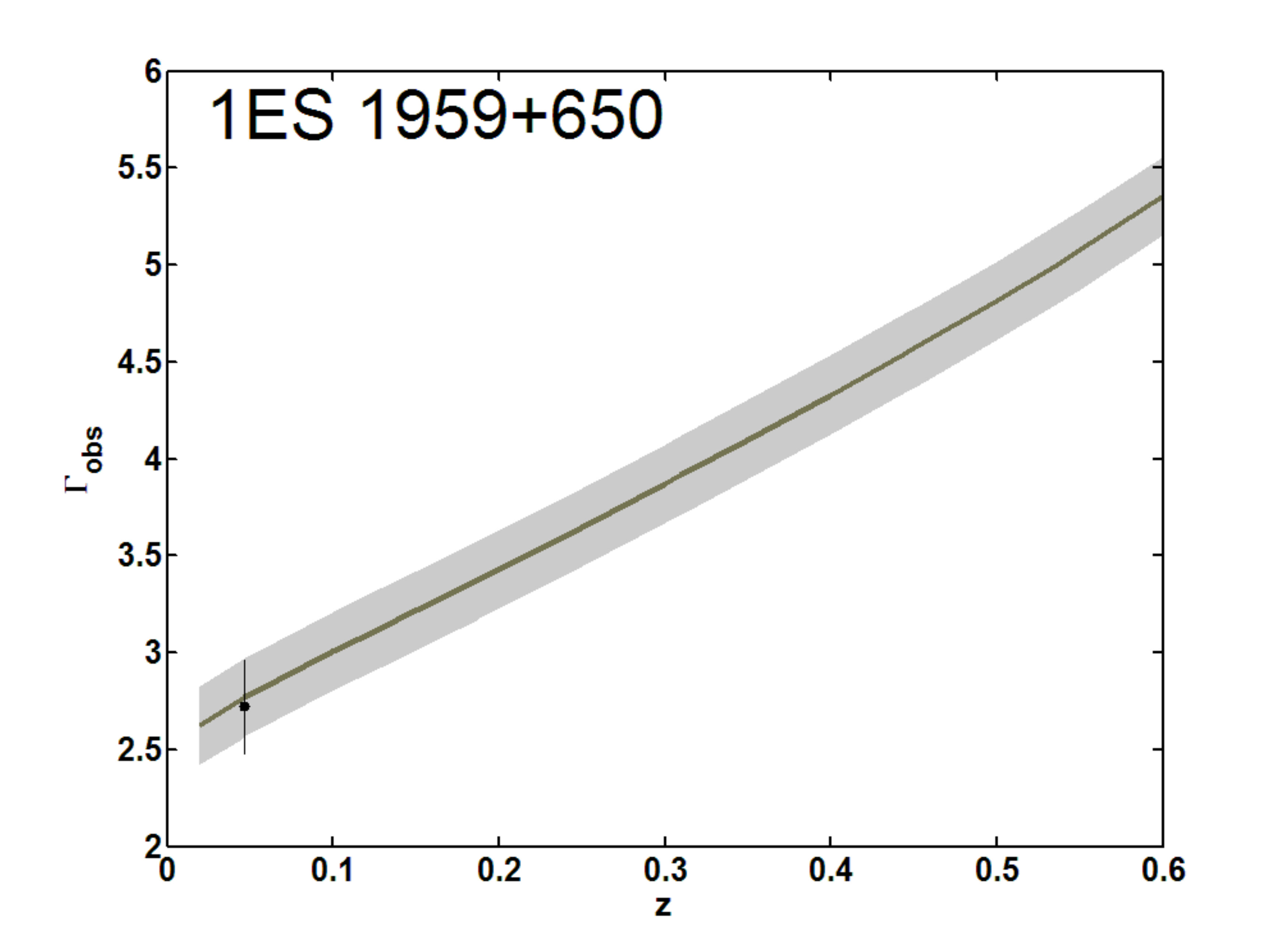}\includegraphics[width=.49\textwidth]{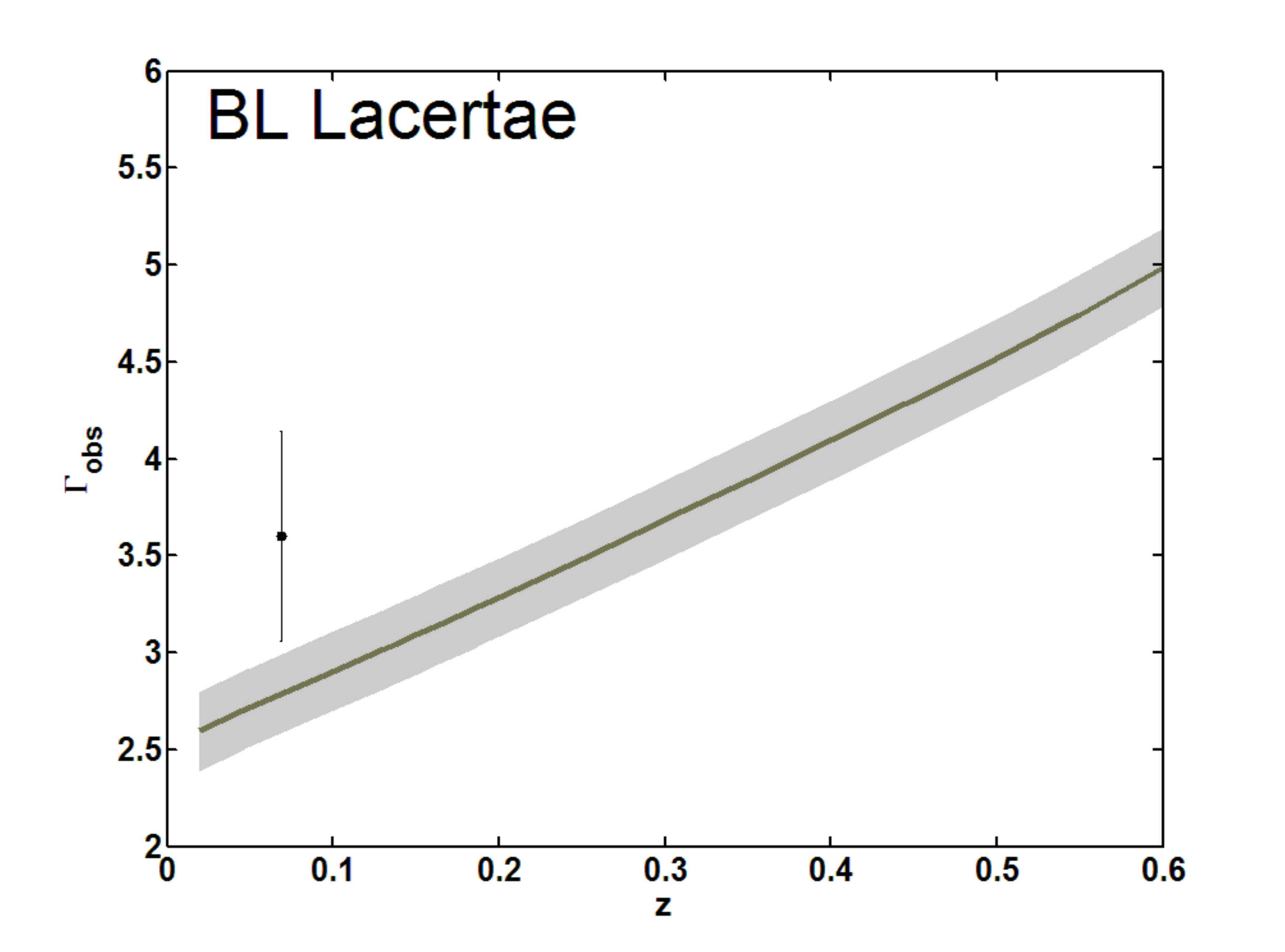}
\includegraphics[width=.49\textwidth]{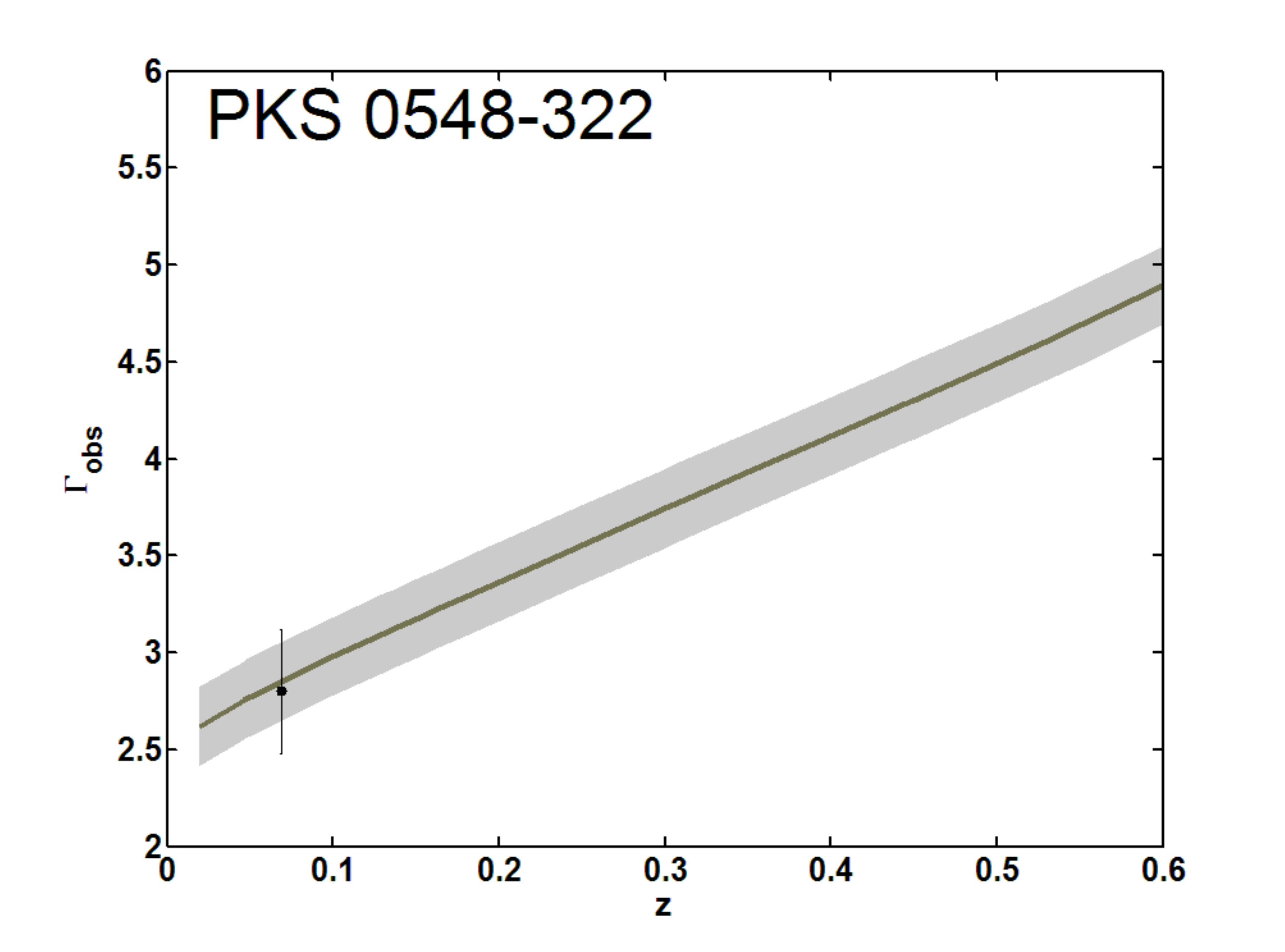}\includegraphics[width=.49\textwidth]{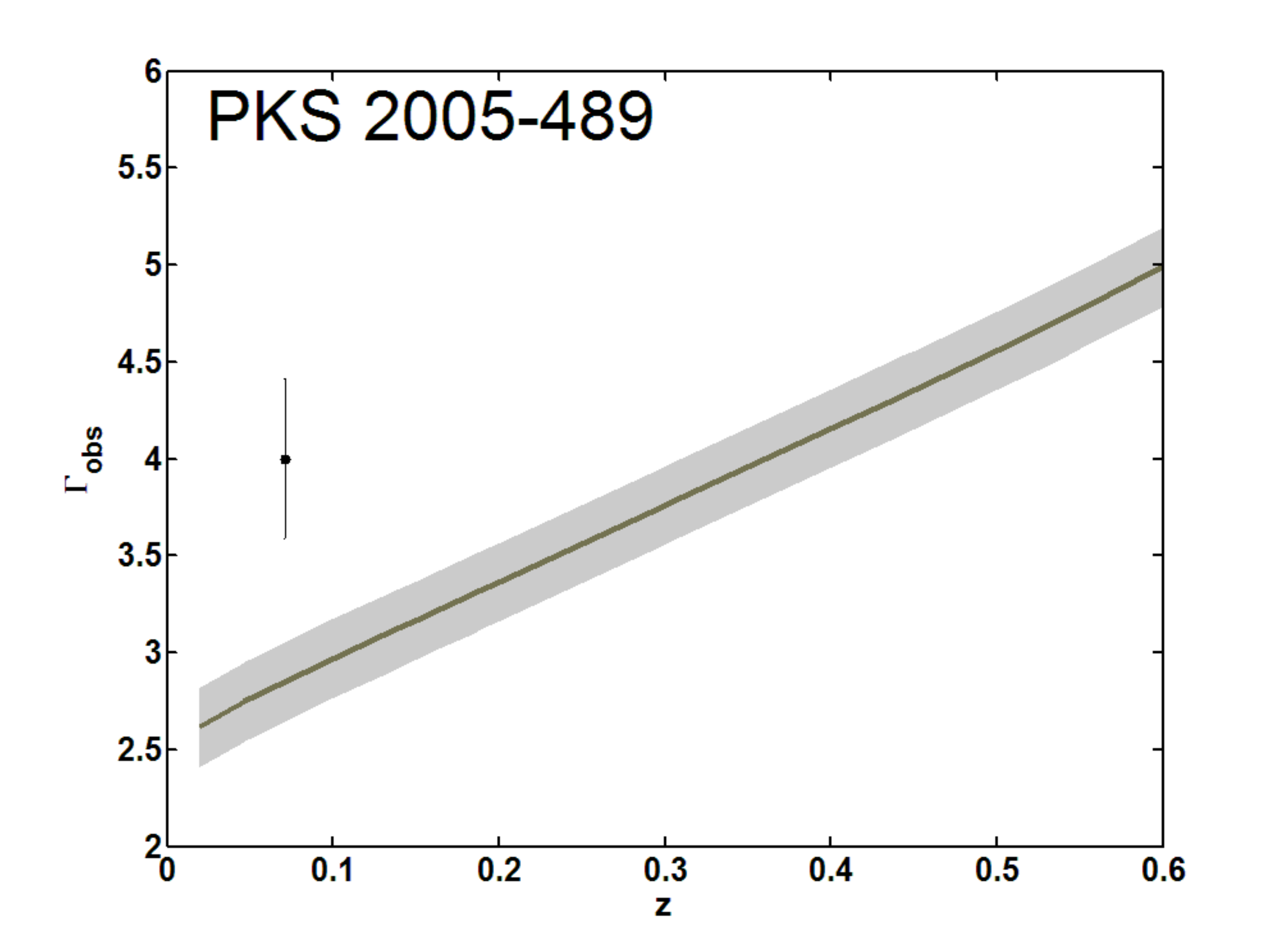}
\includegraphics[width=.49\textwidth]{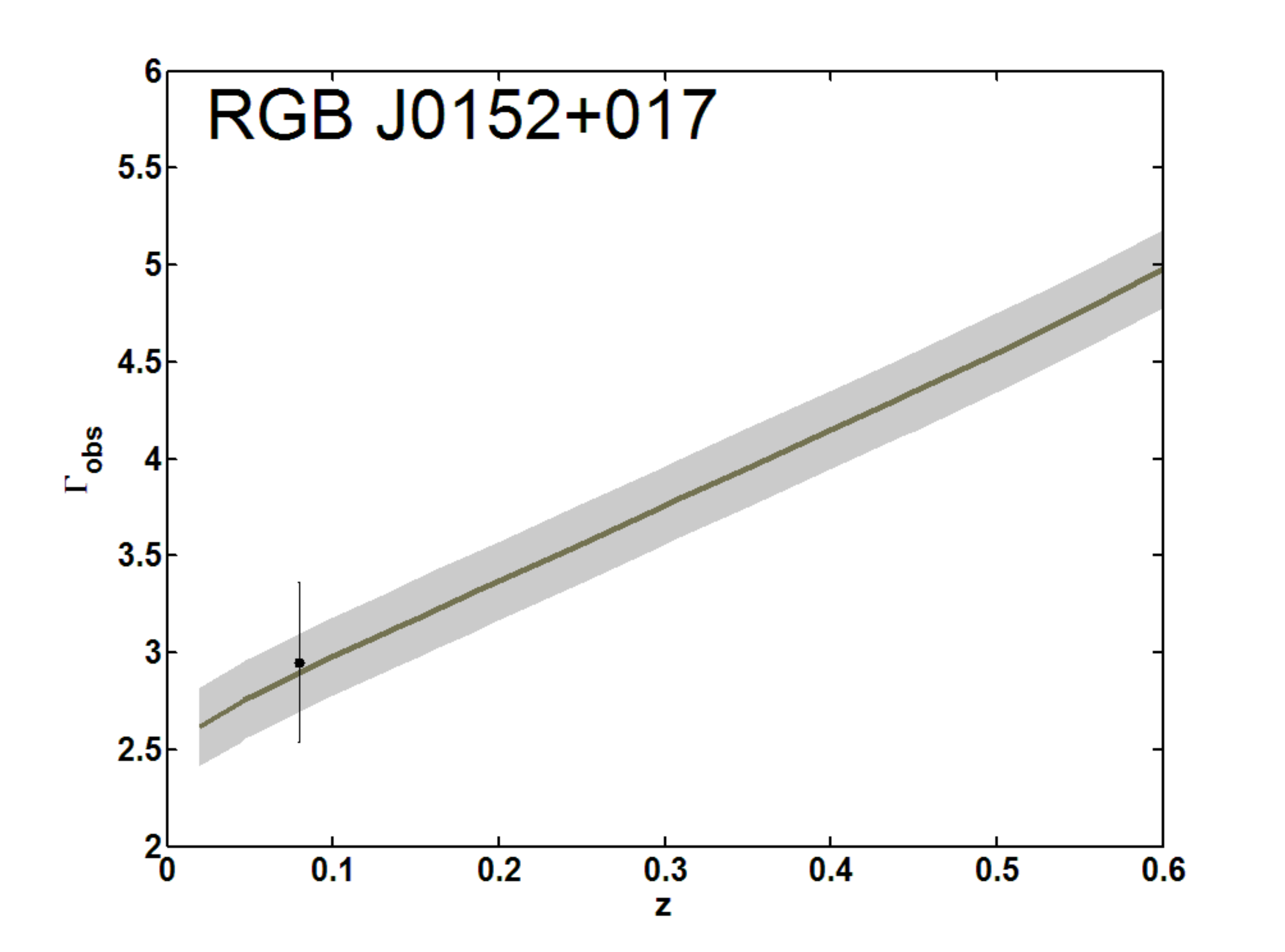}\includegraphics[width=.49\textwidth]{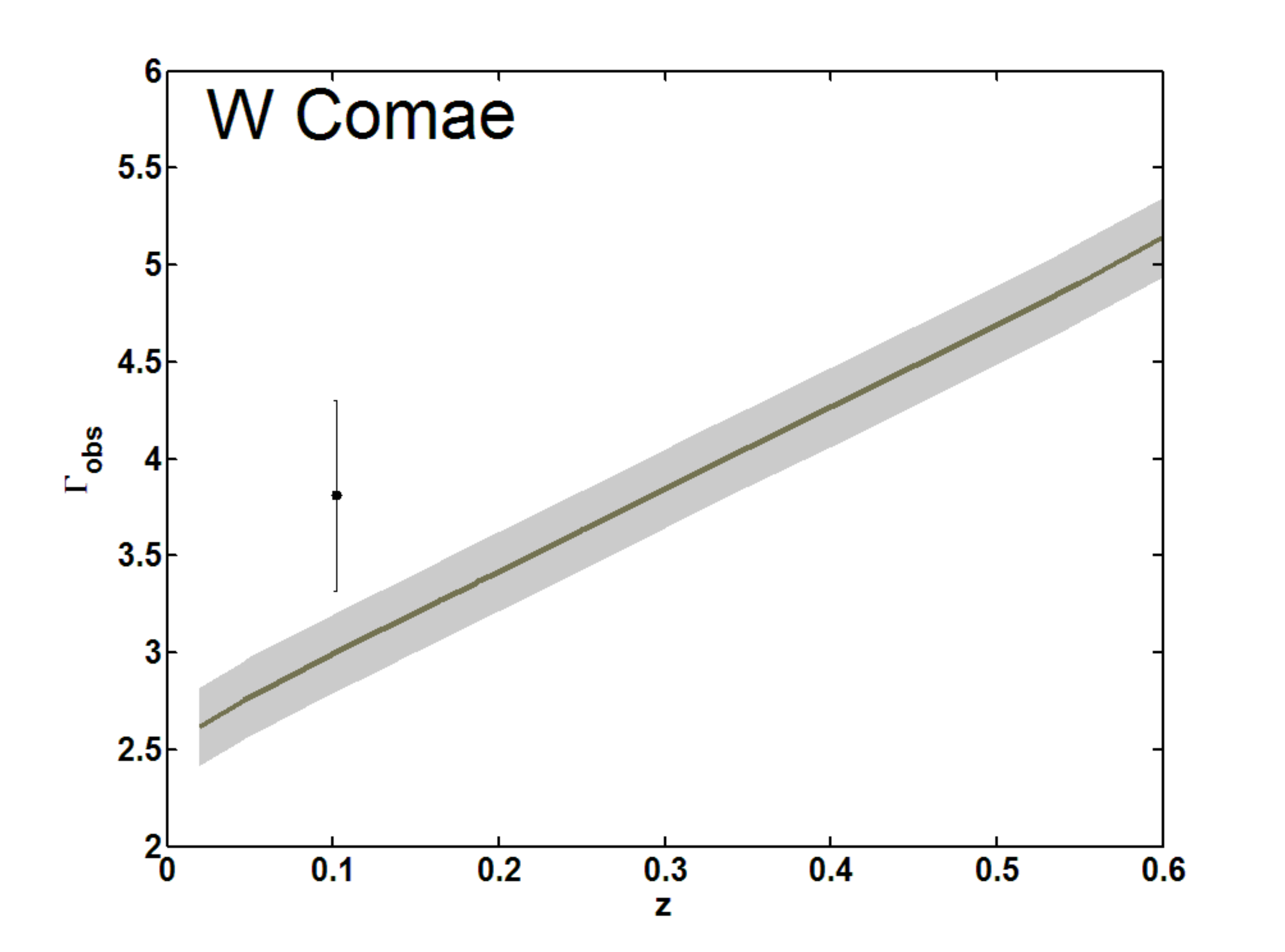}
\end{center}
\caption{\label{pendenze2.pdf} 
Behaviour of $\Gamma^{\rm DARMA}_{\rm obs}$ for the blazars 1ES 1959+650, BL Lacertae, PKS 0548-322, PKS 2005-489, RGB J0152+017 and 
W Comae. The solid black line corresponds to $\Gamma^{\rm DARMA}_{\rm em} = 2.51$ and the grey strip represents the range $2.31 < \Gamma^{\rm DARMA}_{\rm em} < 2.71$.}
\end{figure}

\begin{figure}      
\begin{center}
\includegraphics[width=.49\textwidth]{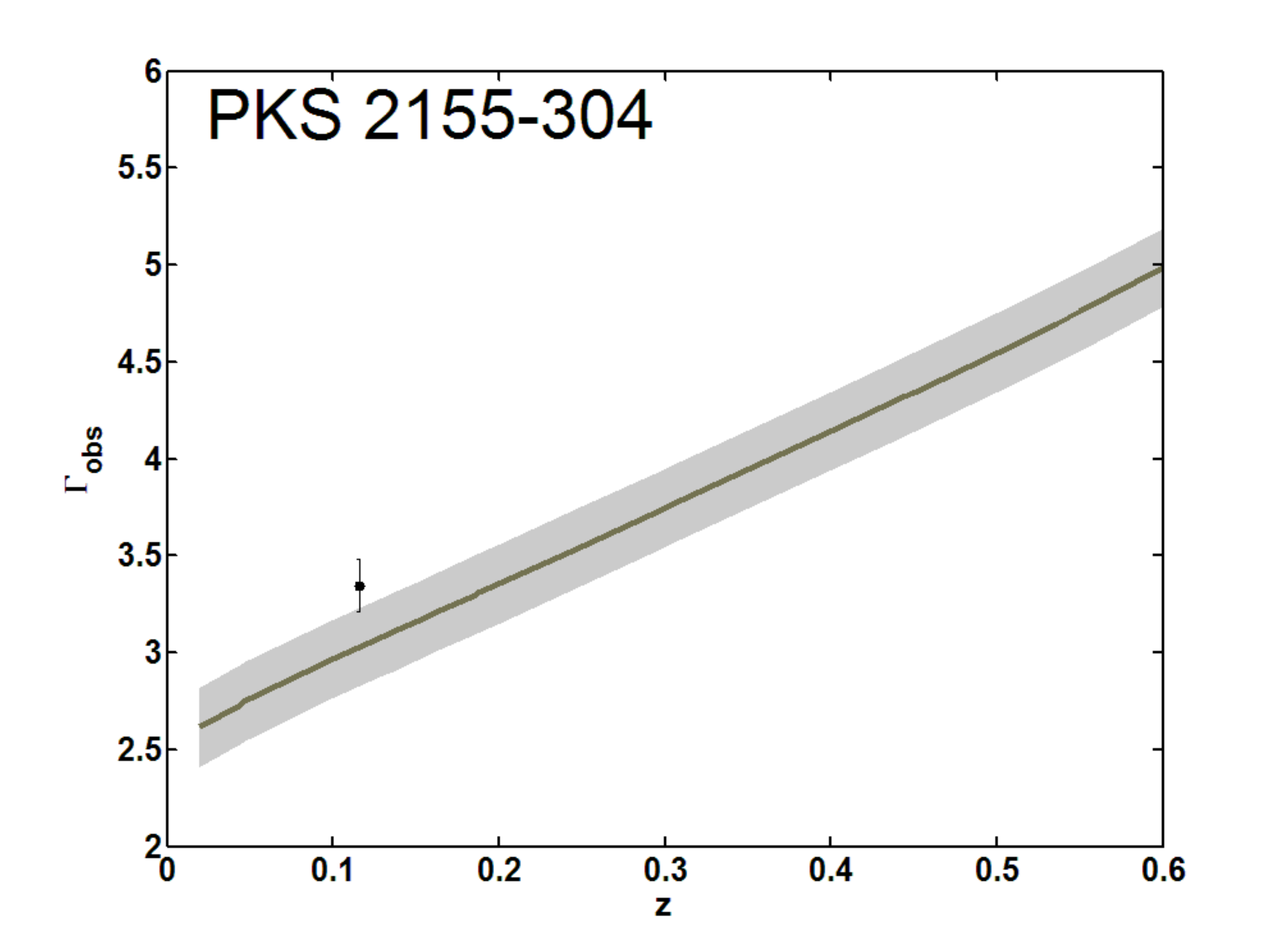}\includegraphics[width=.49\textwidth]{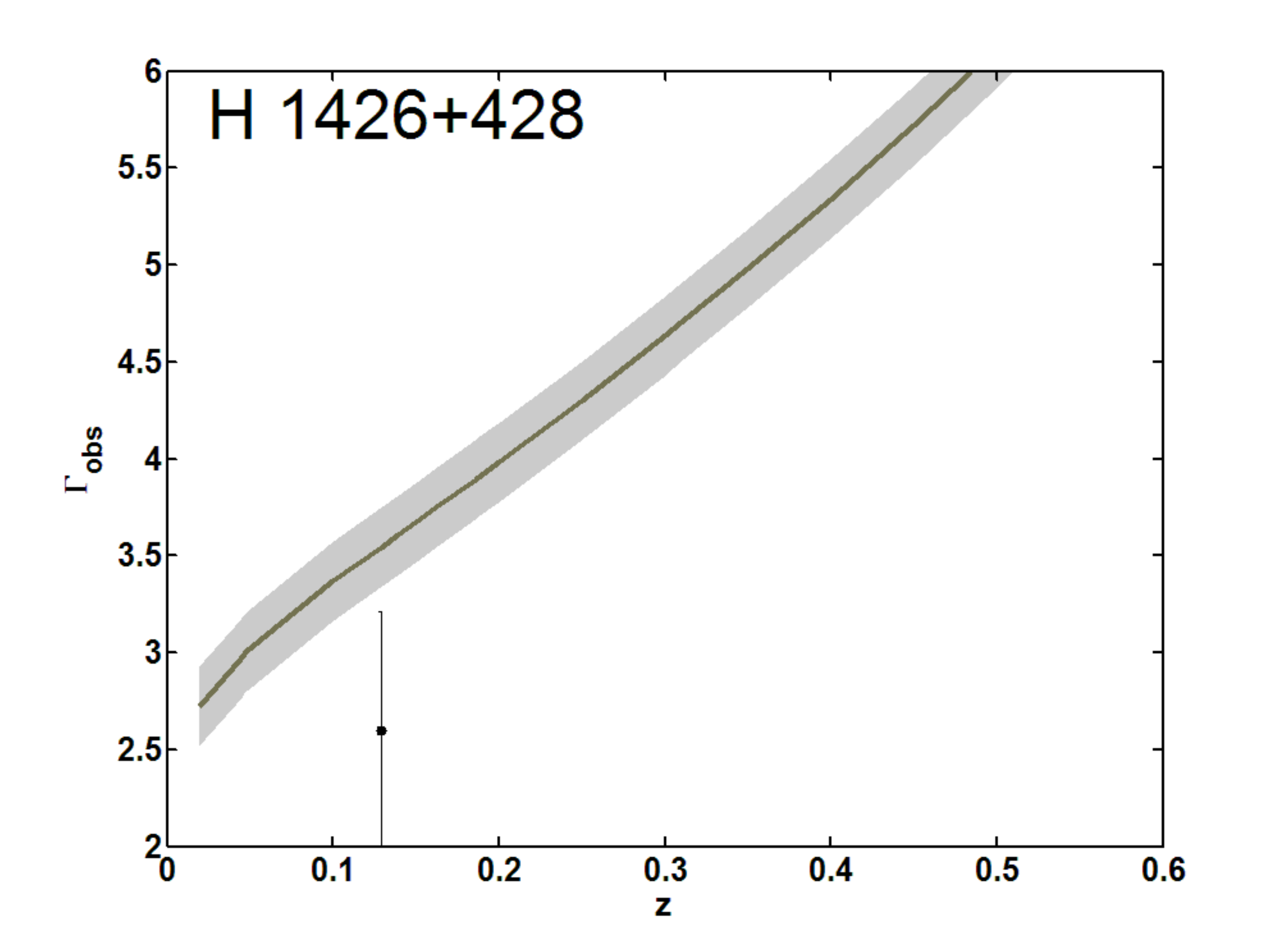}
\includegraphics[width=.49\textwidth]{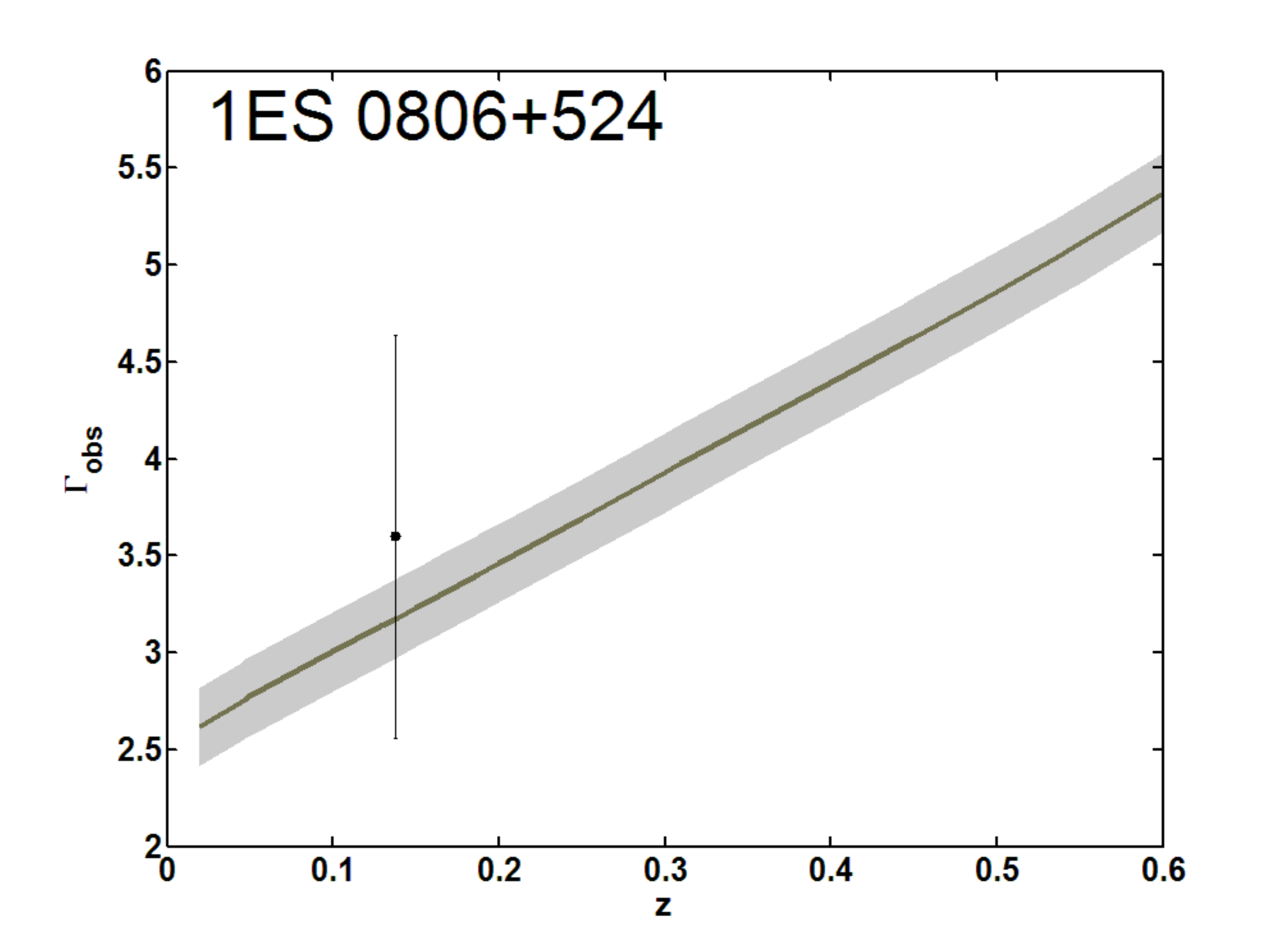}\includegraphics[width=.49\textwidth]{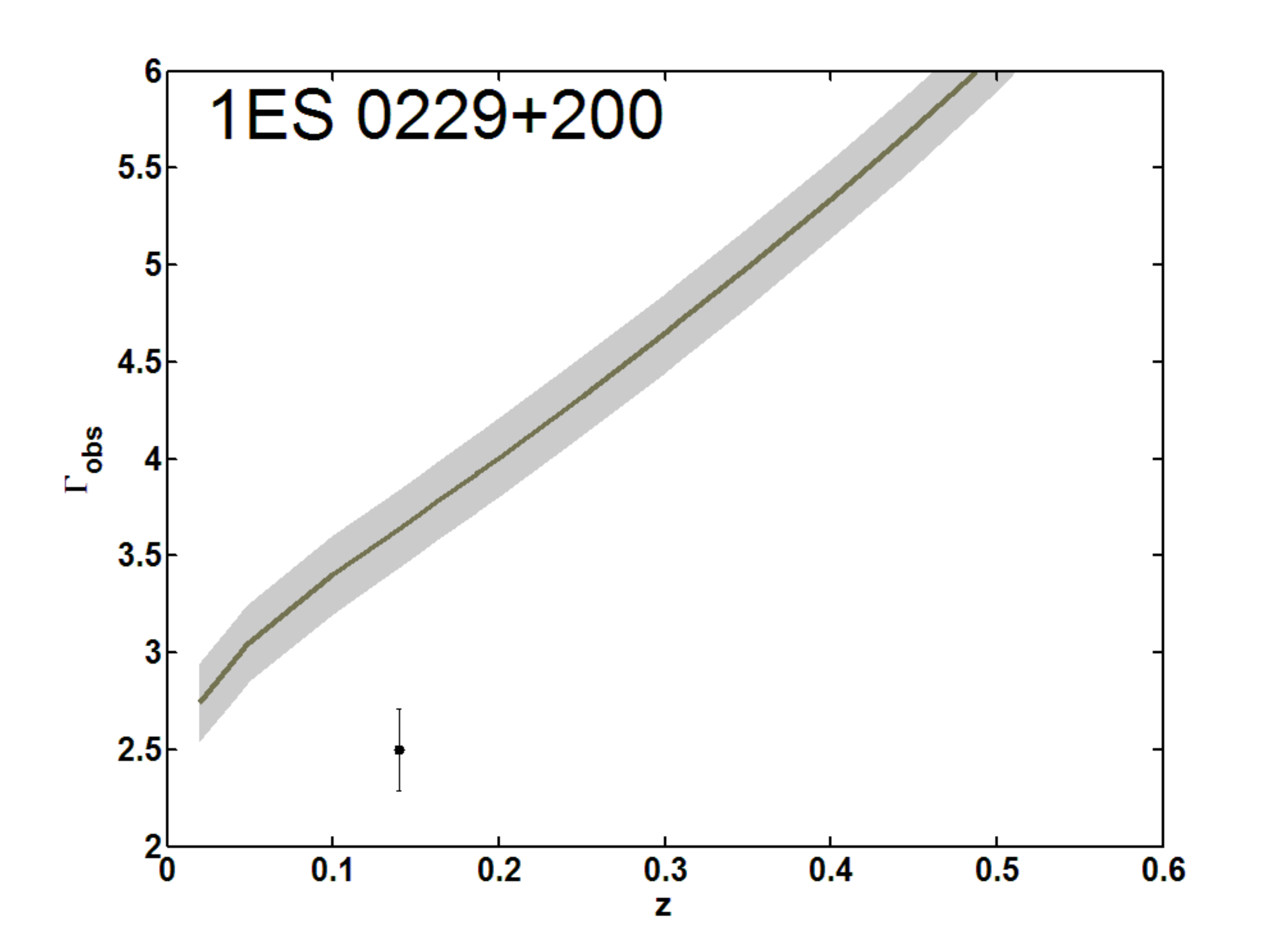}
\includegraphics[width=.49\textwidth]{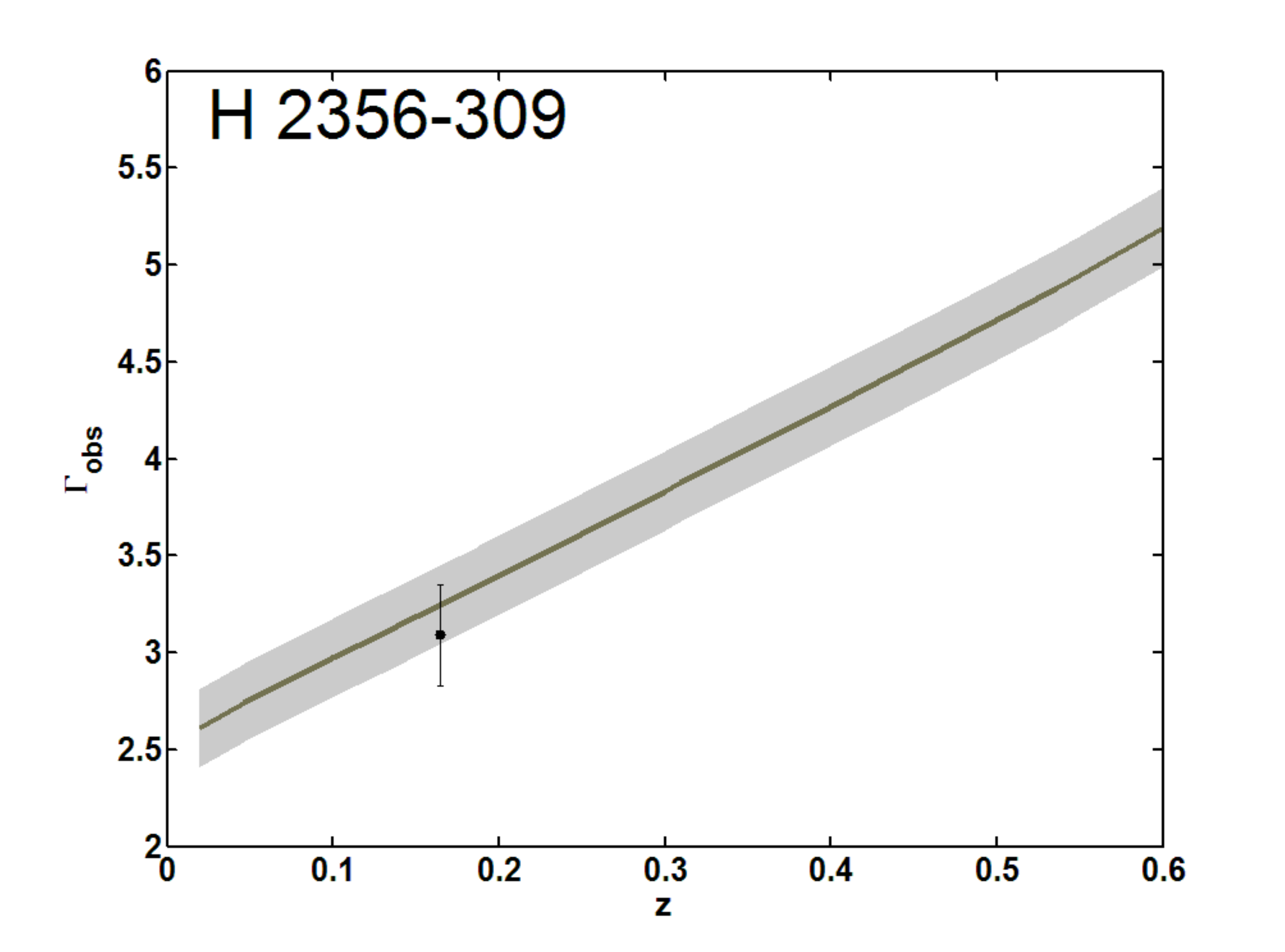}\includegraphics[width=.49\textwidth]{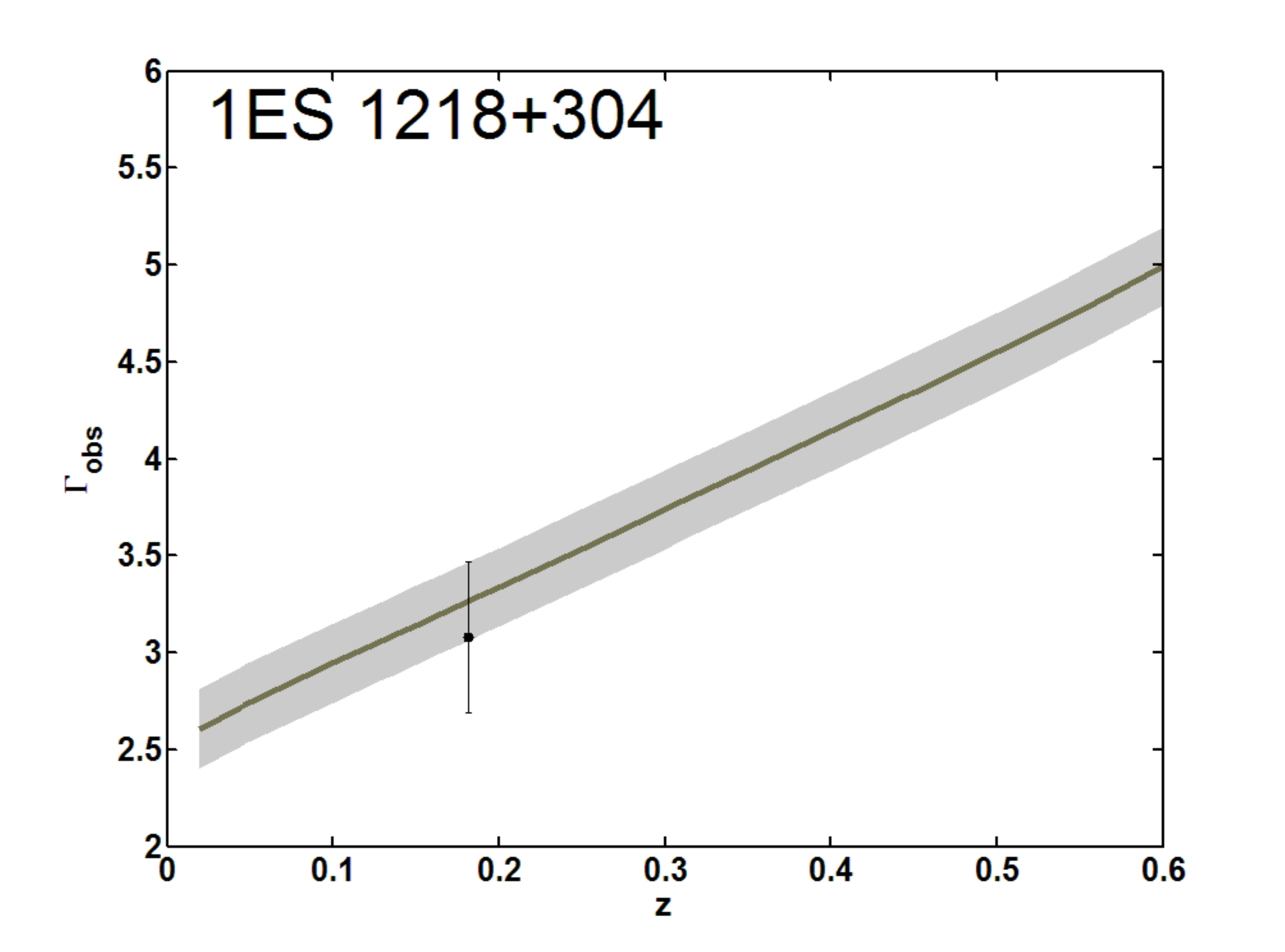}
\end{center}
\caption{\label{pendenze3.pdf} 
Behaviour of $\Gamma^{\rm DARMA}_{\rm obs}$ for the blazars PKS 2155-304, H 1426+428, 1ES 0806+524, 1ES 0229+200, H 2356-309 and 
1ES 1218+304. The solid black line corresponds to $\Gamma^{\rm DARMA}_{\rm em} = 2.51$ and the grey strip represents the range $2.31 < \Gamma^{\rm DARMA}_{\rm em} < 2.71$.}
\end{figure}

\begin{figure}                             
\begin{center}
\includegraphics[width=.49\textwidth]{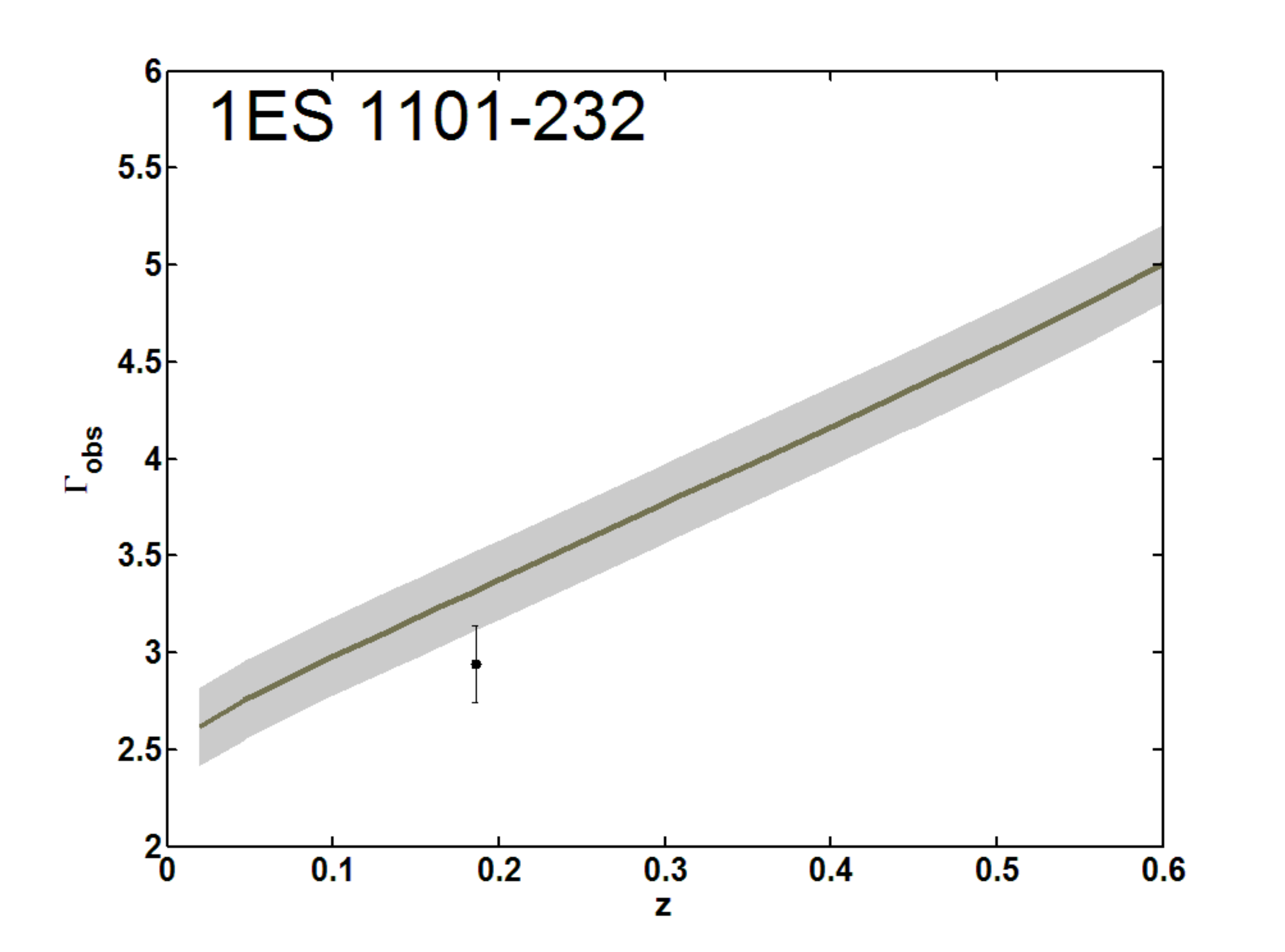}\includegraphics[width=.49\textwidth]{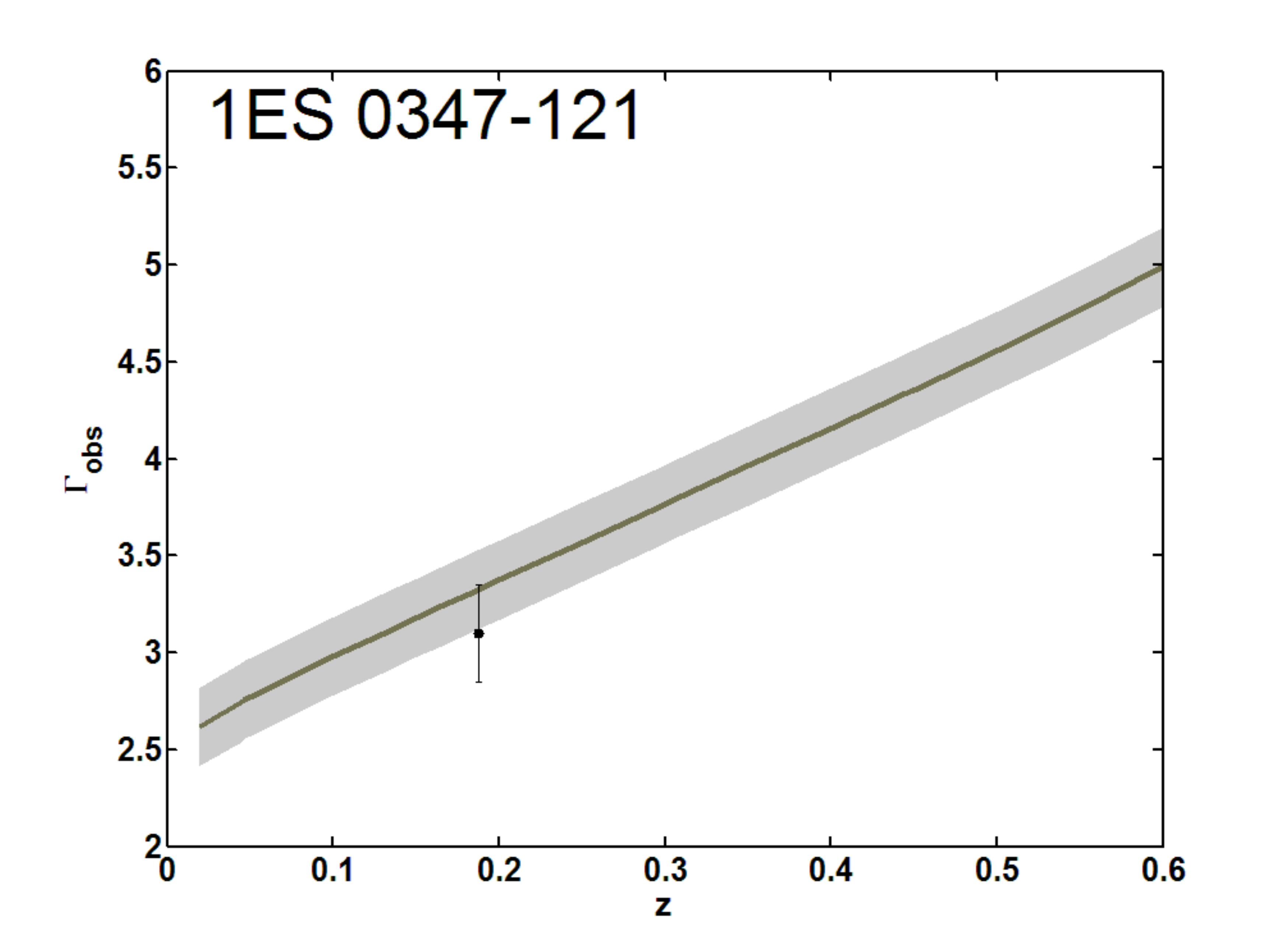}
\includegraphics[width=.49\textwidth]{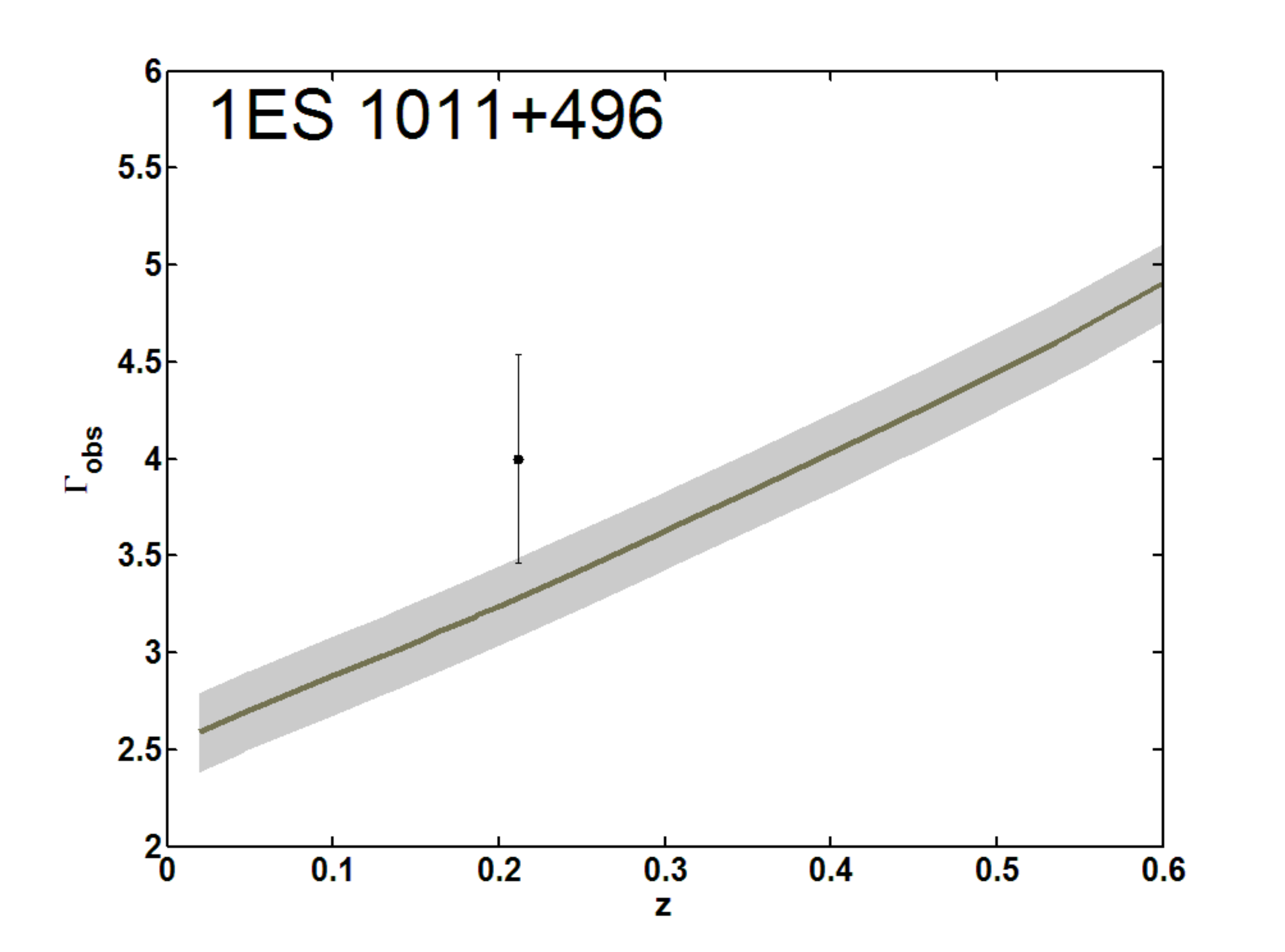}\includegraphics[width=.49\textwidth]{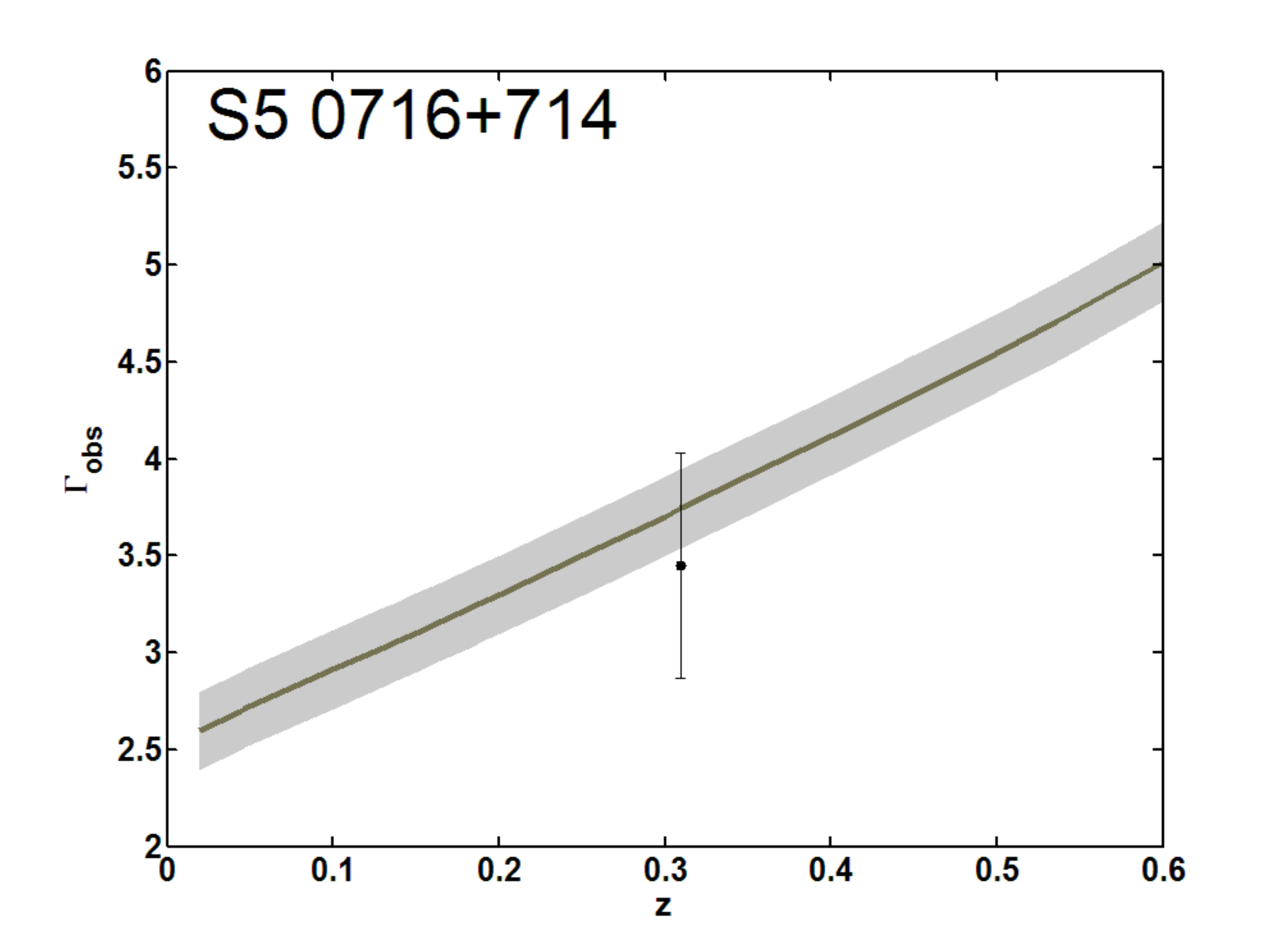}
\includegraphics[width=.49\textwidth]{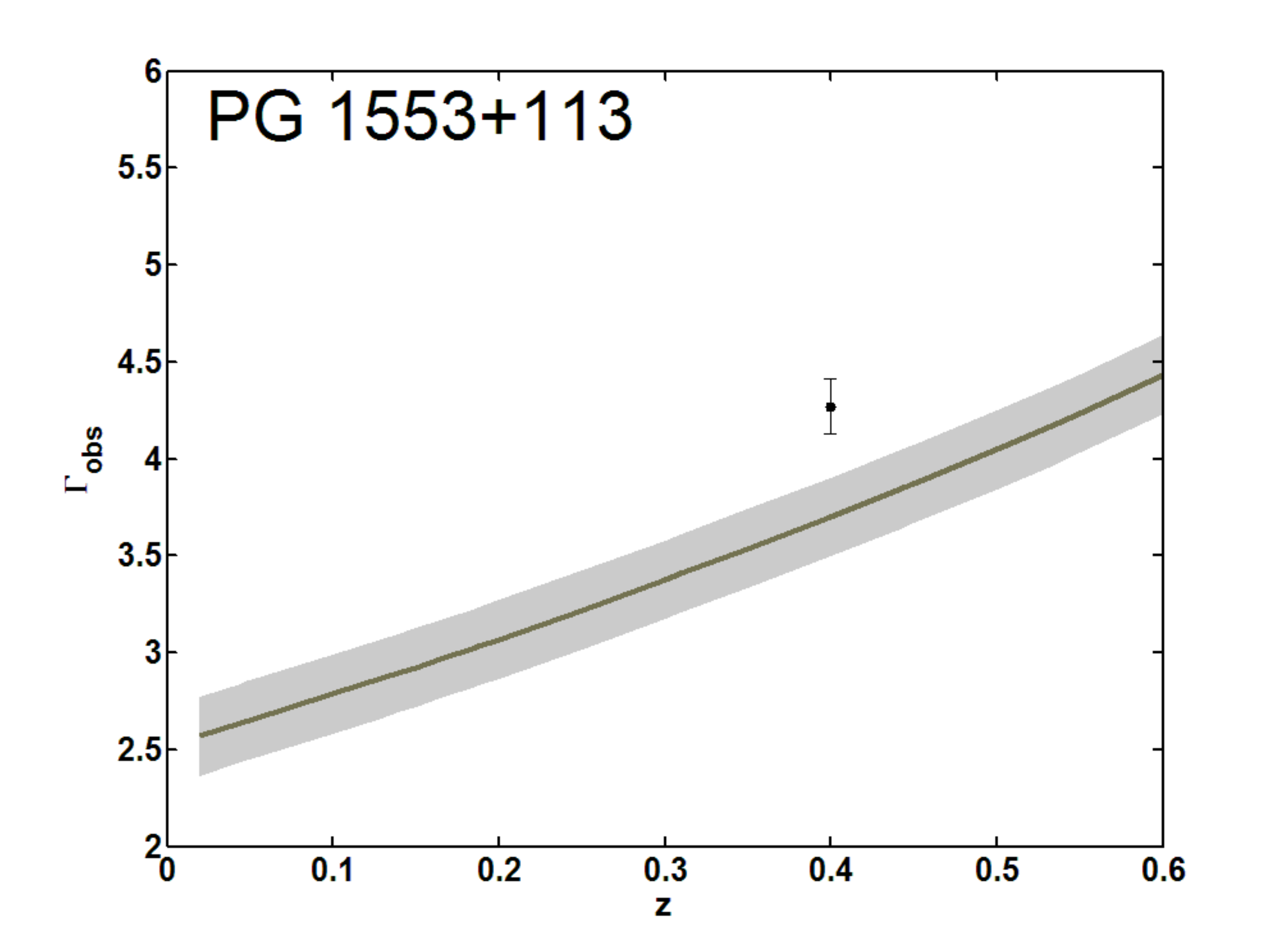}\includegraphics[width=.49\textwidth]{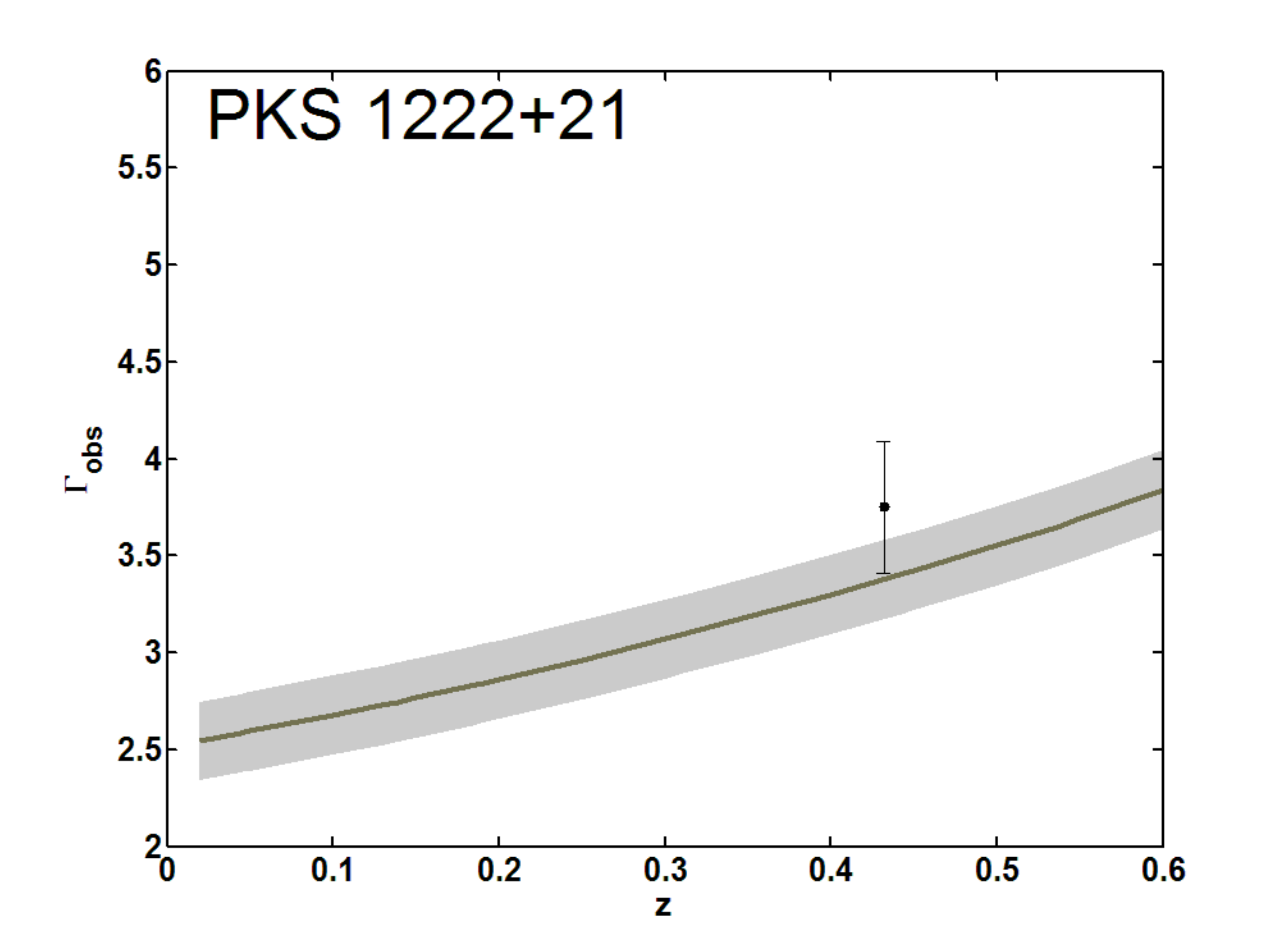}
\end{center}
\caption{\label{pendenze4.pdf} 
Behaviour of $\Gamma^{\rm DARMA}_{\rm obs}$ for the blazars 1ES 1101-232, 1ES 0347-121, 1ES 1011+496, S5 0716+714, PG 1553+113 and 
PKS 1222+21. The solid black line corresponds to $\Gamma^{\rm DARMA}_{\rm em} = 2.51$ and the grey strip represents the range $2.31 < \Gamma^{\rm DARMA}_{\rm em} < 2.71$.}
\end{figure}

\begin{figure}                
\begin{center}
\includegraphics[width=.49\textwidth]{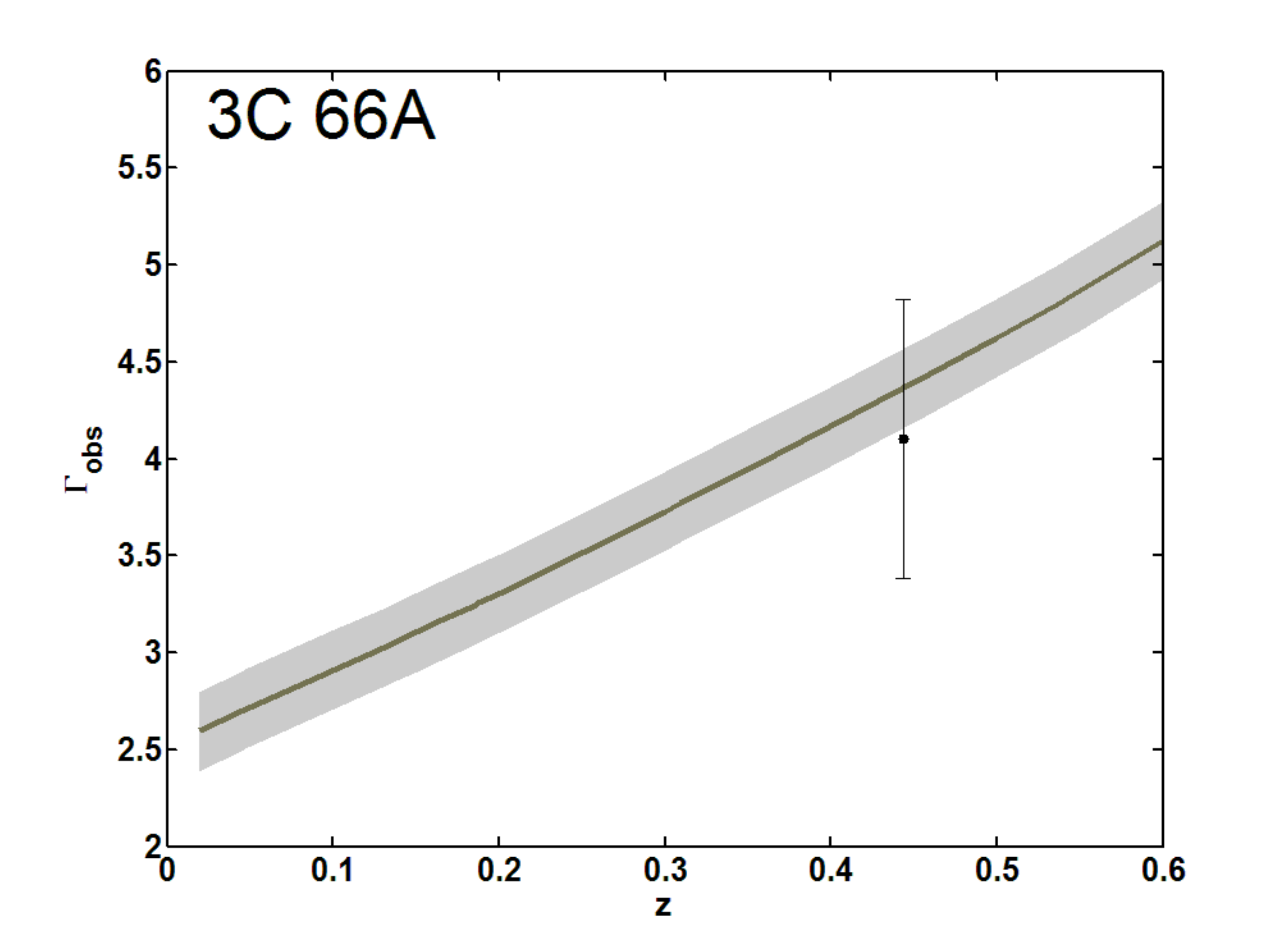}\includegraphics[width=.49\textwidth]{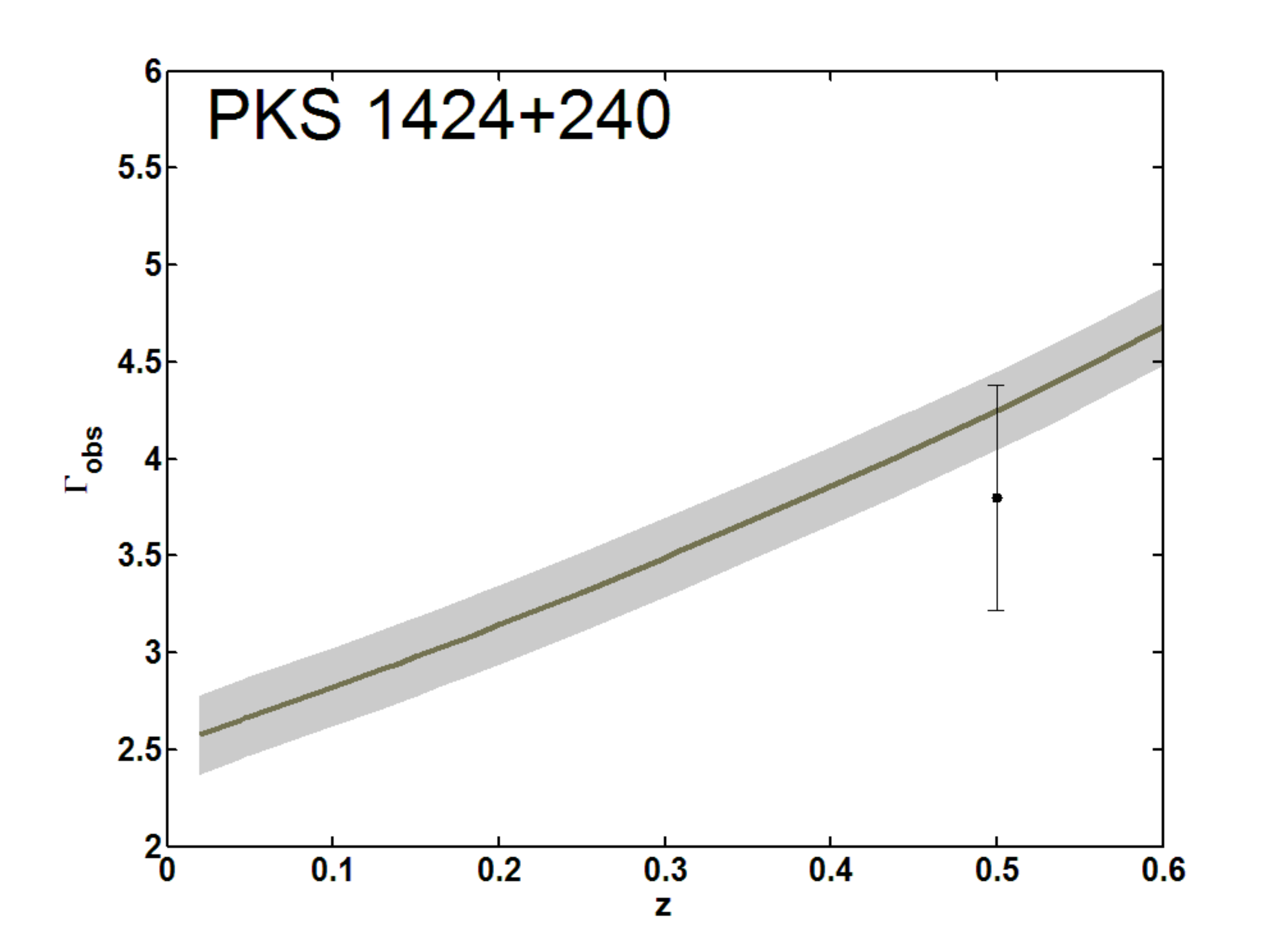}
\includegraphics[width=.49\textwidth]{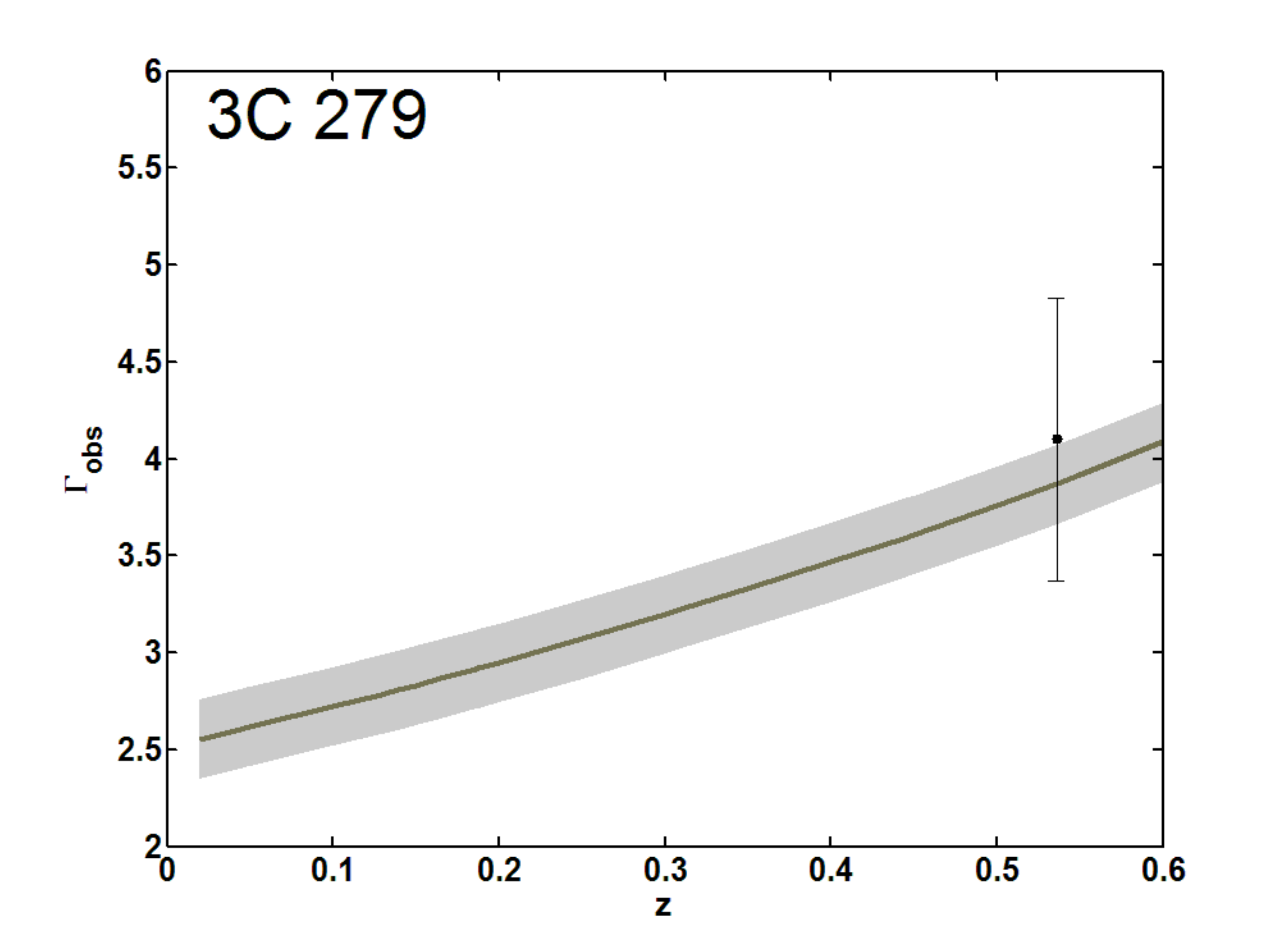}
\end{center}
\caption{\label{pendenze5.pdf} 
Behaviour of $\Gamma^{\rm DARMA}_{\rm obs}$ for the blazars 3C 66A, PKS 1424+240 and 3C 279. 
The solid black line corresponds to $\Gamma^{\rm DARMA}_{\rm em} = 2.51$ and the grey strip represents the range $2.31 < \Gamma^{\rm DARMA}_{\rm em} < 2.71$.}
\end{figure}

{  We are now ready to check this view by performing a fit to all observed VHE blazars. This is shown in Figures \ref{pendenze1.pdf} to \ref{pendenze5.pdf}, where the solid black line corresponds to $\Gamma_{\rm obs}^{\rm exp} (z)$ while the grey strip represents the range $\Gamma_{\rm obs}^{\rm exp} (z) \pm 0.2$. We stress that $\Gamma_{\rm obs}^{\rm exp} (z)$ is different for different sources, owing to the different observed energy range.} 

{  A look at those Figures shows that by assuming that {\it all} VHE blazars have $\Gamma_{\rm em}$ in a range of $2.51 \pm 0.2$ allows to fit observations of 19 sources out of a total of 27 ones. The role of photon-ALP oscillations is to partially offset EBL absorption, and thus the DARMA scenario departs from conventional physics only to the extent that EBL attenuation becomes important; for $z \geq 0.1$ 12 sources out of a total of 16 ones are successfully fitted, and for $z \geq 0.138$ the fit is successful for 10 blazars out of a total of 11.}

\section{CONCLUSIONS}

Very light ALPs are a generic prediction of many attempts to extend the Standard Model along different directions towards a more satisfactory fundamental theory of all elementary-particle interactions including gravity. We have systematically investigated the DARMA scenario in which the mechanism of photon-ALP oscillations triggered by large-scale magnetic fields is regarded as a means to effectively reduce the EBL attenuation affecting blazar observations above 
$100 \, {\rm GeV}$. Our assumptions can be summarized as follows:
\begin{itemize}
\item Large-scale magnetic fields exist with a cellular morphology characterized by a coherence lenth in the $1 - 10 \, {\rm Mpc}$ range and a strength not 
much smaller than the available upper bound $B_0 < 6 \, {\rm nG}$.
\item ALPs have to be very light in order to ensure that the strong-mixing regime is realized. The upper bound on their mass depends on the adopted value of the $a \gamma \gamma$ coupling constant $B_0/M$ -- see Table \ref{tab:axi} -- but in any case the condition $m < 5 \cdot 10^{- 10} \, {\rm eV}$ has to be met. This prevents the axion needed to solve the strong CP problem from playing any role in the present context.
\item The parameter $M$ is consistent but fairly close to the strongest upper bound $M > 10^{11} \, {\rm GeV}$ coming from observations of supernova SN1987a. We remark that this bound is however affected by a large uncertainty and exceeds by one order of magnitude the robust bound $M > 10^{10} \, {\rm GeV}$ coming both from theoretical considerations of star cooling and from the negative result of the CAST collaboration.
\end{itemize}

We predict that a boost factor of 10 in the photon survival probability with respect to conventional physics takes place for all VHE blazars observed so far well below the upper detection threshold of the planned CTA and HAWC water Cherenkov $\gamma$-ray observatory. Moreover, the energy $E_{10}$ at which such a boost factor occurs decreases as the source distance increases and becomes e.g. as low as $2 \, {\rm TeV}$ for the blazar 3C 279 at $z = 0.536$. Hence, our prediction can certainly be tested with the above planned detectors and possibly also with currently operating IACTs H.E.S.S., MAGIC, CANGAROO\,III, VERITAS as well as with the Extensive Air Shower arrays ARGO-YBJ and MILAGRO. 

We find it a remarkable fact that the DARMA scenario also offers a new interpretation of the observed VHE blazars, according to which the values of 
$\Gamma_{\rm em}$ for far-away VHE blazars are in the same ballpark of nearby ones and the large spread in the values of $\Gamma_{\rm obs}$ is mainly traced to the wide spread in the source distances. 

As is well known, weakly interacting massive particles (WIMPs) can be detected either indirectly through astrophysical effects or directly at the Large Hadron Collider (LHC). The situation of ALPs characteristic of the DARMA scenario is in a sense similar. Besides being detectable indirectly through the astrophysical effects discussed in this paper, they lend themselves to a direct detection {  either in the GammeV~\cite{gammaev} experiment at FERMILAB or more likely with planned photon regeneration experiment ALPS at DESY~\cite{ringwaldfuture}}, or else with large xenon scintillation detectors developed for dark matter searches~\cite{avignone}.  

{\it Note added in proof.} {  After submission of the present paper, we have become aware that our predicted lower-than-expected transparency of the Universe in the VHE band is supported by an independent result~\cite{meyer} which rests upon a new statistical analysis of all VHE blazars based on a Kolmogorov-Smirnov test in conjunction with the minimal EBL model~\cite{kneiske2010}.}

\section*{Acknowledgments}

Many enlightening conversations with our collaborators Oriana Mansutti and Massimo Persic are gratefully acknowledged. We also thank Dieter Horns and Fabrizio Tavecchio for informative discussions and Aldo Treves for several remarks.

\section{APPENDIX A}

We solve here the mathematical problem of finding the transfer matrix ${\cal U} (y, y_0; 0)$ associated with the reduced Sch\"odinger-like equation
\begin{equation}
\label{k3l1app}
\left( i \frac{d}{d y} + {\cal M} \right) \, \psi (y) = 0~,
\end{equation}
with
\begin{equation}
\label{k3w1app}
\psi (y) \equiv \left(
\begin{array}{c}
A_x(y) \\
A_z(y) \\
a(y) \\
\end{array}
\right)
\end{equation}
as in the text, and mixing matrix of the form
\begin{equation}
\label{a9app}
{\cal M} = \left(
\begin{array}{ccc}
s & 0 & 0 \\
0 & t & v \\
0 & v & u \\
\end{array}
\right)~,
\end{equation}
where the coefficients $s$, $t$, $u$ and $v$ are supposed to be complex numbers. 

We start by diagonalizing ${\cal M}$. Its eigenvalues are
\begin{equation}
\label{a91212a1app}
{\lambda}_{1} = s~,
\end{equation}
\begin{equation}
\label{a91212a2app}
{\lambda}_{2} = \frac{1}{2} \left( t + u - \sqrt{\left(t - u \right)^2 + 4 \, v^2} \right)~, 
\end{equation}
\begin{equation}
\label{a91212a3app}
{\lambda}_{3} = \frac{1}{2} \left( t + u + \sqrt{\left(t - u \right)^2 + 4 \, v^2} \right)~, 
\end{equation}
and it is straightforward to check that the corresponding eigenvectors can be taken to be
\begin{equation}
\label{k3w1Aappa1}
X_1 = \left(
\begin{array}{c}
1 \\
0 \\
0 \\
\end{array}
\right)~,
\end{equation}
\begin{equation}
\label{k3w1Aappa2}
X_2 = \left(
\begin{array}{c}
0 \\
v \\
{\lambda}_2 - t \\
\end{array}
\right)~,
\end{equation}
\begin{equation}
\label{k3w1Aappa3}
X_3 = \left(
\begin{array}{c}
0 \\
v \\
{\lambda}_3 - t  \\
\end{array}
\right)~.
\end{equation}
Correspondingly, any solution of Eq. (\ref{k3l1app}) can be represented in the form
\begin{equation}
\label{a91212a1appQ}
\psi (y) = c_1 \, X_1 \, e^{i {\lambda}_1 \, \left(y - y_0 \right)} + c_2 \, X_2 \, e^{i {\lambda}_2 \, \left(y - y_0 \right)} + c_3 \, X_3 \, e^{i {\lambda}_3 \, 
\left(y - y_0 \right)}~,
\end{equation}
where $c_1$, $c_2$, $c_3$ and $y_0$ are arbitrary constants. As a consequence, the solution with initial condition 
\begin{equation}
\label{k3w1appW}
\psi (y_0) \equiv \left(
\begin{array}{c}
A_x(y_0) \\
A_z(y_0) \\
a(y_0) \\
\end{array}
\right)
\end{equation}
emerges from Eq. (\ref{a91212a1appQ}) for
\begin{equation}
\label{a91212a1appq1}
c_{1} = A_x(y_0)~,
\end{equation}
\begin{equation}
\label{a91212a1appq2}
c_{2} = \frac{{\lambda}_3 - t}{v ({\lambda}_3 - {\lambda}_2 )} \, A_z(y_0) - \frac{1}{{\lambda}_3 - {\lambda}_2} \, a(y_0)~,
\end{equation}
\begin{equation}
\label{a91212a1appQ3}
c_{3} = - \, \frac{{\lambda}_2 - t}{v ({\lambda}_3 - {\lambda}_2 )} \, A_z(y_0) + \frac{1}{{\lambda}_3 - {\lambda}_2} \, a(y_0)~.
\end{equation}
It is a simple exercize to recast the considered solution into the form
\begin{equation}
\label{k3lasqapp}
\psi (y) = {\cal U} (y, y_0; 0) \, \psi (y_0) 
\end{equation}
with
\begin{equation}
\label{mravvq2abcapp}
{\cal U} (y, y_0; 0) = e^{i {\lambda}_1 (y - y_0)} \, T_1 (0) + e^{i {\lambda}_2 (y - y_0)} \, T_2 (0) + e^{i {\lambda}_3 (y - y_0)} \, T_3 (0)~, 
\end{equation}
where we have set
\begin{equation}
\label{mravvq1app}
T_1 (0) \equiv
\left(
\begin{array}{ccc}
1 & 0& 0 \\
0 & 0 & 0 \\
0 & 0 & 0
\end{array}
\right)~,
\end{equation}
\begin{equation} 
\label{mravvq2app}
T_2 (0) \equiv
\left(
\begin{array}{ccc}
0 & 0 & 0 \\ 
0 & \frac{{\lambda}_3 - t}{{\lambda}_3 - {\lambda}_2} & - \, \frac{v}{{\lambda}_3 - {\lambda}_2} \\
0 & \frac{({\lambda}_2 - t) ({\lambda}_3 - t)}{v ({\lambda}_3 - {\lambda}_2)} &  - \, \frac{{\lambda}_2 - t}{{\lambda}_3 - {\lambda}_2}
\end{array}
\right)~,
\end{equation} 
\begin{equation}
\label{mravvq3app}
T_3 (0) \equiv 
\left(
\begin{array}{ccc}
0 & 0 & 0 \\ 
0 & - \, \frac{{\lambda}_2 - t}{{\lambda}_3 - {\lambda}_2} & \frac{v}{{\lambda}_3 - {\lambda}_2} \\
0 & - \, \frac{({\lambda}_2 - t) ({\lambda}_3 - t)}{v ({\lambda}_3 - {\lambda}_2)} &  \frac{{\lambda}_3 - t}{{\lambda}_3 - {\lambda}_2}
\end{array}
 \right)~,
\end{equation}
from which it follows that the desired transfer matrix is just ${\cal U} (y, y_0; 0) $ as given by Eq. (\ref{mravvq2abcapp}).

\section{APPENDIX B}

It proves very useful for illustrative purposes to have the approximate behaviour of $\tau_{\gamma}(E_0, z)$ in an analytic form. This goal can be achieved by taking advantage from the fact that $\sigma_{\gamma \gamma}(E,\epsilon,\varphi)$ is maximized when condition (\ref{eq.sez.urto-1}) is met (we restrict ourselves to head-on collisions for simplicity).

Accordingly, the crudest attempt to estimate the dominant contribution to the optical depth would be to approximate the $\epsilon (z)$ integration by the product 
of ${\sigma}_{\gamma \gamma}^{\rm max}$ times $n_{\gamma}(\epsilon(z), z) $ as evaluated for that particular value of $\epsilon (z)$ selected by condition (\ref{eq.sez.urto-1}) for fixed $E$. This amounts to insert the Dirac delta $\delta ( \epsilon (z^{\prime})/{\rm eV} - 500 \, {\rm GeV}/E (z^{\prime}) )$ into the r.h.s. of Eq. (\ref{eq:tau}), which leads to 
\begin{equation}
\label{lungh19122010}
\tau_{\gamma}(E_0, z) \simeq 2.25 \cdot 10^3 \int_0^{z} {\rm d} z ~ \frac{n_{\gamma} \left(\epsilon (z) , z \right)}{\left(1 + z \right) 
\left[ 0.7 + 0.3 \left(1 + z \right)^3 \right]^{1/2}} \  {\rm cm}^3 \, {\rm eV}~,
\end{equation}
with
\begin{equation}
\label{lungh19122010S}
\epsilon (z) \simeq \frac{1}{1 + z} \left( \frac{500 \, {\rm GeV}}{E_0} \right) {\rm eV}~. 
\end{equation}
Unfortunately, experience with this problem shows that the resulting $E_0$-dependence of $\tau_{\gamma}(E_0, z)$ is too steep, and since $P_{\gamma \to \gamma}^{\rm CP} (E_0,z)$ depends exponentially on $\tau_{\gamma}(E_0, z)$ this approximation is doomed to failure.

A more satisfactory conclusion emerges by exploiting a popular approximation~\cite{coppiaharonian} which amounts to replace the $\epsilon (z)$ integration in Eq. (\ref{eq:tau}) by the product of ${\sigma}_{\gamma \gamma}^{\rm max}$ times $\epsilon (z) \, n_{\gamma}(\epsilon(z), z) $ at the particular value of $\epsilon (z)$ dictated by condition (\ref{eq.sez.urto-1}) for fixed $E$. We find in this way
\begin{eqnarray}
\label{lungh19012011}
\tau_{\gamma}(E_0, z) \simeq 2.25 \cdot 10^3 \left( \frac{500 \, {\rm GeV}}{E_0} \right) \int_0^{z}  \frac{{\rm d} z}{\left(1 + z \right)^2 \left[ 0.7 + 0.3 \left(1 + z \right)^3 \right]^{1/2}} \ \times  \\
\nonumber 
 \times \, n_{\gamma} \left( \frac{\left(500 \, {\rm GeV} / E_0 \right) {\rm eV}}{1 + z }, z \right) \, {\rm cm}^3 \, {\rm eV}~,       \ \ \ \ \ \ \ \ \ \ \ \ \ \ \ \ \ \ \ \ \ \ 
\end{eqnarray}
where Eq. (\ref{lungh19122010S}) has been used.

\begin{figure}
\centering
\includegraphics[width=.70\textwidth]{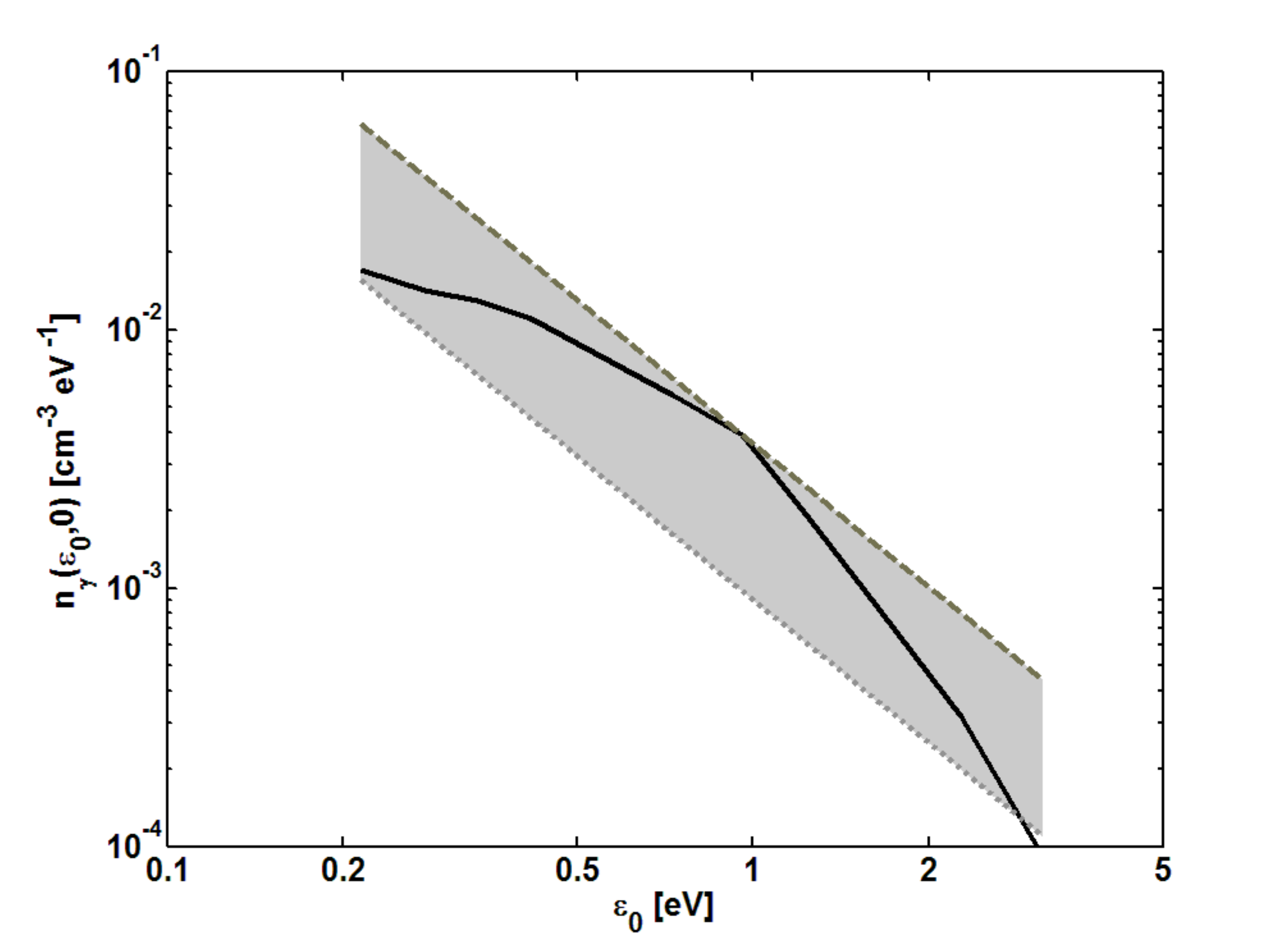}
\caption{\label{EBL.pdf} 
The spectral photon number density in the present Universe $n_{\gamma}( \cdot ,0)$ is plotted versus the energy $\epsilon_0$  in the energy range $0.25 - 2.5 \, {\rm eV}$. The solid line represents the result of the FRV model. The dotted and dashed lines correspond to the lower ($\alpha = 0.9$) and upper 
($\alpha = 3.6$) limit, respectively, of our power-law approximation defined in Eq. (\ref{pndq1x}).}
\end{figure}

Owing to Eq. (\ref{lungh19012011}), the derivation of an approximate analytic behaviour of the optical depth requires an approximate analytic expression for $n_{\gamma} (\epsilon (z), z)$. Unfortunately,  the FRV model does not give an analytic form for $n_{\gamma}(\epsilon_0,0)$ but provides a plot of $\epsilon_0 \, n_{\gamma} (\epsilon_0, 0 )$ versus $\epsilon_0$ (see their Figure 4). The corresponding plot of $n_{\gamma} (\epsilon_0, 0 )$ as a function of $\epsilon_0$ is reproduced by the solid line in Figure \ref{EBL.pdf} for the EBL energy range relevant for the observed blazas, namely $0.25 \, {\rm eV} < \epsilon_0 < 2.5 \, {\rm eV}$. A look at Figure \ref{EBL.pdf} shows that within the considered energy range the spectral energy distribution of the EBL departs from a power-law behaviour, owing to the emission bump resulting from the integrated emission of the low-mass stellar population that remained close to the main-sequence over cosmological times. For this reason, we approximate the FRV result for $n_{\gamma} (\epsilon_0,0)$ with the shadowed linear strip shown in Figure \ref{EBL.pdf} enveloping the exact behaviour, which is ex\-pres\-sed by the following power-law representation 
\begin{equation}
\label{pndq1x}
n^{\rm app}_{\gamma}(\epsilon_0, 0) \simeq 10^{-3} \, \alpha  \left(\frac{ {\rm eV}}{\epsilon_0} \right)^{1.85} \, {\rm cm}^{-3} \, {\rm eV}^{-1}~,
\end{equation}
with the constant $\alpha$ in the range $0.9 \leq {\alpha} \leq 3.6$ so as to enclose the FRV curve. 

Evolutionary effects in the EBL spectral energy distribution can be taken into account as follows. Besides redshifting all energies in proportion of $1+z$, the cosmic expansion dilutes the EBL by a factor $(1+z)^3$ similarly to what happens for the CMB. But in addition the EBL spectral photon number density changes because of the intrinsic evolution of the galactic population over cosmic times. A quantitative analysis~\cite{raueazin} shows that the EBL photon number density acquires an extra factor $(1+z)^{-1.2}$ as long as $z < 1$, which is appropriate to our case. On the whole, the spectral photon number density $n^{\rm app}_{\gamma}(\epsilon (z) ,z)$ of the EBL at redshift $z$ is related to $n^{\rm app}_{\gamma}(\epsilon_0,0)$ by
\begin{equation}
\label{pndq1}
n^{\rm app}_{\gamma}(\epsilon (z) ,z) \, {\rm d}{\epsilon} (z)  \simeq \left(1 + z \right)^{1.8} \, n^{\rm app}_{\gamma}(\epsilon_0,0) \, {\rm d}{\epsilon}_0~,
\end{equation}
which yields
\begin{equation}
\label{pndq1a}
n^{\rm app}_{\gamma}(\epsilon (z) ,z) \simeq \left(1 + z \right)^{0.8} \, n^{\rm app}_{\gamma} \left( \frac{\epsilon (z)}{1 + z}, 0 \right)~,
\end{equation}
namely
\begin{equation}
\label{pndq1a05012011}
n^{\rm app}_{\gamma} \left( \frac{\left(500 \, {\rm GeV} / E_0 \right) {\rm eV}}{1 + z }, z \right) \simeq \left(1 + z \right)^{0.8} \, n^{\rm app}_{\gamma} \left( \frac{\left(500 \, {\rm GeV} / E_0 \right) {\rm eV}}{\left(1 + z \right)^2}, 0 \right)~,
\end{equation}
thanks to Eq. (\ref{lungh19122010S}). In particular, Eq. (\ref{pndq1x}) leads to
\begin{equation}
\label{pndq1a19012011}
n_{\gamma}^{\rm app} \left( \frac{\left(500 \, {\rm GeV} / E_0 \right) {\rm eV}}{1 + z }, z \right) \simeq 10^{-3} \, \alpha 
\left(\frac{E_0}{500 \, {\rm GeV}} \right)^{1.85} \left(1 + z \right)^{4.5} \, {\rm cm}^{- 3} \, {\rm eV}^{- 1}~.
\end{equation}

\begin{figure}
\centering
\includegraphics[width=.70\textwidth]{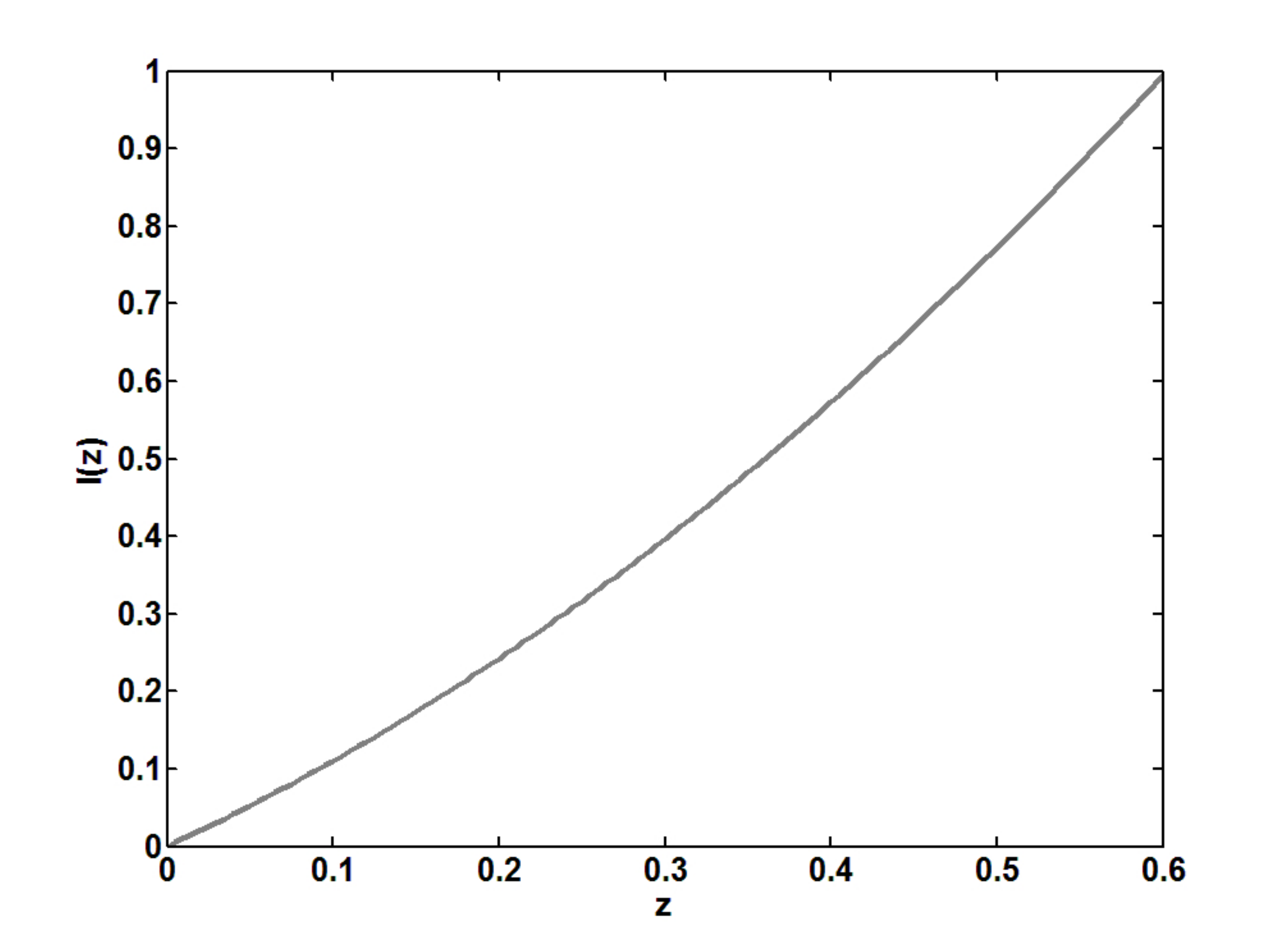}
\caption{\label{I(z).pdf} 
Plot of the behaviour of $I(z)$.}
\end{figure}

The approximate evaluation of the optical depth $\tau^{\rm app}_{\gamma}(E_0,z)$ amounts to insert Eq. (\ref{pndq1a19012011}) into Eq. (\ref{lungh19012011}). Correspondingly, we get
\begin{equation}
\label{pndq1x20122010Q} 
\tau^{\rm app}_{\gamma}(E_0, z) \simeq 2.25 \, \alpha \left(\frac{E_0}{500 \, {\rm GeV}} \right)^{0.85} \, I(z)~,
\end{equation}
where we have set
\begin{equation}
\label{pndq1x20122010Q1}
I(z) \equiv \int_0^{z} {\rm d} z^{\prime} ~ \frac{(1 + z^{\prime})^{2.5}}{\left[ 0.7 + 0.3 \left(1 + z^{\prime} \right)^3  \right]^{1/2}}~.
\end{equation}
This integral has been evaluated numerically and its behaviour is depicted in Figure \ref{I(z).pdf}, which shows that up to $z \simeq 0.1$ it goes linearly with $z$ but it increases more rapidly for larger redshifts.

\begin{figure}
\centering
\includegraphics[width=.70\textwidth]{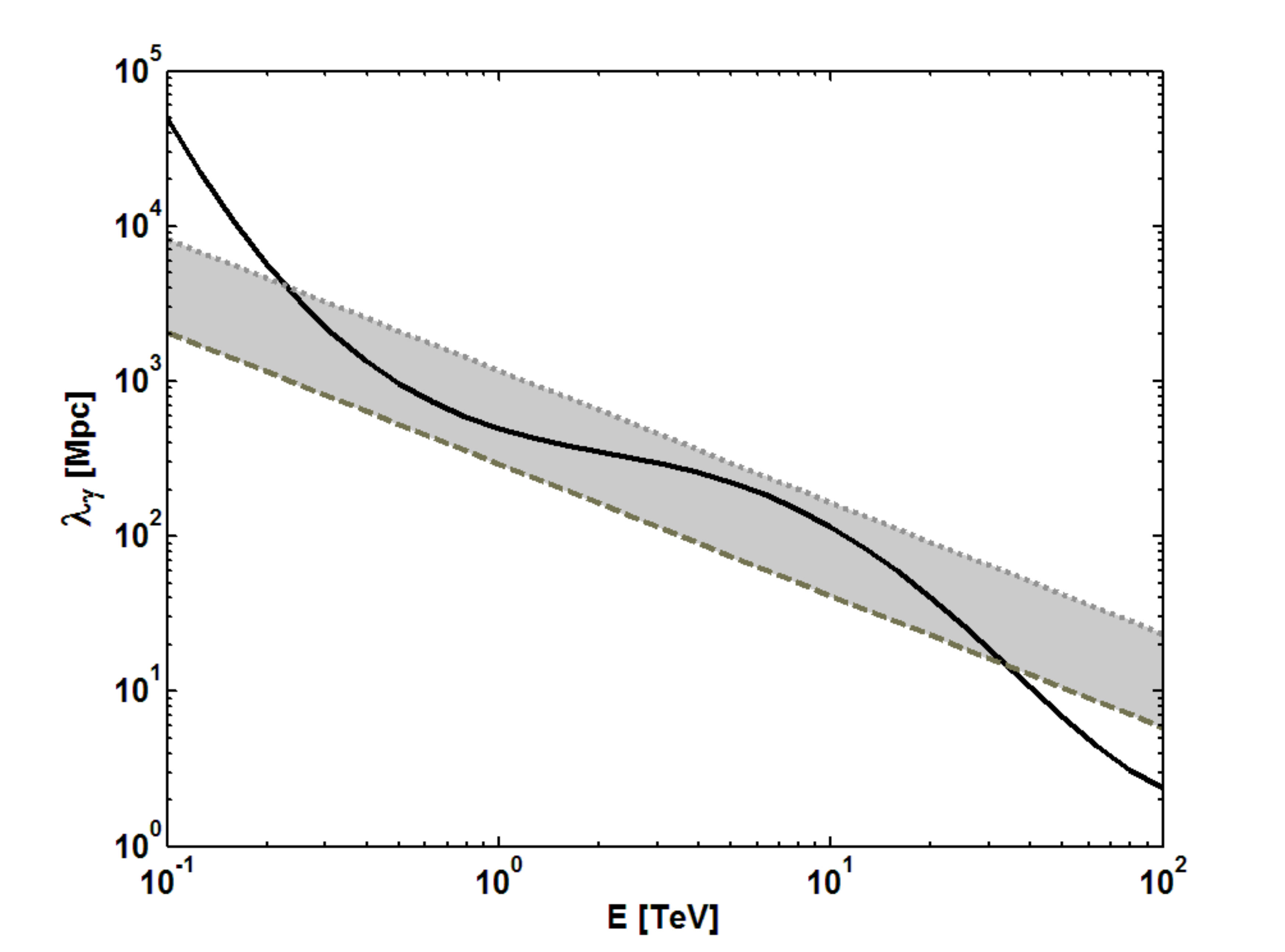}
\caption{\label{MFPapp.pdf} 
The approximate pair-production mean free path ${\lambda}_{\gamma}^{\rm app}$ of a VHE photon is plotted versus its energy $E$ and it is represented by the shadowed area as the parameter $\alpha$ varies in the range 0.9 -- 3.6. The dotted and dashed lines correspond to $\alpha = 0.9$ and $\alpha = 3.6$, respectively. Superimposed is the exact result obtained within the FRV model and shown in Figure \ref{fig:france}.}
\end{figure}

Before leaving this issue it is worthwhile to show how Eq. (\ref{lungh26122010S}) is recovered in the limit of small $z$ where cosmological effects become irrelevant. In such a situation the source distance is $D = c z /H_0$, which allows us to write the optical depth as
\begin{equation}
\label{pndq1x10012011QA} 
\tau^{\rm app}_{\gamma}(E, D) = 5.25 \cdot 10^{- 4} \, \alpha \left(\frac{E}{500 \, {\rm GeV}} \right)^{0.85} \left(\frac{D}{{\rm Mpc}} \right)~,
\end{equation}
with the replacement $E_0 \to E$. So, we see that $\tau^{\rm app}_{\gamma}(E, D) \propto D$ in agreement with Eq. (\ref{lungh26122010S}), which entails that in the present approximation the mean free path for $\gamma \gamma \to e^+ e^-$ is given by
\begin{equation}
\label{lungh04012011SS}
{\lambda}_{\gamma}^{\rm app}(E) = 1.90 \cdot 10^{3} \, {\alpha}^{- 1}  \left(\frac{500 \, {\rm GeV}}{E} \right)^{0.85} \, {\rm Mpc}~.
\end{equation}
This quantity is plotted in Figure \ref{MFPapp.pdf} as a function of $E$, where it is represented by the shadowed region between the dotted line corresponding to $\alpha = 0.9$ and the dashed line corresponding to $\alpha = 3.6$. The solid curve yields ${\lambda}_{\gamma} (E)$ as evaluated exactly within the FRV model and exhibited in Figure \ref{fig:france}. We see that the present approximation is indeed consistent with the result of the FRV model for 
$0.2 \, {\rm TeV} < E < 2 \, {\rm TeV}$.

\begin{thebibliography}{9}
\bibitem{automatic}
We recall that by automatic symmetry it is meant any global symmetry which is present in the most general renormalizable Lagrangian invariant under the gauge group.
\bibitem{remarkA}   
Because Eq. (\ref{a006012011A1}) has to reduce to Eq. (\ref{a006012011A}) in the limit $z \to 0$, the constant $K$ is the same in both equations.
\bibitem{remark2}
We will see later that the situation is more involved, but at this stage we can safely ignore additional complications.
\bibitem{upperb} 
Because $\omega_{\rm pl}$ increases with $z$, we have evaluated it at $z = 1$ to be conservative. Similarly, since also $B$ increases with $z$, we have conservatively taken $B_0$ rather than $B$ in the r.h.s. of condition (\ref{upperbound24012011a}).
\end{thebibliography}
\end{document}